\documentclass[journal,12pt,onecolumn,draftclsnofoot]{IEEEtran}
\usepackage{amsmath,amsfonts,amsthm}
\usepackage{algorithmic}
\usepackage{algorithm}
\usepackage{array}
\usepackage{textcomp}
\usepackage{stfloats}
\usepackage{url}
\usepackage{verbatim}
\usepackage{graphicx}
\usepackage{cite}
\usepackage{bm}
\usepackage[nolist]{acronym}
\begin{acronym}[ACRONYM]
\acro{3GPP}{the 3rd Generation Partnership Project}
\acro{3D}{three dimensional}
\acro{5G}{fifth generation}
\acro{6G}{sixth generation}
\acro{ADC}{analog-to-digital converter}
\acro{AI}{artificial intelligence}
\acro{AR}{augmented reality}
\acro{AoA}{angle-of-arrival}
\acro{AoD}{angle-of-departure}
\acro{AWGN}{additive white Gaussian noise}
\acro{BS}{base station}
\acro{CF}{crest factor}
\acro{CFO}{carrier frequency offset}
\acro{CNN}{convolutional neural network}
\acro{CSI}{channel state information}
\acro{CU}{central unit}
\acro{CP}{cyclic prefix}
\acro{CRLB}{Cram\'er-Rao Lower Bound}
\acro{DAC}{digital-to-analog converter}
\acro{DL}{downlink}
\acro{DoF}{degrees of freedom}
\acro{DPD}{digital predistortion}
\acro{DU}{distributed unit}
\acro{D-MIMO}{distributed multiple-input multiple-output}
\acro{EER}{energy efficiency ratio}
\acro{EFIM}{equivalent Fisher information matrix}
\acro{EIRP}{effective isotropic radiated power}
\acro{ELAA}{electrically large aperture array}
\acro{EM}{electromagnetic}
\acro{FD}{full duplex}
\acro{FIM}{Fisher information matrix}
\acro{GDOP}{geometric dilution of precision}
\acro{gNB}{gNodeB}
\acro{GNSS}{global navigation satellite system}
\acro{HMI}{human-machine interaction}
\acro{IQ}{in-phase and quadrature}
\acro{IP}{incidence point}
\acro{JCS}{Joint Communication and Sensing}
\acro{IBFD}{inband full-duplex}
\acro{ISAC}{integrated sensing and communication}
\acro{JRC}{joint radar and communication}
\acro{JRC2LS}{joint radar communication, computation, localization, and sensing}
\acro{ICI}{inter-carrier interference}
\acro{IOO}{indoor open office}
\acro{IoT}{Internet of Things}
\acro{IRN}{infrastructure reference node}
\acro{ISG}{industry specification group}
\acro{JCAS}{joint communications and sensing}
\acro{KPI}{key performance indicator}
\acro{KVI}{key value indicator}
\acro{LCA}{life cycle analysis}
\acrodefplural{LCA}{life cycle analyses}
\acro{LIS}{large intelligent surface}
\acro{LoS}{line-of-sight}
\acro{LNA}{low noise amplifier}
\acro{MAC}{medium access control}
\acro{MIMO}{multiple-input multiple-output}
\acro{ML}{machine learning}
\acro{mmWave}{millimeter-wave}
\acro{MTT}{multi-target tracking}   
\acro{NF}{network function}
\acro{NLoS}{non-line-of-sight}
\acro{NR}{new radio}
\acro{NTN}{non-terrestrial network}
\acro{OFDM}{orthogonal frequency-division multiplexing}
\acro{OFDMA}{orthogonal frequency-division multiple access}
\acro{OTFS}{orthogonal time-frequency-space}
\acro{PAPR}{peak-to-average power ratio}
\acro{PRS}{positioning reference signal}
\acro{PHY}{physical layer}
\acro{QoS}{Quality of Service}
\acro{RAN}{radio access network}
\acro{RAT}{radio access technology}
\acro{RedCap}{reduced capacity}
\acro{RF}{radio frequency}
\acro{RIS}{reconfigurable intelligent surface}
\acro{RMSE}{root mean square error}
\acro{RTK}{real-time kinematic}
\acro{RTT}{round-trip-time}
\acro{RU}{radio unit}
\acro{SI}{self-interference}
\acro{SINR}{signal-to-noise-plus-interference ratio}
\acro{SCA}{successive convex approximation}
\acro{SDG}{sustainable development goal} 
\acro{SDP}{semi-definite programming} 
\acro{SDR}{semi-definite relaxation} 
\acro{SLAM}{simultaneous localization and mapping}
\acro{SNR}{signal-to-noise ratio}
\acro{SIT}{sustainability, inclusiveness, and trustworthiness}
\acro{SOCP}{second-order cone programming}
\acro{SOTA}{state of the art}
\acro{SPEB}{squared position error bound}
\acro{STAR}{simultaneous transmit-and-receive}
\acro{SL}{sidelink}
\acro{TDD}{time-division duplexing}
\acro{ToA}{time-of-arrival}
\acro{TDoA}{time-difference-of-arrival}
\acro{TR}{time-reversal}
\acro{TRP}{transmission and reception point}
\acro{TXRX}[TX/RX]{transmitter/receiver}
\acro{TX}{transmitter}
\acro{RX}{receiver}
\acro{UE}{user equipment}
\acro{UN}{United Nations}
\acro{multi-RTT}{multi-cell round-trip-time}
\acro{ULA}{uniform linear array}
\acro{UL}{uplink}
\acro{UL-TDOA}{uplink time-difference-of-arrival}
\acro{DL-TDOA}{downlink time-difference-of-arrival}
\acro{UMi}{3D-urban micro}
\acro{UMa}{3D-urban macro}
\acro{UWB}{ultra-wide band}
\acro{FR1}{frequency range 1}
\acro{FR2}{frequency range 2}
\acro{WLAN}{wireless local-area network }
\acro{XL-MIMO}{extremely large-scale MIMO}
\acro{FOV}{field-of-view}
\acro{PF}{particle filter}
\acro{RBPF}{Rao-Blackwellized particle filter}
\acro{PHD}{probability hypothesis density}
\acro{OID}{optimal importance density}
\acro{EKF}{extended Kalman filter}
\acro{V2X}{vehicle-to-everything}
\end{acronym}
\usepackage{enumitem}
\usepackage[dvipsnames,table]{xcolor}
\usepackage{tikz}

\usepackage{subfigure}

\newtheorem{example}{Example}

\newcommand{\hw}[1]{\textcolor{red}{XXX Henk - #1}}

\newcommand{\fk}[1]{\textcolor{teal}{XXX Furkan - #1}}
\newcommand{\mv}[1]{\textcolor{brown}{XXX Mikko - #1}}

\newcommand{\dd}[1]{\textcolor{cyan}{XXX Davide - #1}}

\newcommand{\red}[1]{{\color{red}#1}}

\usepackage{tcolorbox}
\usepackage{microtype}

\begin{document}

\newcommand{\yy}{\mathbf{y}}
\newcommand{\xx}{\mathbf{x}}
\newcommand{\zz}{\mathbf{z}}
\newcommand{\ff}{\mathbf{f}}
\newcommand{\ww}{\mathbf{w}}

\newcommand{\HH}{\mathbf{H}}
\newcommand{\DD}{\mathbf{D}}

\newcommand{\FF}{\mathbf{F}}
\newcommand{\FFrf}{\FF_{\rm{RF}}}
\newcommand{\FFbb}{\FF_{\rm{BB}}}
\newcommand{\WW}{\mathbf{W}}
\newcommand{\WWrf}{\WW_{\rm{RF}}}
\newcommand{\WWbb}{\WW_{\rm{BB}}}
\newcommand{\Nrft}{N_{\rm{RF,T}}}
\newcommand{\Nrfr}{N_{\rm{RF,R}}}
\newcommand{\Ns}{N_s}

\newcommand{\Um}{\mathcal{U}}
\newcommand{\thetad}{\Delta \theta}
\newcommand{\mtCN}{{\mathcal{CN}}}

\newcommand{\FFdir}{\FF^{\rm{dir}}}
\newcommand{\FFdird}{\FF^{\rm{dir/der}}}
\newcommand{\snr}{{\rm{SNR}}}

\newcommand{\deltaf}{ \Delta f }
\newcommand{\deltat}{ \Delta t }

\newcommand{\thetab}{ \bm{\theta} }
\newcommand{\phib}{ \bm{\phi} }

\newcommand{\phiazl}{ \phi_{\rm{az},\ell} }
\newcommand{\phiell}{ \phi_{\rm{el},\ell} }

\newcommand{\thetaazl}{ \theta_{\rm{az},\ell} }
\newcommand{\thetaell}{ \theta_{\rm{el},\ell} }

\newcommand{\arx}{\mathbf{a}_\rmrx}
\newcommand{\atx}{\mathbf{a}_\rmtx}

\newcommand{\Arx}{{\mathbf{A}}_{\mathrm{rx}}}
\newcommand{\Atx}{{\mathbf{A}}_{\mathrm{tx}}}
\newcommand{\Atxbar}{\bar{\mathbf{A}}_{\mathrm{tx}}}

\newcommand{\td}{\mathrm{td}}

\newcommand{\atxdt}{\dot{\mathbf{a}}_{{\rmtx}} }

\newcommand{\rmtx}{{\rm{T}}}
\newcommand{\rmrx}{{\rm{R}}}

\newcommand{\Tsym}{ T_{\rm{sym}} }

\newcommand{\Ntx}{N_\rmtx}
\newcommand{\Nrx}{N_\rmrx}

\newcommand{\complexset}[2]{ \mathbb{C}^{#1 \times #2}  }
\newcommand{\realset}[2]{ \mathbb{R}^{#1 \times #2}  }

\newcommand{\txrm}{{{\rm{T}}}}
\newcommand{\rxrm}{{{\rm{R}}}}
\newcommand{\comrm}{{{\rm{com}}}}
\newcommand{\radrm}{{{\rm{rad}}}}
\newcommand{\thn}[1]{ {#1^{\rm{th} } } }

\newcommand{\complexsett}{ \mathbb{C}  }
\newcommand{\nset}{ \mathbb{N}  }
\newcommand{\zset}{ \mathbb{Z}  }

\newcommand{\trp}{^\top}
\newcommand{\herm}{^\mathrm{H}}

\newcommand{\boldzero}{{ {\boldsymbol{0}} }}
\newcommand{\boldone}{{ {\boldsymbol{1}} }}
\newcommand{\boldonet}{{ {\boldsymbol{1}}^{T} }}
\newcommand{\Imatrix}{{ \boldsymbol{\mathrm{I}} }}

\newcommand{\aaa}{\mathbf{a}}
\newcommand{\cc}{ \mathbf{c} }
\newcommand{\bb}{ \mathbf{b} }
\newcommand{\nn}{ \mathbf{n} }

\newcommand{\diag}[1]{ {\rm{diag}}\left(#1\right)  }
\newcommand{\ddiag}[1]{ {\rm{ddiag}}\left(#1\right)  }
\newcommand{\blkdiag}[1]{ {\rm{blkdiag}}\left(#1\right)  }

\newcommand{\Tcp}{ T_{\rm{cp}} }

\newcommand{\deltafc}{ \Delta \fc }
\newcommand{\fc}{ f_c }

\newcommand{\ynm}{ y_{n,m} }
\newcommand{\xnm}{ x_{n,m} }

\newcommand{\XX}{ \mathbf{X} }
\newcommand{\YY}{ \mathbf{Y} }
\newcommand{\ZZ}{ \mathbf{Z} }

\newcommand{\YYcom}{ \YY_\comrm }
\newcommand{\HHcom}{ \HH_\comrm }

\newcommand{\sml}{ s_{m,\ell} }
\newcommand{\sm}{ s_{m} }

\newcommand{\rect}[1]{ { \rm{rect} }\left(#1\right) }

\renewcommand{\vec}[1]{\ensuremath{{\mathbf{#1}}}}
\newcommand{\vx}[0]{\vec{x}}
\newcommand{\vm}[0]{\vec{m}}
\newcommand{\vz}[0]{\vec{z}}
\newcommand{\vy}[0]{\vec{y}}
\newcommand{\vp}[0]{\vec{p}}
\newcommand{\vu}[0]{\vec{u}}
\newcommand{\vv}[0]{\vec{v}}
\newcommand{\vH}[0]{\vec{H}}
\newcommand{\vI}[0]{\vec{I}}
\newcommand{\vecsymbol}[1]{\ensuremath{\boldsymbol{#1}}}
\newcommand{\vmu}[0]{\vecsymbol{\mu}}
\newcommand{\vnu}[0]{\vecsymbol{\nu}}

\newcommand{\bA}[0]{\mathbf{A}}
\newcommand{\bC}[0]{\mathbf{C}}
\newcommand{\Mtx}{M_\rmtx}
\newcommand{\Mrx}{M_\rmrx}

\newcommand{\blueline}{\tikz[baseline]{\draw[blue,solid,line width = 1.0pt](0,0.8mm) -- (3.0mm,0.8mm)}}
\newcommand{\bluetriangle}{\tikz[baseline]{\draw[blue,solid,line width = 0.5pt] 
(0.0mm,1.6mm) -- (2.0mm,1.6mm) --  (1.0mm,0.0mm) -- (0.0mm,1.6mm)}}
\newcommand{\blackcross}{\tikz[baseline]{\draw[black,solid,line width = 1.0pt] 
(0.0mm,0.0mm) -- (1.6mm,1.6mm) (0.0mm,1.6mm) -- (1.6mm,0.0mm)}}
\newcommand{\redplus}{\tikz[baseline]{\draw[red,solid,line width = 0.5pt]
(0mm,0.8mm) -- (1.6mm,0.8mm)  (0.8mm,1.6mm) -- (0.8mm,0mm) }}

\title{The Integrated Sensing and Communication Revolution for 6G: Vision, Techniques, and Applications}
 \author{Nuria Gonz\'{a}lez-Prelcic,~\IEEEmembership{Senior Member,~IEEE,} Musa Furkan Keskin,~\IEEEmembership{Member,~IEEE,} Ossi Kaltiokallio,~\IEEEmembership{Member,~IEEE,}  Mikko Valkama,~\IEEEmembership{Fellow,~IEEE,} Davide Dardari,~\IEEEmembership{Senior Member,~IEEE,} Xiao Shen, Yuan Shen,~\IEEEmembership{Senior Member,~IEEE,} Murat Bayraktar,~\IEEEmembership{Student Member,~IEEE,} and Henk Wymeersch,~\IEEEmembership{Fellow,~IEEE}
\thanks{The work by N. Gonz\'{alez}-Prelcic and M. Bayraktar was supported in part by  the National Science Foundation under grant no. 2147955 and is supported in part by funds from federal agency and industry partners as specified in the Resilient \& Intelligent NextG Systems (RINGS) program.
The work by M. F. Keskin and H. Wymeersch was supported by the Swedish Research Council (VR grant 2022-03007) and by Hexa-X-II, part of the European Union’s Horizon Europe research and innovation programme under Grant Agreement No 101095759.
The work by D. Dardari was supported by the HORIZON-JU-SNS-2022 project TIMES (Grant no. 101096307) and the Italian National Recovery and Resilience Plan (NRRP) of NextGenerationEU, partnership on ``Telecommunications of the Future” (PE00000001 - program ``RESTART”) cofunded by the EU.
The work by O. Kaltiokallio and M. Valkama was supported in part by the Academy of Finland under the grants \#352754 and \#359095, and in part by the Business Finland under the project ``6G--ISAC”.
The work by Y. Shen and X. Shen was supported by the National Natural Science Foundation of China under Grant 62271285 and Shanghai AI Laboratory.\\
N. Gonz\'{alez}-Prelcic and M. Bayraktar are with the
Department of Electrical and Computer Engineering, University of California San Diego, USA (ngprelcic@ucsd.edu;mbayraktar@ucsd.edu).\\
M. F. Keskin and H. Wymeersch are with the Department of Electrical Engineering, Chalmers University of Technology, Sweden (furkan@chalmers.se;henkw@chalmers.se).\\
O. Kaltiokallio and  Mikko Valkama are with the Unit of Electrical Engineering, Tampere University, Finland
(ossi.kaltiokallio@tuni.fi; mikko.valkama@tuni.fi).\\
D. Dardari is with the Department of Electrical, Electronic, and Information Engineering "Guglielmo Marconi", University of Bologna, and WiLAB-CNIT, Italy (davide.dardari@unibo.it).\\
X. Shen and Y. Shen are with the Department of Electronic Engineering and the Beijing National Research Center for Information Science and Technology, Tsinghua University, Beijing 100084 China (shenx20@mails.tsinghua.edu.cn; shenyuan\_ee@tsinghua.edu.cn).
}
}

\maketitle

\begin{abstract}
Future wireless networks will integrate sensing, learning and communication to provide new services beyond communication and to become more resilient. Sensors at the network infrastructure, sensors on the user equipment, and the sensing capability of the communication signal itself provide a new source of data that connects the physical and radio frequency environments. A wireless network that harnesses all these sensing data can not only enable additional sensing services, but also become more resilient to channel-dependent effects like blockage and better support adaptation in dynamic environments as networks reconfigure. 
In this paper, we provide a vision for integrated sensing and communication (ISAC) networks and an overview of how signal processing, optimization and machine learning techniques can be leveraged to make them a reality in the context of 6G. We also include some examples of the performance of several of these strategies when evaluated using a simulation framework based on a combination of ray tracing measurements and mathematical models that mix the digital and physical worlds. 
\end{abstract}



\section{Introduction}

Early work on integrated sensing and communication targeted the development of 
cooperation strategies to enable spectral coexistence between communication and sensing \cite{Darpa2016}. At the same time, there was also interest in sharing hardware among radar and communication systems to reduce cost, weight and size, what motivated the 
initial designs of joint sensing and communication systems \cite{Tavik2005}. Nowadays, the number of potential avenues to integrating sensing and communication and their related benefits have exploded \cite{Fan:22}. On the one hand, communication operation at higher frequencies with large arrays and bandwidths has led to waveforms and signal processing algorithms in the transceiver which are naturally well-suited for sensing. 
On the other hand, the diversification and sophistication of devices in the wireless network have resulted in the creation of wireless networks where the communicating devices are also sensing devices, which opens challenges related to sensor and communication data fusion. Good examples of sensing/communicating devices are the connected vehicles to be supported by cellular networks (already a use case in the 5G standard), equipped with a wide variety of sensors including cameras, radars, or LIDAR. All these technological advancements bring new opportunities for integrating sensing and communication with motivations and benefits that go beyond conventional ones. \ac{ISAC}  has emerged as a renewed research area that aims to develop all these opportunities by exploiting the similarities between the required hardware, the waveforms, the signal processing algorithms and the machine learning strategies to be exploited, or the sensing and communication channels, and define new applications and frontiers for the future wireless communication and sensing systems. 

In the cellular industry, the 6G Roadmap elaborated by the NextG Alliance (an industry initiative to advance North American technology for future cellular networks) considers joint sensing and communication a key technology for 6G \cite{6GRoadmap}. Similar considerations are being made by the European counterpart to NextG, the 6G Smart Networks and Services Industry Association (6G-IA) \cite{6GEurope}. Moreover, the European Telecommunications Standards Institute (ETSI) has also launched an Industry Specification Group (ISG) for ISAC, which is developing a roadmap and prioritization of sensing types and ISAC use cases that can potentially be covered  in future 6G releases of \ac{3GPP} \cite{ETSI-ISAC}. 
The performance requirements envisioned by the industry for some application verticals are, however, stringent and can only be met with continued research that develops advanced solutions.

In this paper, we describe different frameworks for integrating sensing and communications in future-generation cellular systems, discuss the different features to be exploited at different frequency bands, and present an overview of the recent techniques and advances that can make ISAC a reality in 6G. We focus on a communication-centric perspective for ISAC with tight integration of waveform and time and frequency resources for sensing and communication, versus other approaches where integration only appears at the site or spectrum level \cite{WymeerschArXiv2024}. In this communication-centric vision, we also review how sensing can assist the network operation.
Previous overview/tutorial papers do not clearly focus on a communication-centric perspective of ISAC which includes a comprehensive survey of all the ISAC techniques relevant for 6G and beyond. 

\begin{figure}
    \centering
    \includegraphics[width=0.6\columnwidth]{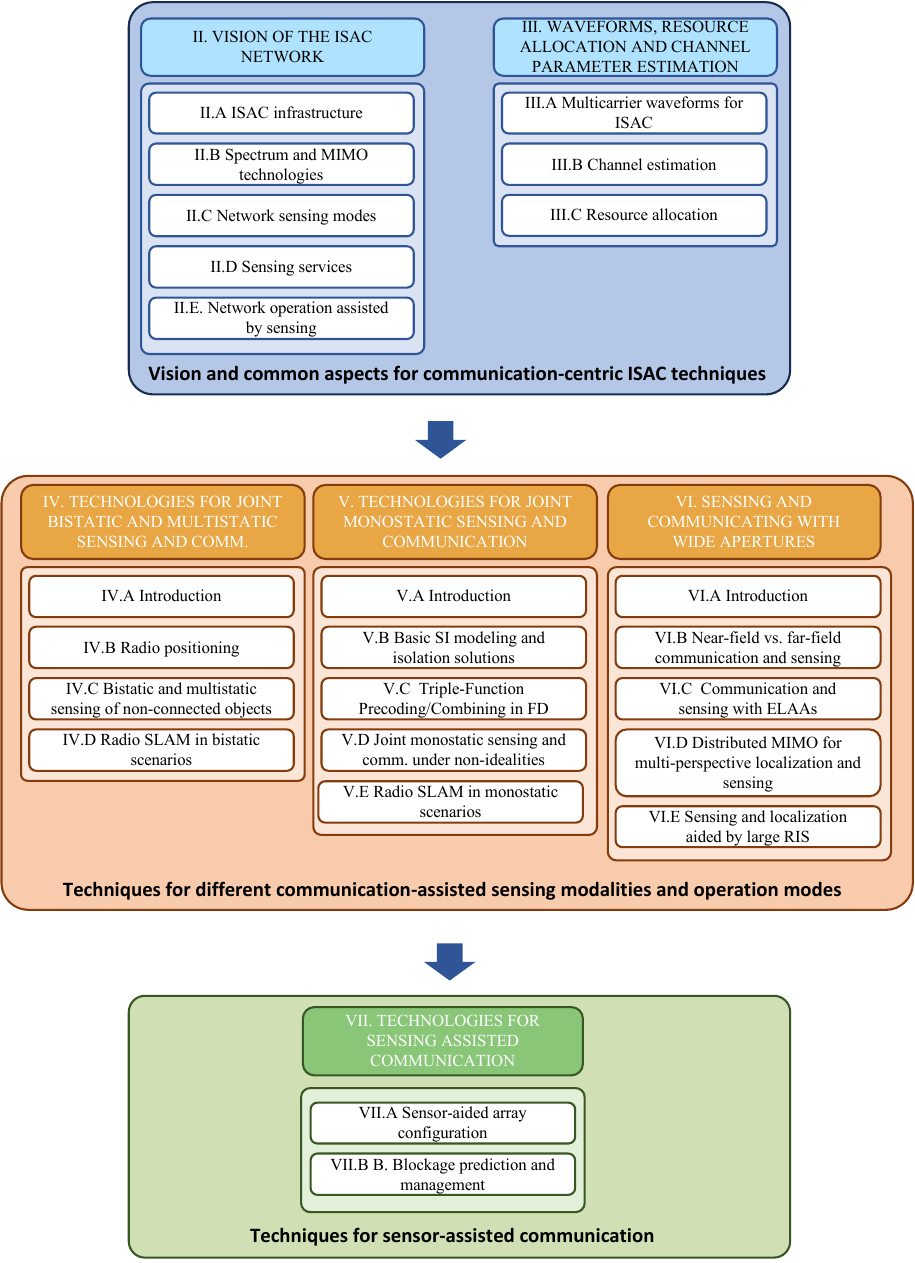}
    \caption{Paper outline, covering the technical sections.}
    \label{fig:paper-outline}
\end{figure}

The structure of the technical sections in this paper is shown in Fig.~\ref{fig:paper-outline}. 
We start by describing our vision of the ISAC network, potential sensing modes, the concept of sensing-aided communication, and  the status of ISAC services --mainly positioning-- developed within the most recent standardization efforts in \ac{3GPP}. Then we focus on ongoing research, reviewing first approaches for joint bistatic and multi-static sensing and communication that provide the sensing information directly from the fronthaul link, downlink, uplink, or even sidelink signal, exploiting different geometric transformations between some or all the channel parameters and the user's position and orientation. These approaches are especially relevant at \ac{mmWave}  and sub-THz bands, where the large arrays and bandwidths provide good angular and delay resolvability, and it is also easier to map the channel parameters to the objects in the propagation environment due to the channel sparsity. Next, we revisit joint monostatic sensing  and communication strategies, where the exploitation of \ac{FD} circuits that provide an appropriate isolation between the transmitter and receiver enables simultaneous transmission of the communication signal and reception of the reflections on potential targets, which can be processed in a radar-like operation to provide position and velocity information.  In a different section, we also discuss how novel wide aperture technologies such as large \ac{RIS} and distributed MIMO can provide a potential avenue to further increase position estimation accuracy by exploiting information from additional relevant paths.  To complete the perspective of the sensing capabilities of the cellular network, we will describe several approaches for radio-\ac{SLAM}, the process of simultaneously locating the user and creating a map of the environment. Finally, we will describe the opposite setting,  sensor-aided communication, where  sensing information (e.g., user's position) can be leveraged to enhance network operation, for example significantly reducing the overhead associated with link configuration and reconfiguration or enabling early blockage detection. Throughout the paper, we will make the case that time is right for communication and sensing to be considered together, and why communication and sensing will likely be THE defining physical layer feature of 6G. 

\section{Vision of the ISAC network}

\subsection{ISAC infrastructure}
\begin{figure*}[t!]
\begin{center}
			 \includegraphics[width=\textwidth]{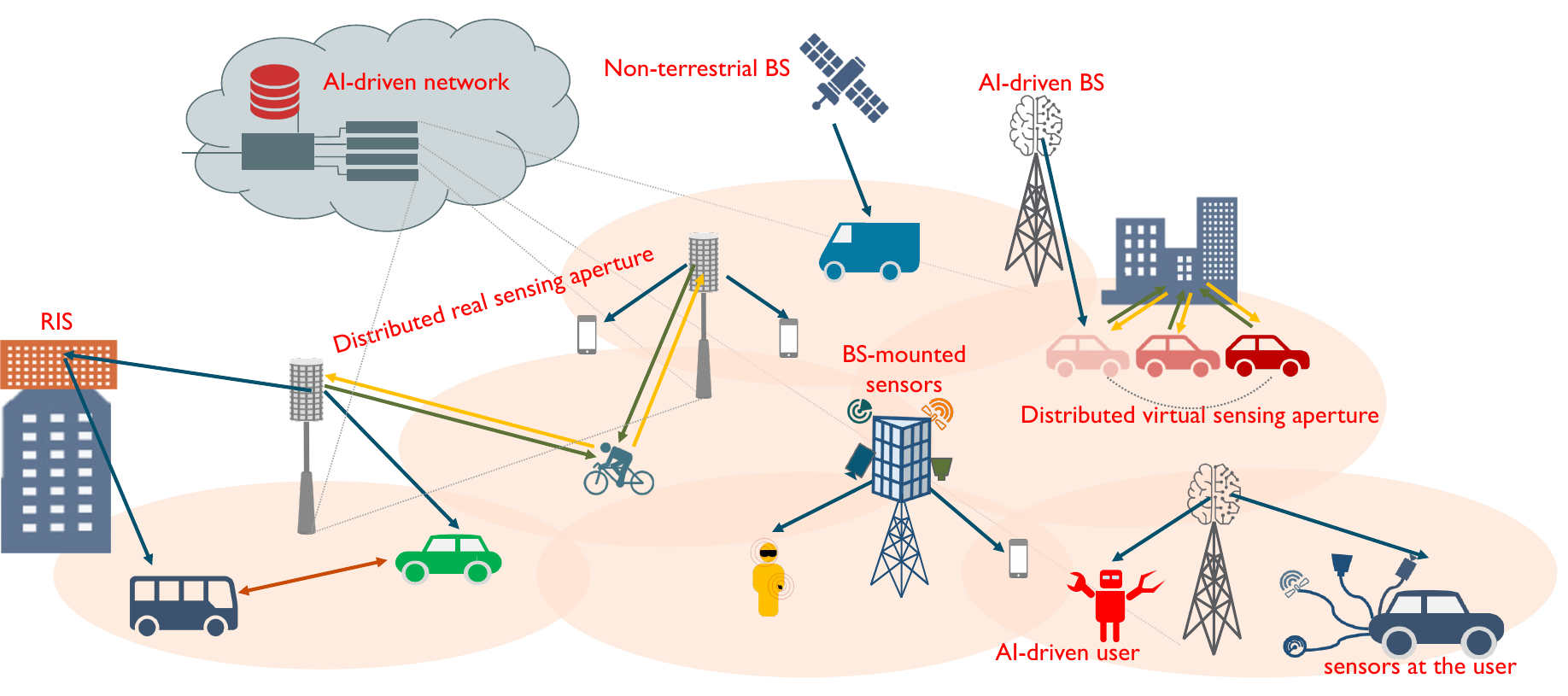}
\caption{Illustration of the different elements in the infrastructure of an ISAC network.}
\label{fig:ISLAC-infrastructure}
\end{center}
\end{figure*}

The ISAC network provides an integrated combination of sensing and communication. It offers sensing and communication as services to applications that are run in, around and by the network. A smart meter may subscribe to communication as a means to send back meter measurements, a bicycle commuter may subscribe to sensing to enhance their situational awareness, while an automated vehicle may subscribe to sensing and communication as part of an automated driving package. An ISAC network, and pertinent components of infrastructure, are illustrated in Fig.~\ref{fig:ISLAC-infrastructure}. We summarize the different components of the infrastructure here. 

The foundation of an ISAC network is a diverse and heterogeneous cellular communication infrastructure. There is conventional terrestrial \ac{BS} infrastructure in the form of macro/micro/pico cells, which are typically mounted on towers, rooftops or lightpoles. To expand coverage and increase sensing accuracy and probability of target detection, important for higher frequencies like mmWave, there are low-power \ac{RIS} to generate favorable reflections between the \ac{BS}s and the \ac{UE}. There are also relays or \ac{BS}s that connect to the cellular infrastructure using the same spectrum as the network, with integrated access and backhaul. The infrastructure need not be terrestrial; ISAC also supports non-terrestrial components where the \ac{BS}s are untethered to the ground in the form of satellites or unmanned aerial vehicles. Wired networks, typically realized via cable, fiber or point-to-point microwave links form another piece of infrastructure in the cellular network. These fixed communication links are used as part of backhaul to network \ac{BS}s together with the core network, and also to implement front-haul, where a \ac{BS} is realized in two pieces as a remote radio head connecting antennas to a distantly located baseband unit. 

The sensing part of an ISAC network is realized with several infrastructural components. Sensors in the form of cameras, radars, and LIDARs are present at some of the \ac{UE} in the network. These sensors use spectrum that is different from the communication spectrum. But the sensors need not just be on the \ac{UE}s as there is a compelling case to co-locate sensing with the infrastructure to offer a birds-eye-view of the environment \cite{AliComMag2020}. Sensing is also facilitated by reusing the communication signal for radio sensing. A network with integrated sensing and communication reuses the existing communication waveforms possibly with more capable hardware (for example a \ac{FD} \ac{BS} to enable monostatic sensing), and additional network components to fuse data collected in the environment. 

The learning portion of an ISAC network is realized through the combination of data and computation. Data refers to the storage of past sensor data and communication performance data, which is collected over time. Computation refers to the training of machine learning models on the data, the updating of models based on new data, and the execution of inference operations using the trained models. As shown in Fig.~\ref{fig:ISLAC-infrastructure}, the data and computation are distributed in different components of the network including the \ac{UE}, the \ac{BS} (so-called network edge), and in the core network. The computation capabilities will vary significantly at these different components, as will the extent of data stored and shared with other network components. 

One aspect that makes ISAC networks interesting is the vastly different communication, sensing, computation and learning capabilities of the devices in the network. It includes devices that directly serve people like smart phones, watches and glasses; low-capability connected devices like smart meters and location tags; and high-capability connected devices like automotive vehicles, aerial vehicles and robots. 
 
Fundamentally, the ISAC network infrastructure is itself smart. Many elements of the infrastructure and the network itself can be driven by protocols and algorithms which exploit both models and data. The learning (shown via data and computation in Fig.~\ref{fig:ISLAC-infrastructure}) should be viewed not just as a sensing service provided to devices in the network, but also as a fundamental component of the network's self-optimization capability. For example, the AI-driven \ac{BS} could use data to optimize how it balances sensing and communication needs with users in its coverage area. The network central processor could use data to reconfigure how all the different infrastructure components work together to serve broader sensing and communication functions over a larger geographic area.

\subsection{Spectrum and MIMO technologies}
In addition to the infrastructure, the ability to perform sensing is strongly related to the resolution (i.e., the ability to resolve multipath) provided in different dimensions, in particular bandwidth (providing delay resolution) and array aperture (providing angle resolution). For that reason, in this section, we briefly review the different frequency bands and \ac{MIMO} architectures.

\subsubsection{Spectrum Considerations}
In 6G, there are several bands under discussion. The combination of the frequency band and the available bandwidth are important from the sensing perspective.
\begin{itemize}
    \item \emph{\Ac{FR1}:} This band spans from around 400 MHz to 7 GHz. In this band, bandwidths between 5 MHz and 100 MHz can be supported. 
    The main benefits of the low carrier frequency are a low path loss leading to large coverage, and small Doppler frequencies, supporting high mobility. On the other hand, the small bandwidth leads to poor delay resolution. 
    Moreover, the propagation tends to be less geometric (i.e., the channel does not have a clear geometric relation to the environment and is more statistical in nature), due to weak shadowing and complex multipath propagation. 
    \item \emph{\Ac{FR2}:} This band spans from around 24 GHz to 70 GHz, with supported bandwidths ranging from 50 to 400 MHz. Due to the higher path loss, use of this band must be combined with directional arrays. This implies that while resolution is good, coverage is limited and only applications with moderate mobility can be supported. In terms of the propagation,  shadowing is more pronounced  leading to fewer propagation paths and a more geometric channel. 
    \item \emph{Upper mid-band:} This band lies between 7 GHz and 24 GHz, and is sometimes referred to as the golden band or even FR3. This band has not been studied extensively, but is expected to provide a good trade-off between data rate and wide coverage. Initial studies claim that it is possible to maintain the same area coverage as in FR1 while achieving a significant improvement in throughput thanks to the exploitation of extremely large arrays  and increased bandwidth \cite{Holma2021whitepaper}.
    \item \emph{Sub-THz:} The sub-THz bands spans from 100 GHz to 300 GHz. This band is envisioned for extremely high data rates in nearly static conditions. Thanks to large bandwidths and large arrays for fine, high-gain beams, resolution is expected to be high, but range is likely very short (tens of meters). The channel is characterized by  diffusive, rather than  specular reflections, as the wavelength gets close to the roughness of materials. This provides opportunities of sub-THz for imaging and mapping applications. 
\end{itemize}
In summary, each band features clear benefits and drawbacks for sensing. Consequently, judicious selection and aggregation of different bands will be important to support the wide variety of sensing services. Multi-band networks, with the possibility to combine or switch between a variety of bands (ranging from sub-6 GHz to THz and visible light), are promising in this respect \cite{SaeidiComMag2024}, but require further study in terms of transceiver and antenna design, propagation, and resource allocation \cite{Aboagye2024TCOM}. Finally, we emphasize that geometric models will be needed to evaluate sensing capabilities within and across the 6G bands, relying on ray-tracing or common data-bases, rather than conventional stochastic channel models. Such a common data-set is important not just for standardization, but also for academic research. 




\subsubsection{MIMO architectures}
Multiantenna communication is a distinctive feature in current cellular networks at both FR1 and FR2. MIMO architectures are however radically different at different frequency bands, due to different hardware constraints, antenna scales, and channel bandwidths. At \ac{FR1}, it is possible to operate with small arrays and one radio frequency (RF) chain per antenna element, so that all the signal processing operations are performed in digital. At mmWave frequencies, power consumption considerations and circuit technologies introduce different hardware constraints \cite{mmWavetutorial2016,MIMOPrecodingComMag2014}. For example, space limitations and excessive power consumption when operating with high resolution converters prevent from using an RF chain per antenna. This has led to specific MIMO architectures to operate at mmWave, which include analog beamforming, hybrid precoding and combining and low resolution architectures that keep one RF chain per antenna but significantly reducing the number of bits in the \ac{ADC}s and/or \ac{DAC}s. The MIMO architecture heavily impacts the received signal model and the techniques used to extract the channel parameters, later used for localization or sensing. In addition, the design of precoders and combiners for joint sensing and communication purposes also depends on the specifc MIMO architecture.

In an analog architecture, beamforming is performed in the analog domain by configuring a set of phase shifters. Configuring these phases  for analog beamforming requires several stages of beam training at both sides of the link.  Assuming a number of $\Ntx$ transmit antennas and $\Nrx$ receive antennas, the beamforming operation is represented by multiplication by a beamforming vector $\ff \in \mathcal{F}^{\Ntx\times1}$ at the transmitter and a weight beamforming vector $\ww \in \mathcal{W}^{\Nrx\times1}$ at the receiver, with $\mathcal{F}$ and $\mathcal{W}$ the sets of possible phases 
at the transmitter and receiver. The performance that can be achieved with analog beamforming, both for communication and sensing, is limited by the lack of amplitude tunability and the number of bits used to quantize the phases. In general, it is not a suitable approach for multistream, multiuser, or multitarget scenarios. Moreover, the spatial processing performed with analog beamforming is frequency flat, since the the same set of phase shifters are used for the entire band. The design of analog beamformers for joint sensing and communication purposes heavily depends on the sensing mode. For example, Section~\ref{sec:monostatic} provides the details of an analog beamforming design for monostatic network sensing that incorporates sensing and communication metrics in addition to the phase shifter constraints and self-interference mitigation requirements.  

\begin{figure}[t!]
        \begin{center}
        \subfigure[]{
			 \label{fig:monostatic}
			 \includegraphics[width=0.45\textwidth]{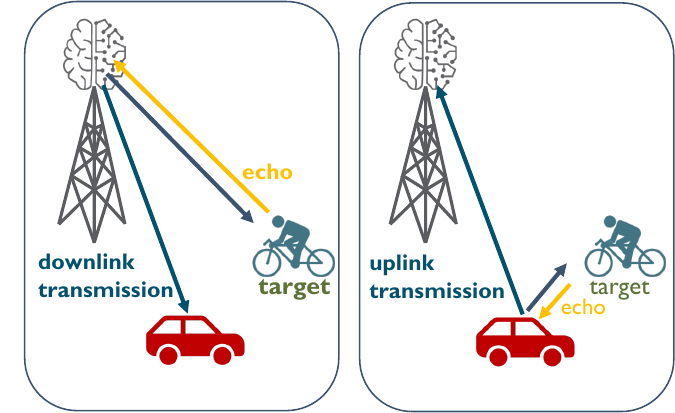}
		}
        \subfigure[]{
			 \label{fig:bistatic}
			 \includegraphics[width=0.45\textwidth]{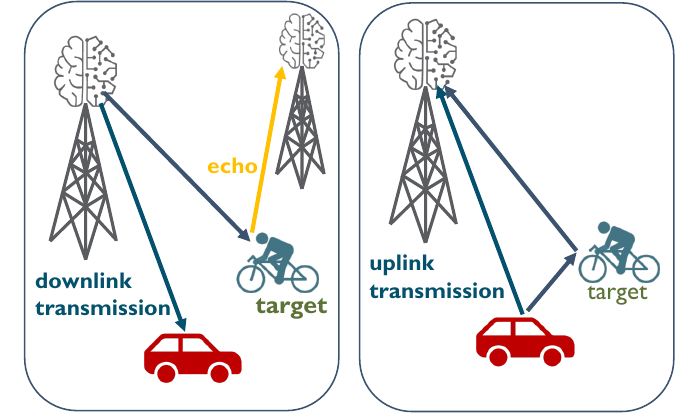}
		}
  %
	\end{center}
        \caption{Illustration of different network sensing modes. (a) Monostatic sensing: the transmitter and receiver are co-located, share a common clock, and knowledge of the transmitted signal so the echoes can be processed in radar-like operation. It can be performed by the BS or a mobile user. (b) Bistatic sensing: the transmitter and receiver are physically separated. It can be performed in the uplink or the downlink, and the transmitter and receiver can be two BSs, or one BS and one user.}  
\end{figure}

A hybrid precoding architecture provides an interesting  performance/complexity trade-off. In this case, the precoding  and combining operations are divided between the analog and digital domains, with a number of RF chains much lower than the number of antennas. In other words, $\Nrft < \Ntx$ and $\Nrfr < \Nrx$. This way, the precoding matrix can be represented by $\FF=\FFrf\FFbb$, where $\FFrf$ is the analog precoding matrix and $\FFbb$ is the digital or baseband precoding matrix. Analogously, $\WW=\WWrf\WWbb$, with $\WWrf$ the analog combiner and $\WWbb$ the digital combiner. The additional digital precoding/combining stage enables frequency selective spatial processing.
Moreover, the higher number of degrees of freedom in a hybrid design allows mutibeam solutions, making it suitable for multistream, multiuser, or multitarget scenarios.
The sensing and communication performance provided by hybrid designs is close to that obtained with all-digital solutions. However, the hardware constraints of the analog counterparts complicate the channel parameter estimation process and the optimization problems to be solved to design the hybrid precoders and combiners.

An alternative to the analog and hybrid architectures is the low resolution architecture. It is a fully digital architecture where low resolution \ac{DAC}s and \ac{ADC}s are employed to reduce power consumption and cost.  Performance is compromised  because of  the high quantization noise. 
The investigation of specific designs and its performance in the context of ISAC systems is very limited \cite{Kumari2020ICASSP,Kaushik2022ICC} and will not be further discussed in this paper.

Operation in the upper-mid-band will likely be driven by BSs equipped with extremely large arrays and a hybrid MIMO architecture with a very high number of RF chains \cite{Holma2021whitepaper}. MIMO configurations will likely vary within this band, to accommodate different channel features and bandwidths as moving from lower to higher carrier frequencies.  Multiband array designs will integrate different types of antenna arrays for each band \cite{kang2023arxiv}. Open challenges include multiband array configuration both for communication and sensing.


\subsection{Network sensing modes} 

The purpose of this section is to clarify the different types of sensing and relate them to other concepts used throughout this paper. 

\subsubsection{Sensing, Positioning, Localization}
Sensing in 6G networks is a highly overloaded term. Sensing comprises receiving a radio signal or a set of radio signals and processing these radio signals to extract information relevant for a service. The received radio signals in general depend on the geometric state of the transmitter, receiver, and the environment (e.g., radar sensing), though not all sensing services rely on this geometry (e.g., pollution monitoring). Hence, localizing of connected users and passive objects relies on sensing information. In \ac{3GPP}, positioning refers to localizing \acp{UE} from uplink, downlink, or sidelink transmissions. Localization then extends positioning to also include the estimation of the position of a passive object / target (as in device-free localization), which includes also the detection of the presence of these objects. In this paper, we will use localization and positioning interchangeably, and when we refer to sensing, we intend radar-like sensing, whereby we detect objects and determine their state. 

Sensing measurements and information derived from them can be fused with sensors external to 6G, such as cameras, lidar, or radar to provide a more detailed or complementary view. For example, the cellular system could provide additional sensing information to automotive sensors, allowing vehicles to see around corners.

\subsubsection{Mono-, Bi-, and Multistatic Sensing}

Sensing is conventionally broken down intro three types, though many other forms of sensing exist, which are not covered in this paper. 
\begin{itemize}
    \item \emph{Monostatic sensing:} In monostatic sensing, illustrated in Fig.~\ref{fig:monostatic}, the transmitter and receiver are co-located, share a common clock, and knowledge of the transmitted signal. Hence, sensing can be based on pilot or data signals. The sensing measurements (e.g., \ac{ToA}, \ac{AoA} and corresponding detected objects) are in relation to the coordinate system of the transmitter. Hence, the transmitter may be a \ac{BS} or a mobile \ac{UE}. In the former case, sensed objects are tracked, based on downlink transmissions, in the frame of reference of the static \ac{BS}. In the latter case, the sensed objects are tracked, based on up/sidelink transmissions, in the frame of reference of the mobile \ac{UE}. In this case, the \ac{UE} can over time build a map of the environment, with respect to its original position and orientation, a process known as \ac{SLAM}. 
    \item \emph{Bistatic sensing:} In bistatic sensing, the transmitter and receiver are physically separate, as shown in Fig.~\ref{fig:bistatic}. If the transmitter or receiver is a \ac{UE}, no synchronization can be assumed and sensing is based on pilot signals. Also, sensing measurements must account for the unknown location of the \ac{UE}, leading to a \ac{SLAM} problem in a global coordinate system. If both transmitter and receiver are \acp{BS}, time or even phase synchronization may be assumed, as well as knowledge of the transmitted data. In case such synchronization is not available, the \ac{LoS} path can serve as a reference for all later multipath components. 
    \item \emph{Multistatic sensing:} In multistatic sensing, there are several transmitters and/or several receivers, all physically separated. Pilot signals or some form of multiplexing is needed in case there are several transmitters, to avoid interference. As in bistatic sensing, different levels of synchronization may be available (time or phase synchronization), leading to different ways to fuse measurements from the different receivers. 
\end{itemize}

\subsection{Sensing services}



\begin{figure*}
\centering
\includegraphics[width= 0.8\linewidth]{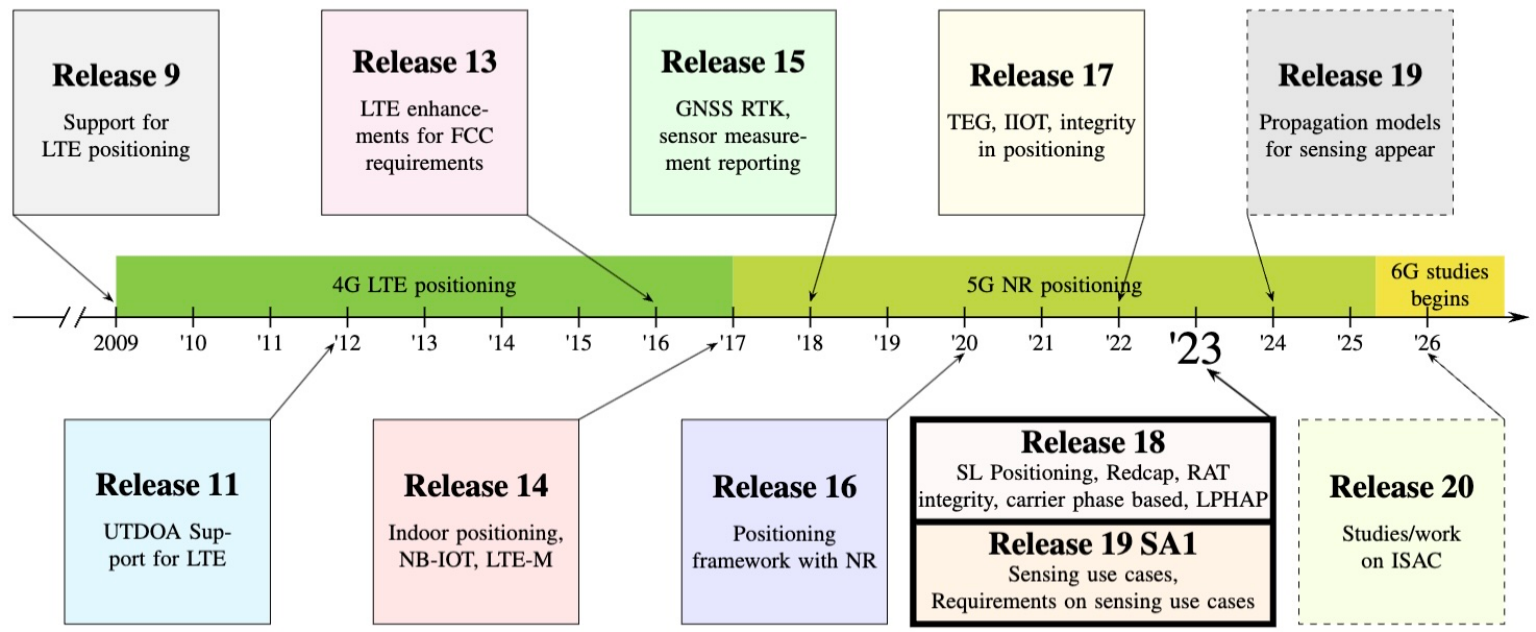}
\caption{Evolution of positioning across 3GPP releases and beginning of sensing. Dashed box show undecided release of 3GPP.}
\label{Fig:posEvol}
\end{figure*}

\subsubsection{Evolution from positioning to sensing, a standardization perspective}

Positioning a  \ac{UE} has been standardized over multiple 3GPP releases.  Fig.~\ref{Fig:posEvol} shows the history of 3GPP positioning Radio Access Network (RAN)  standardization and evolution of foreseeable sensing standardization in 3GPP. As can be seen from the figure, specification for estimating UE's location has been built over many years.  One aspect of this evolution is the change on the accuracy requirement. It started with positioning in release 9 LTE networks, with the aim of meeting the regulatory requirement of 50 m accuracy in positioning a UE. Regulatory requirement was the mainstay for building the positioning specifications in nearly all LTE releases. Positioning signals, measurements, procedure and architecture were specified to meet this requirement. The new radio (NR) in 5G supports larger bandwidth than LTE. The increase in bandwidth in 5G NR also improved positioning accuracy requirements in 5G NR releases. The first 5G NR release had positioning requirements down to 3 m for indoor use cases which tightened further to 1 m in release 17 for industrial indoor \ac{IoT} use cases \cite{dwivedi2021positioning}. 


As a location estimation problem, sensing seems to be an evolution of the positioning with objective of locating a passive scatterer. However, the protocol and architectural landscape of sensing will be significantly different and may not be seen as an evolution of positioning. 
New signals for sensing may be standardized in 6G releases of 3GPP if existing signals in specifications do not meet sensing requirements.

\subsubsection{Wireless network based use cases and requirements} 

3GPP has begun the standardization of sensing with a study on use cases \cite{TR22837} and subsequently building specification on the requirements \cite{TS22137}. Network sensing can enable new services and use cases for various verticals including smart homes, smart factories or \ac{V2X}. There are 32 use cases proposed in \cite{TR22837}. In initial phases of standardization selective use cases will be prioritized. A possible prioritization of use cases can be the following,  

\begin{itemize}
\item \textbf{Smart home/building intrusion detection} - Intrusion detection in buildings or surroundings of smart home.
\item \textbf{Transport use cases} - Examples include intrusion detection of animal/human on highway, sensing aided automotive maneuvering and navigation, parking space determination or blind spot detection.
\item \textbf{Industry use cases} - 
 Detection and tracking of autonomous ground vehicles in factories, autonomous mobile robot collision avoidance, or integrated sensing and positioning in a factory hall.
\item \textbf{Unmanned Aerial Vehicles (UAV) use cases} - Some examples include  UAV flight trajectory tracking and UAV detection near smart grid equipment.
\end{itemize}
\subsubsection{3GPP Propagation modeling for sensing}
  The current version of channel models in 38.901 does not support sensing evaluation in detail and sensing specific parameters and aspects need to be added to the channel models. For example, modelling of sensing targets in terms of their physical scattering surface as radio cross section (RCS) and modelling the mobility of the sensing targets has to be included in a new channel model that supports sensing. In addition, realizations of the current 3GPP channel model never generate reflections fulfilling the geometric relationships
  imposed by the laws of physics, yet these relationships are generally exploited for localization or sensing. Finally, the possibility of tracking an object requires a spatially consistent channel model depicting movement of the object consistently with respect to the evolution of multipath, phases of the signals, etc. Recent work has started to address these challenges \cite{Xiong2023VTC,Zhang2023ComMag}, but the number of contributions is still scarce. 


\subsection{Network operation assisted by sensing}\label{sec:intro-sensing-assisted-communication}

Networks can sense their surroundings to provide sensing data interesting for the users, the cities, or the road infrastructure for example.
Moreover, an ISAC network can also exploit these sensing data to become more resilient. A wireless network that
harnesses such data can improve its adaptability to changes in the propagation environment and become more resilient to channel-dependent effects like blockage \cite{Prelcic2017,AliComMag2020}. The huge amount of information that this type of network can collect is the basis for exploiting
machine learning (ML) algorithms to assist communication. ML can help in creating representations of the
environment that fuse sensing data with digital maps and models for the communication system. Moreover,
ML can also provide intelligent recommendations that exploit sensing data for network adaptation to a dynamic environment \cite{Va2019Access,Va2018TVT,Klautau2019WCL,Graff2023TVT}. 

\begin{figure}[t!]
        \begin{center}
        \subfigure[]{
			 \label{fig:urban-scenario}
			 \includegraphics[width=0.5\textwidth]{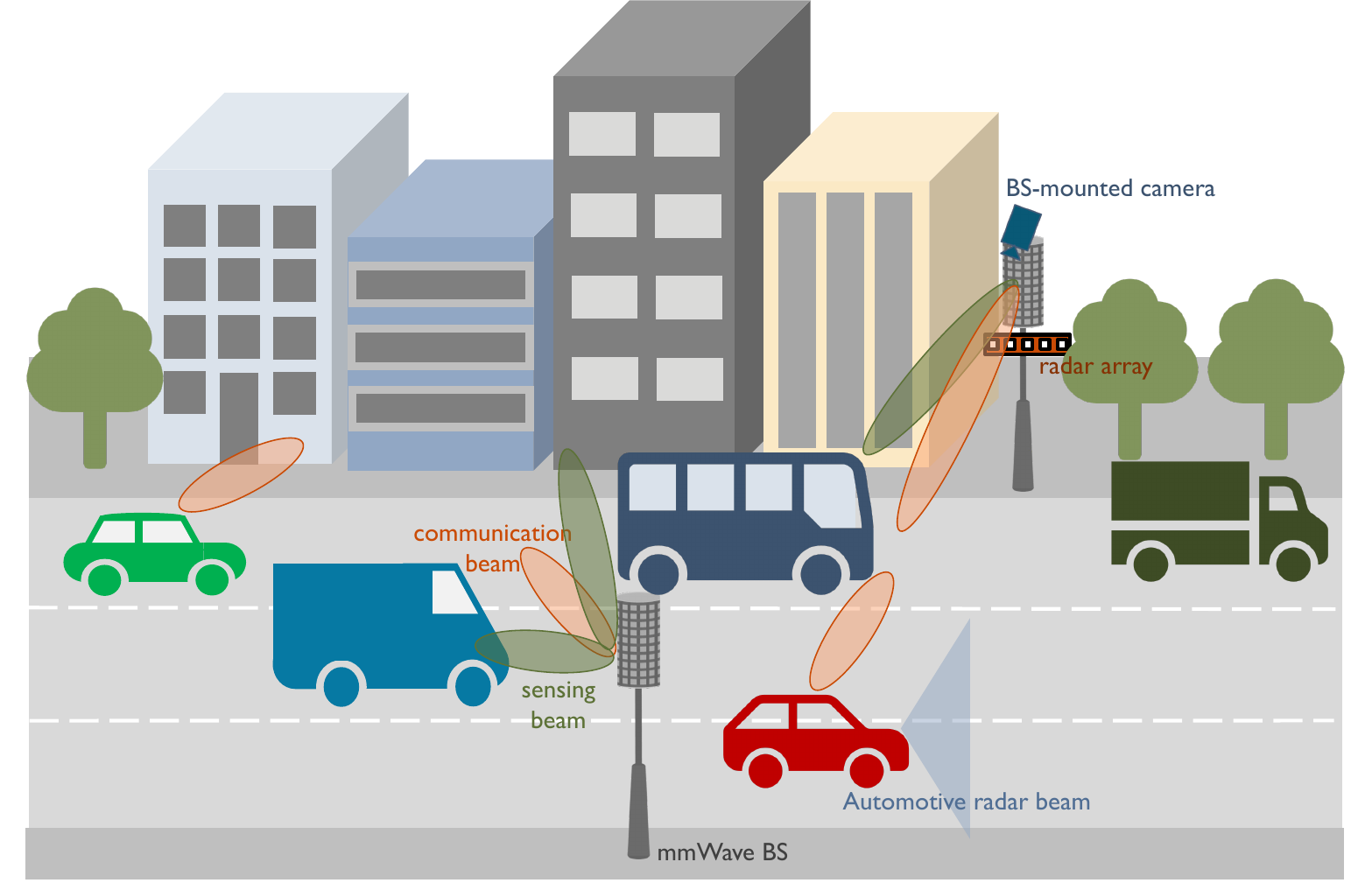}
		}
        \subfigure[]{
			 \label{fig:ISLAC-map}
			 \includegraphics[width=0.5\textwidth]{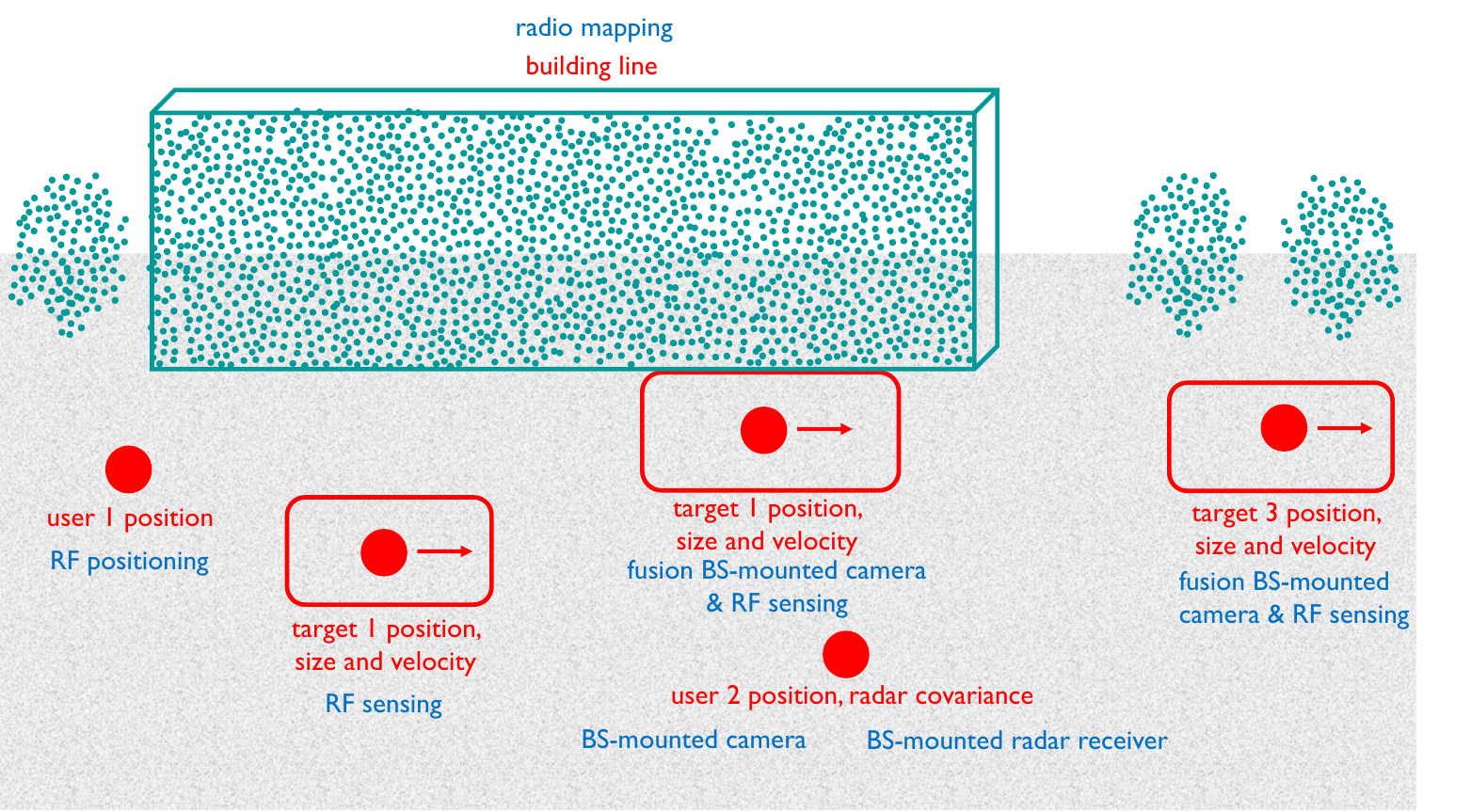}
		}		
  \subfigure[]{
			 \label{fig:ISLACmap-uses-cases}
			 \includegraphics[width=0.5\textwidth]{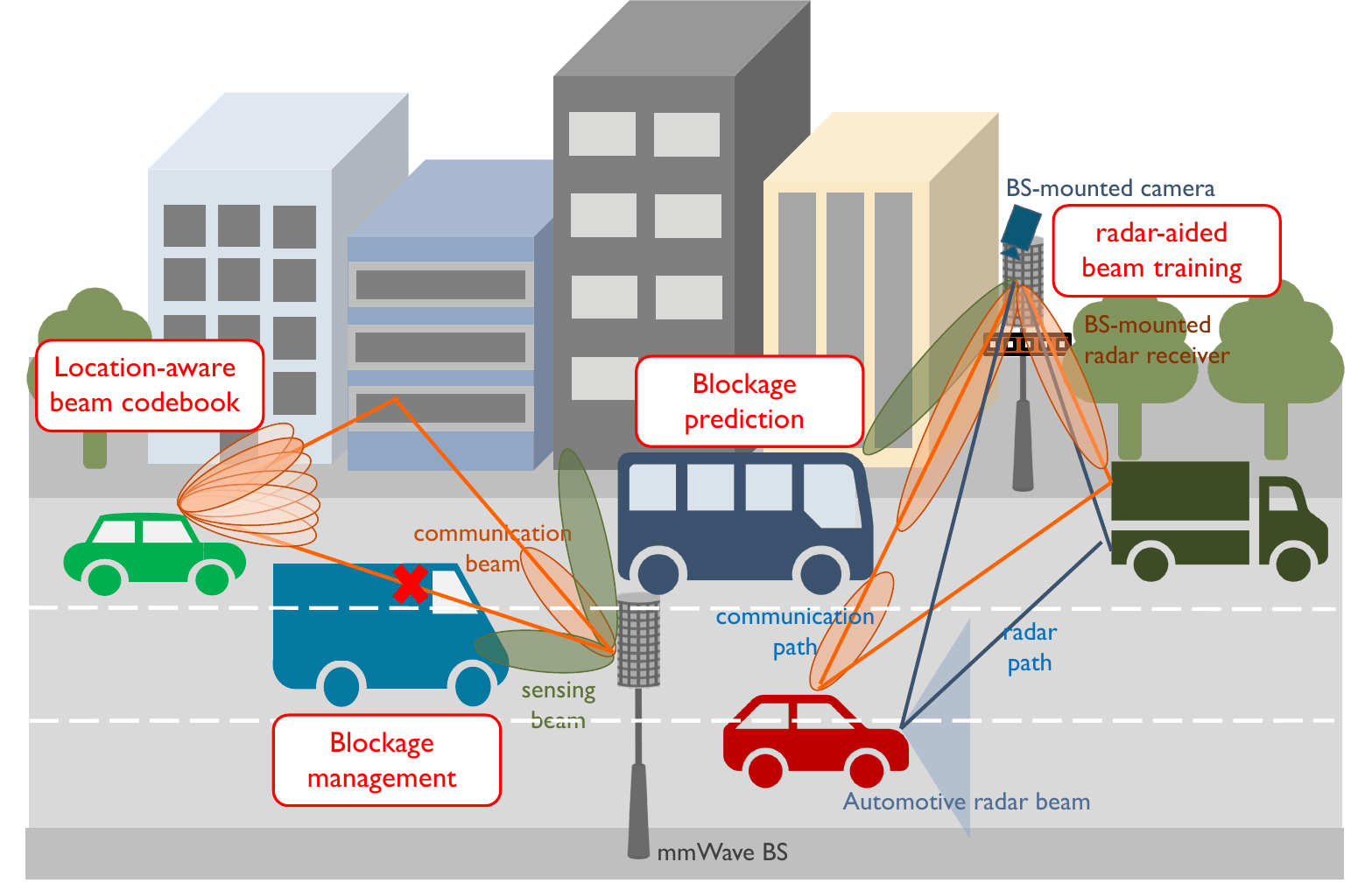}
		}
			\end{center}
  \caption{Illustration of the sensing assisted communication concept: (a) Joint sensing and vehicular communication system supported by a cellular network in an urban scenario. (b) ISAC map corresponding to the urban scenario in (a), showing the detected users, targets and scatterers and the technology employed for detection. (c) Application of the ISAC map for sensing-assisted communication: location aware beam codebook design for user 1, blockage prediction created by target 2, and radar-aided beam training for user 2.}
  \end{figure}
  
We envision a network that can create a joint map that combines measurements of both the physical
world and the radio world, which we call an ISAC-map. Radio maps have been used extensively for
cell network planning based on propagation simulation tools and drive testing \cite{Bi2019WC}. Those maps normally only capture average received signal strength or signal-to-interference radio (SIR)  as a function of location for the purpose of network configuration and densification. The ISAC-map we envision goes well beyond the idea of radio maps, capturing the distribution of objects in the real world and their inferred properties like type, size, trajectory, and so on. This could be obtained using the radio localization and sensing capabilities of the network itself  or the wealth of data that can be obtained with conventional sensors on the \ac{UE} and/or on the infrastructure \cite{Gonzalez-Prelcic2016ITA, AliComMag2020}, for example on lamp poles \cite{Max2020ACM}. The ISAC map will include all sensing information relevant to network operation: location of users, speed and position of blockers, information about static scatterers of the environment, etc. This information can be superimposed on a digital map of the network coverage area to also leverage information about landmarks in the digital map relevant for wireless propagation. Conceptually, an ISAC map is a semantic representation of the propagation environment useful for network operation. In addition, past configurations of the network that provided good performance, for example a reduced beam codebook associated to a given location, can also be fused with the ISAC map. An example of the ISAC map for the urban scenario in Fig~\ref{fig:urban-scenario} is illustrated in Fig.~\ref{fig:ISLAC-map}---including the technology used to obtain the sensing data---while Fig.~\ref{fig:ISLACmap-uses-cases} shows some use cases that exploit the ISAC map for enhancing communication operation. For example, the location of user 1 can be used to create a beam codebook adapted to the statistical behavior of the propagation environment around that location \cite{Va2018TVT}. In addition, information about the moving target 1 can be exploited to predict the blockage that user 1 experiences, so the network can proactively find an alternative path and mitigate its impact. Finally, the automotive radar signal created at user 2 can be tracked by the radar receiver deployed at the \ac{BS}, and the information about the radar channel can be exploited as a prior to reduce the training overhead of establishing the link between user 2 and the \ac{BS}.   
In Section~\ref{sec:sensing-assisted-comm}, we consider in detail two particular aspects of the network operation that can greatly benefit from the exploitation of sensing data in the ISAC map: the configuration of the antenna arrays  required for directional communication and blockage prediction and management.


 

\section{Waveforms, resource allocation and channel parameter estimation in ISAC networks}
 


\subsection{Multicarrier waveforms for ISAC}

\subsubsection{Fundamentals}

Multicarrier modulation forms the basis for the physical-layer waveform in various contemporary and emerging wireless systems. Good examples are WLAN/WiFi networks, digital video broadcasting systems, and the latest generations of mobile cellular networks, i.e., 4G LTE/LTE-Advanced and 5G NR \cite{Toskala5Gbook}. While there are many alternative multicarrier modulation schemes \cite{RenforsBook,FettweisVTC2009}, the so-called \ac{OFDM} principle \cite{Bingham1990COMMAG} is by far the most commonly adopted approach – including all the previously noted commercial systems. Powered by the involved subchannel or subcarrier structure, multicarrier modulation---and \ac{OFDM} in particular---allows for efficient mitigation of channel time-dispersion in the form of computationally efficient channel equalizers. In general, \ac{OFDM} enables flexible and reconfigurable physical-layer, in terms of multicarrier symbol durations while supporting also backwards compatibility and coexistence of LTE and NR. Complementary filtering and windowing \cite{Ylikaakinen2017JSAC,Ylikaakinen2020TSP} can also be added, either at the \ac{TX} or the \ac{RX} or both, in an essentially transparent manner \cite{Levanen2019WCM}, to enhance the waveform spectral containment. Additionally, \ac{OFDM} and its multiple access variant called \ac{OFDMA} are particularly well-suited for MIMO communications, facilitating efficient frequency-dependent precoding or beamforming. \ac{OFDMA} also allows for harnessing efficiently the \ac{CSI} available at the transmitter, in the form of channel fading responses and interference levels, for link adaptation and scheduling in adaptive modulation and coding based multiuser systems, while rate adaptation in power-domain through, e.g., water-filling is also technically feasible.

\ac{OFDM}/\ac{OFDMA} has also its challenges and limitations. One particular implementation concern is related to the highly dynamic envelope of the transmit waveform, commonly quantified through the \ac{CF} or the \ac{PAPR}. Such highly dynamic envelope is problematic from the power amplifiers (PAs) point of view, as the PA power-efficiency is commonly improved if operating closer towards the saturating region. Such operation point, however, also implies highly nonlinear PA behavior, thus efficient PA linearization through \ac{DPD} is commonly needed – especially in cellular BSs. There exist also different precoded OFDM schemes, most notably the DFT-spread OFDM (DFT-s-OFDM), where the precoding across the subcarriers helps to reduce the PAPR---especially with contiguous spectrum allocations. Such DFT-s-OFDM approach is supported in the uplink of LTE/LTE-Advanced and NR. OFDM is also known to be sensitive to oscillator phase noise, carrier frequency offsets, and the Doppler spread of the mobile radio channel---all primarily because of the long symbol duration of the multicarrier system. These hold particularly when interpreted from the data communications and the related demodulation and decoding perspectives. An alternative multicarrier scheme called \ac{OTFS} modulation offers increased robustness, by design, against the Doppler phenomenon \cite{gaudio2020effectiveness}.

When it comes to sensing and localization in the spirit of ISAC, multicarrier waveforms and MIMO-OFDM in particular are attractive for several reasons. In general, multicarrier waveforms allow for flexible injection of known reference signals in time, frequency and space, to facilitate efficient channel parameter estimation. Such is the key aspect, both from the communications receiver as well as the sensing receiver perspectives. Additionally, while the ordinary receiver implementations build commonly on OFDM symbol-wise FFT processing, extending this to two-dimensional (2D) FFT/IFFT pairs over multiple symbol durations provides basis for accurate delay/range and Doppler/velocity estimation. Such processing leads to the basic delay and Doppler resolutions of the form $\Delta \tau = N/ \Delta f$ and $\Delta f_D = \Delta f/M$, respectively, where $N$ and $M$ refer to the transform sizes in frequency and time, respectively, while $\Delta f$ refers to the subcarrier spacing. One may also straightforwardly, e.g., combine the individual range profiles obtained for the different consecutive OFDM symbols. Importantly, wider bandwidths improve the delay estimation and thereon ranging capability, while longer observation intervals in time allow for improved Doppler and thus velocity estimation. 

Especially in the basic 2D-transform based implementations, the involved \ac{CP} length limits directly the sensing range such that all the involved target reflections and dominant scattering components are within the CP duration. With 30\,kHz subcarrier spacing adopted commonly in the current C-band (3.5\,GHz) 5G NR networks, this still leads to target distances in the order of 350\,m. However, when the networks evolve towards mmWave bands, the symbol durations and the corresponding CP lengths are reduced, thus this may become a more obvious limitations if not properly handled. Additionally, the long symbol durations of OFDM waveforms may easily lead to \ac{ICI}, calling for attention in devising OFDM-based ISAC and sensing systems. When properly handled, such phenomenon can also be turned from a foe to a fried, and described and demonstrated later in this article, in Section~\ref{sec:monostatic}.

The ambiguity function of multicarrier waveforms, measuring the capability to separate multiple coexisting targets, e.g., in range or velocity domains, is impacted by the sidelobes stemming from the FFT processing together with the involved cyclic prefix (CP). Additionally, for example the frequency-sparsity of certain known reference signals, such as the positioning reference signal (PRS) allowing for simultaneous yet orthogonal transmission from multiple nodes, may impose further ambiguity challenges. The ambiquity as well as the ultimate target parameter estimation performance can be impacted through waveform optimization, for which the subcarrier structure of MIMO-OFDM forms an excellent basis. Representative example works are for example \cite{Liyanaarachchi2021TWC, RadCom_Proc_IEEE_2011,OFDM_DFRC_TSP_2021,Bica2019ICASSP}. These aspects are also discussed further in this same section, under Subsection~\ref{sec:resource-allocation}.

\subsubsection{Unified Communication/Localization/Sensing Signal Model with MIMO-OFDM} 
In this part, we provide the unified MIMO-OFDM receive signal model that covers communications, localization and sensing (including both monostatic and bistatic configurations), to be employed throughout the rest of the paper (see Table~\ref{tab_notations} for the notations). Extending the model in \cite{Venugopal2017} to the case of time-varying channels, the received signal $\yy_{n,m} \in \complexset{\Nrfr}{1}$ at subcarrier $n$ and symbol $m$ can be written as\footnote{The \ac{SI} term in monostatic sensing is omitted from \eqref{eq_rec} for ease of exposition, yet it will be duly considered in \eqref{eq_rec_FD}.}
\vspace*{-1mm}
\begin{equation}\label{eq_rec}
    \yy_{n,m} = \WWrf\herm \HH_{n,m} \FFrf \FFbb[n,m] \xx_{n,m} + \zz_{n,m}   ~,
\end{equation}
where $\zz_{n,m}$  is the \ac{AWGN}, $\xx_{n,m} \in \complexset{\Ns}{1}$ contains the transmit symbols of $\Ns$ data streams at subcarrier $n$ and symbol $m$, $\FFbb[n,m] \in \complexset{\Nrft}{\Ns}$ is the digital baseband precoder at subcarrier $n$ and symbol $m$, $\FFrf \in \complexset{\Ntx}{\Nrft}$ is the analog RF precoding matrix applied in the time domain for the entire bandwidth, $\WWrf \in \complexset{\Nrx}{\Nrfr}$ denotes the analog combining matrix at the RX, and $\HH_{n,m} \in \complexset{\Nrx}{\Ntx}$ is the channel at subcarrier $n$ and symbol $m$, given by\footnote{To provide a more generic channel model, it is possible to account for the impact of filters involved in pulse shaping, analog-to-digital (A/D) conversion, and matched filtering through the incorporation of complex coefficients on a per-subcarrier, per-path basis, as shown in \cite[(4)]{Venugopal2017}. Nevertheless, in some practical implementations, subcarriers located within the roll-off region of the combined filter in the frequency domain may be left unused. This approach ensures that the frequency response of the filter remains flat over the active subcarriers, and justifies the adoption of the simplification in \eqref{eq:channel}.}
\begin{align}
    \HH_{n,m} = \sum_{\ell=0}^{L-1} \alpha_{\ell}  e^{-j 2 \pi n \deltaf \tau_{\ell} } e^{j 2 \pi m \Tsym \nu_{\ell} } \arx(\phib_{\ell}) \atx^T(\thetab_{\ell}) ~.\label{eq:channel}
\end{align}

In \eqref{eq:channel}, $\atx(\thetab) \in \complexset{\Ntx}{1}$ and $\arx(\phib) \in \complexset{\Nrx}{1}$ denote the array steering vectors at the TX and RX, $\alpha_{\ell}$, $\tau_{\ell}$,  $\nu_{\ell}$, $\phib_{\ell} = [\phiazl, \phiell]$ and $\thetab_{\ell} = [\thetaazl, \thetaell]$ denote the complex channel gain, delay (including clock offset), Doppler shift (including carrier frequency offset), AOA and AOD of the $\ell$-th path/target, respectively. For localization, we assume that $\ell = 0$ indicates the LOS path, implying that $\alpha_0$ involves the impact of one-way attenuation of the LOS path, while $\alpha_{\ell}$ for $\ell >0$ includes the combined attenuation of the first and second legs of the $\ell$-th reflected/scattered path and the corresponding reflection/scattering coefficient. For sensing, $\alpha_{\ell}$ covers the radar cross section (RCS) of the $\ell$-th target and the two-way attenuation in monostatic sensing (bistatic RCS of the $\ell$-th target and the combined attenuation of the first and second legs associated with the $\ell$-th target, in bistatic sensing). 

We note that for communications, $\HH_{n,m}$ in \eqref{eq:channel} can be usually modeled as frequency selective  yet time-invariant (i.e., not doubly-selective as in localization and sensing), since the channel coherence time is such that the impact of Doppler can be neglected \cite{80211_Radar_TVT_2018}. For localization and sensing, high-mobility applications might necessitate even more comprehensive Doppler modeling that accounts for not only \textit{slow-time} (i.e., inter-symbol) phase shifts represented by $e^{j 2 \pi m \Tsym \nu_{\ell} }$, but also \textit{fast-time} (i.e., intra-symbol) phase progressions. Although fast-time effects can be neglected in low- and medium-mobility scenarios (e.g., target/UE radial velocities below $30 \, \rm{m/s}$) with standard 5G NR FR2 parameters \cite{TR_38211}, they can lead to ICI in high-mobility scenarios \cite{MIMO_OFDM_ICI_JSTSP_2021} and must be taken into account explicitly (see \eqref{eq_y_ici} for further details).

\subsection{Channel estimation}
\subsubsection{Why channel estimation for localization or sensing is different}
Technologies for both positioning and sensing usually involve the estimation and exploitation of some or all of the multipath channel parameters described in \eqref{eq:channel}. Channel estimation is more challenging, however, when the estimated parameters are used for localization  or sensing. First, the required estimation accuracy is higher than that required when the only objective is the design of the communication system. For example, the precoder or combiner designs for communications based on channel estimates are relatively robust to small variations in the AoA or AoDs, while a high accuracy localization algorithm exploiting angular measurements will require very precise estimations (as an example, if we target a localization error of 1 m for a user 50 meter away from a BS, the angle estimation accuracy should be approximately 1 degree). This pushes the limits of the estimation algorithms, increasing complexity and length of the training sequence, which impacts the overall overhead of the system. Second, while for communications channel estimation is usually performed in the frequency domain without need of explicitly extracting the delays, these are key parameters for localization and sensing; moreover, many localization/sensing techniques need precise estimation of the absolute delays, which requires the consideration of an additional parameter in the estimation process, the clock offset between the TX and RX \cite{Palacios2022EUSIPCO}. Third, for communication, it can be assumed that the channel is not varying within the coherence time $T_c=1/\sigma_{D}$, with $\sigma_{D}$ the Doppler spread, so the doubly selective channel model in \eqref{eq:channel} can be simplified to a time invariant frequency selective channel model, where the impact of the Doppler frequencies can be neglected, i.e. $\sigma_\text{D}\Tsym\ll 1$, with $\sigma_\text{D}$ the Doppler spread \cite{Palacios2022EUSIPCO}. In contrast, for sensing, the channel has to be observed over a longer period of time, so the Doppler shifts can also be estimated and exploited for velocity estimation. The joint estimation of these space-time-frequency parameters leads to higher computational complexity solutions than the usual channel estimator for communications-only systems. Finally, an additional element to be considered in some practical systems during the channel estimation process is the impact of pulse shaping, lowpass filtering after downconversion and matched filtering. In this case, the channel model in \eqref{eq:channel} has to be modified to introduce a time domain function which represents all the filtering stages which impact the baseband equivalent model \cite{Venugopal2017}. This function contributes to the entanglement of the channel parameters and complicates its estimation. In summary, channel estimation for joint sensing and communication requires very high resolution and accuracy, which increases complexity and training overhead.

\subsubsection{Techniques for channel estimation}
Recent work on channel estimation has attempted to provide low complexity solutions, very high resolution parameter estimation, or both. Different frequency bands and system architectures lead to different features and structure in the MIMO channel matrices, and many of the channel estimation algorithms have been specifically designed to exploit particular features. Most of the techniques share, however, a common process to sound the channel. First, a number of pilot sequences are transmitted using a given number of training precoders and combiners as spatial filters, and the corresponding received sequences, which follow \eqref{eq_rec}, are collected. These training precoders/combiners have to be designed to sound the channel in the spatial dimension.  The collected measurements are later exploited in the estimation process to extract the multipath parameters. The estimation techniques considered in the literature on MIMO communication or MIMO joint sensing and communication can be classified into three main categories: based on maximum likelihood estimation \cite{Chen2002ICC,Carvajal2013,Fleury1999,Fleury2002,shahmansoori2018}, exploiting compressed sensing \cite{lee2014Globecom,Venugopal2017, Javier2018TWC,Wu2019WCL}, or subspace-based estimators. Some types of techniques might be more suitable for a particular frequency band than others, as discussed in the next paragraphs.

When operating at sub-6 GHz or with relatively small or moderate size antenna arrays, techniques that exploit the idea of maximum likelihood estimation become a solution that can provide high resolution. The conventional ML estimator is optimal \cite{Chen2002ICC} but results in high complexity. Alternative techniques based on expectation maximization (EM) are also effective to provide high accuracy channel estimates but their complexity is still high \cite{Carvajal2013}.  In contrast, the space-alternating generalized expectation maximization (SAGE) algorithm \cite{Fleury1999} and its variations---such as in \cite{Fleury2002, shahmansoori2018}---can provide super resolution at a moderate complexity. 

Channel estimation at mmWave is more challenging than at low frequencies \cite{mmWavetutorial2016}. First, channel estimation is usually performed before array configuration. Since the precoders and combiners at this stage have not been adapted to the channel yet, the directional beam patterns  of the TX and RX are not aligned, and the estimation has to be performed at low or very low  SNR. Second, since a hybrid MIMO architecture is commonly used at mmWave, the channel is observed through the lens of the analog combiner, without direct access to the outputs of every antenna. This way, the analog combiner acts as a compression stage for the receive signal. Finally, the large antenna arrays used at both ends of the link heavily increase the dimensionality of the channel matrices, making unfeasible many of the techniques used at lower frequencies or with smaller arrays. 
Literature on channel estimation at mmWave exploits the sparse nature of the channel to develop suitable solutions at this frequency band. These solutions assume a frequency selective channel model and may provide sufficient information for localization as a byproduct of communication. Different compressed sensing based   techniques---including greedy sparse recovery and nuclear norm or atomic norm minimization---have been proposed in the recent literature \cite{lee2014Globecom,Venugopal2017, Javier2018TWC,Wu2019WCL}. For example, greedy solutions considering frequency selective channels can operate either in the time domain or the frequency domain. In both cases, the channel estimation problem can be formulated as the recovery of a sparse vector. 
For the frequency domain approaches, the dictionaries
are built as a Kronecker product of the array steering vectors at the transmitter and at the
receiver evaluated on a grid for the angle of departure (AoD) and the angle of arrival (AoA) \cite{Javier2018TWC}. In the time domain approaches, the delay domain also has to be considered when building the dictionary. In this case, and assuming \ac{ULA}s at both ends, the received signal for the $k-$th training frame can be written as \cite{Venugopal2017}
\begin{equation}
\mathbf{y}_{k}={\bf{\Phi}}^{(k)}_{\td}\left(\mathbf{I}\otimes {\Atxbar} \otimes \Arx \right)\bf{\Gamma}\mathbf{h}_{\text{vec}} + \mathbf{z}_{k}, \label{eq:rx-sparse}
\end{equation}
where $\mathbf{h}_\text{vec}$  is the sparse vector  containing the time domain complex channel gains after vectorization of the channel matrix, ${\bf{\Phi}}^{(k)}_{\td}$ is the sensing matrix built from the $k$-th training precoder, the $k$-th training combiner and the pilot symbols, $\Atxbar$ is the conjugate of the dictionary for the angle of departure, which contains the transmit steering vectors evaluated on a grid of potential AoDs,  $\Arx$ s the dictionary for the angle of arrival, containing the receive steering vectors evaluated on a grid of potential AoAs, and 
$\Gamma$ is  a dictionary that represents the sparsity in the delay domain.
\begin{figure}[t!]
    \centering
    \includegraphics[width=0.7\textwidth]{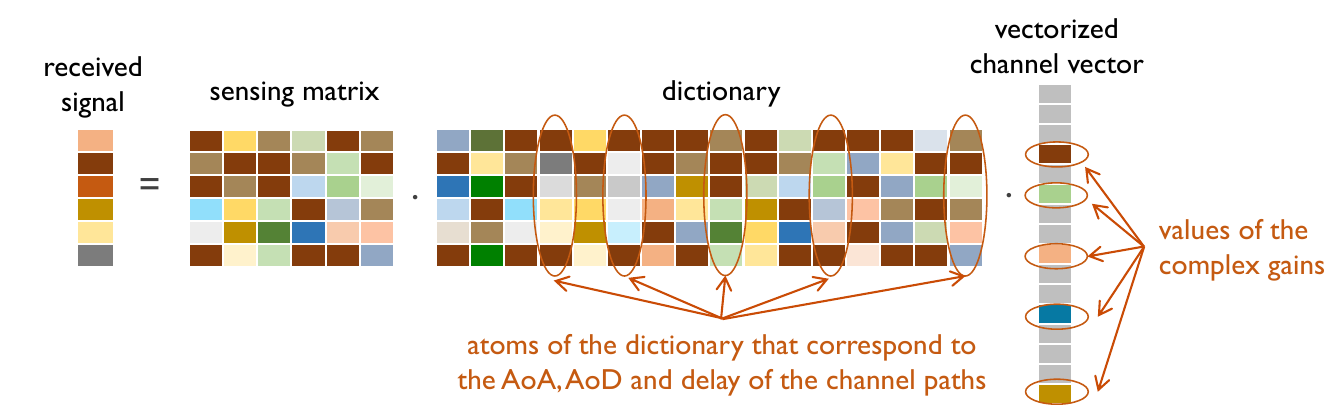}
    \caption{Illustration of the channel estimation algorithms exploiting sparsity. The MIMO channel matrices are expanded in terms of a sparsifying dictionary and vectorized to create the sparse vector to be estimated. Identifying the AoD/AoA and delay boils down to the selection of the columns (atoms) in the dictionary which represent the sparse vector.}
    \label{fig:CS-based-channel-est}
\end{figure}
The observation stacking all the measurements for a number $K$ of training frames can be denoted $\mathbf{y}_{td}$, while the overall sensing matrix that stacks ${\bf{\Phi}}^{(k)}_{\td}$ for all $k$ is denoted as ${\bf{\Phi}}_{\td}$.
The overall sparsifying dictionary is defined as 
\begin{equation}
\mathbf{\Psi}_{\td} = \left({\mathbf{I}}\otimes\ {\Atxbar}\otimes \Arx \right){\bf{\Gamma}}.
\end{equation}
Each column in $\mathbf{\Psi}_{\td}$ corresponds to a given combination of AoD, AoA and delay. Estimating these parameters
is equivalent to identifying the support of the sparse vectorized channel, as illustrated in Fig.~\ref{fig:CS-based-channel-est}. Finding the support and the gains in $\mathbf{h}_{\text{vec}}$ is equivalent to solving the following problem
\begin{align}
\min \Vert \mathbf{h}_{\text{vec}} \Vert_1 \quad\text{such~that}\quad \Vert \mathbf{y}_{td}  - {\bf{\Phi}}_{\td} {\bf{\Psi}}_{\td}\mathbf{h}_{\text{vec}} \Vert_2 \leq \epsilon, \label{eqn:cs_fund_cvx}
\end{align} 
which is the $\ell_1$ relaxation of a spare recovery problem. For both frequency and time domain estimations, highly overcomplete dictionaries for the angular and delay domains have to be exploited to achieve high resolution \cite{Venugopal2017,Javier2018TWC}, making some of the approaches proposed for communications impractical for sensing or localization. This is because a more stringent resolution requirement results in a larger dictionary, which may lead to prohibitive computational complexity or memory requirements. In this context, to reduce complexity, new greedy solutions have been recently proposed to operate with a multidimensional dictionary built as the product of independent and smaller dictionaries instead of a large dictionary based on a Kronecker product \cite{Palacios2022EUSIPCO,Palacios2023GC}. 

Subspace-based techniques have been proposed to estimate the doubly selective MIMO channel with high resolution at different frequency bands, providing Doppler shifts information in addition to delays and angles to enable sensing applications beyond localization. For example, ESPRIT-based channel estimation \cite{Jiang2021WCNC,Wen2020ICASSP,Zhang2021JSTSP} provides good resolution for localization and sensing at moderate complexity. The main limitation of state-of-the-art techniques based on ESPRIT for channel parameter estimation is that they can only operate when the channel model does not include any filtering effect as in \eqref{eq:channel}. An alternative approach that combines the strengths of beamspace ESPRIT for angular estimation  with a dictionary-based sparse recovery solution that targets delay estimation, and can operate when the channel model includes the filtering effect, has been proposed in \cite{Chen2024arxiv}.

\subsubsection{Spatial designs for channel estimation}
The training precoders and combiners used to sound the channel and build the observations for channel estimation  could be directly created from the beam codebooks used in the communication network. However, an enhanced design for training can help to reduce the overhead of channel estimation or to increase the accuracy of the estimation. For example, to reduce the number of measurements when estimating the channel by exploiting a sparse recovery algorithm, it is interesting to design the training precoders and combiners---that lead to a specific ${\bf{\Phi}}_{\td}$---so that the product ${\bf{\Phi}}_{\td} {\bf{\Psi}}_{\td}$ exhibits a low mutual coherence \cite{Bayraktar2022Asilomar,Ge2022TWC}. It is also possible to consider the accuracy of the estimation as the metric that drives the design of the spatial filters. For example, the works in \cite{spatialSignal_TVT_2022,ris_beamdesign_JSTSP_2022,e2e_positioning_wcl_2023} design a new codebook for accurate angle estimation in a downlink localization scenario. The example below shows the significantly better performance provided by the new design.

\begin{example}
We consider a 5G/6G downlink localization scenario with the parameters $\fc = 28 \, \rm{GHz}$, $\deltaf = 120 \, \rm{kHz}$, $N = 1024$ and $M = 20$. For ease of illustration, we consider a downlink multiple-input single-output (MISO) scenario with LOS-only propagation in a two-dimensional setup, where $\Ntx = 16$ and $\Nrx = 1$. In this case, using \eqref{eq_rec} and \eqref{eq:channel}, the received signal at the single-antenna UE is given by
\begin{align} \label{eq_y_signal_design}
    y_{n,m} = \alpha  e^{-j 2 \pi n \deltaf \tau } e^{j 2 \pi m \Tsym \nu }  \atx^T(\theta) \ff_m x_{n,m} + z_{n,m} ~,
\end{align}
where $\ff_m \in \complexset{\Ntx}{1}$ is the RF beamformer at the BS, with controllable amplitude and phase per antenna (i.e., analog active phased array \cite{phasedArray_proc_2016}) and $z_{n,m} \sim \mtCN(0, \sigma^2 )$ denotes additive noise. The BS transmits unit-amplitude pilots $x_{n,m}$ over $N$ subcarriers and $M$ symbols, and the UE aims to estimate the AoD $\theta$ from $y_{n,m}$ in \eqref{eq_y_signal_design}. The goal herein is to design the precoder $\FF  = \left[ \ff_0 \, \ldots \, \ff_{M-1} \right] \in \complexset{\Ntx}{M}$ that maximizes the accuracy of AoD estimation under an a-priori knowledge on $\theta$ (i.e., how to optimally allocate the pilot resources over time to achieve the highest accuracy in AoD estimation). This a-priori knowledge is quantified by an AoD uncertainty interval $\Um = [\theta - \thetad, ~ \theta + \thetad]$. We evaluate the performance of two codebooks used to construct $\FF$: 
\begin{itemize}
    \item \textbf{Conventional Codebook:} conventional \textit{directional} codebook employed in 5G NR mmWave systems \cite{beamSweep_mmWave,dwivedi2021positioning}, given by
    \begin{align} \label{eq_ffdir}
        \FFdir = [ \atx(\theta_1) ,\, \ldots \,, \atx(\theta_{2G}) ]^{*} ~.
    \end{align}
    \item \textbf{New Codebook:} recently proposed \textit{directional/derivative} codebook \cite{spatialSignal_TVT_2022,ris_beamdesign_JSTSP_2022,e2e_positioning_wcl_2023} (similar to \textit{sum/difference} beams used in monopulse radar \cite{monopulse_review}) given by
    \begin{align} \label{eq_ffdir_der}
        \FFdird = [ \atx(\theta_1), \, \ldots \, , \atx(\theta_G), \, \atxdt(\theta_1), \, \ldots \, \atxdt(\theta_G) ]^{*} ~.
    \end{align}
\end{itemize}
Here, $\atxdt(\theta) = \partial \atx(\theta) / \partial \theta$, $\{ \theta_g \}_{g=1}^{2G}$ represent uniformly sampled grid points from $\Um$, and each column of $\FFdir$ and $\FFdird$ is normalized to have unit norm.

Fig.~\ref{fig_dir_der} showcases the beampatterns of both directional and derivative beams. The incorporation of derivative beams $\atxdt(\theta)$ alongside standard directional beams $\atx(\theta)$ is motivated by the need for the UE to detect subtle deviations around the intended direction $\theta$ and also supported by the theoretical CRLB analysis \cite{spatialSignal_TVT_2022,ris_beamdesign_JSTSP_2022}. The sharp curvature around $\theta$ in the beampattern of $\atxdt(\theta)$ ensures that slight perturbations in angle result in significant changes in amplitude, which enables precise mapping of angles based on complex amplitude measurements. Hence, high-accuracy AoD estimation and localization in 5G/6G systems can be achieved by a judicious combination of directional and derivative beams. It is worth emphasizing that these localization-optimal beams are different from those used in communications (i.e., directional beams for sweeping an angular region of interest as in \eqref{eq_ffdir}).


To evaluate the AoD estimation performance of the codebooks in \eqref{eq_ffdir} and \eqref{eq_ffdir_der}, we construct $\FF$ in \eqref{eq_y_signal_design} by selecting its columns from these codebooks. The time sharing of the columns of $\FFdir$ over $M$ symbols follows a uniform strategy, while that of $\FFdird$ is optimized based on the CRLB criterion \cite{spatialSignal_TVT_2022}. Fig.~\ref{fig_rmse_vs_aod} shows the AoD \ac{RMSE} performances with respect to the AoD of the UE for $\thetad = 1^\circ$ at $\snr = \lvert \alpha \rvert^2 \Ntx / \sigma^2 = 0 \, \rm{dB}$, using the maximum-likelihood estimator \cite[Eq.~(11)]{e2e_positioning_wcl_2023}. We observe substantial improvements in AoD estimation accuracy with the use of $\FFdird$ compared to the traditional 5G codebook $\FFdir$, suggesting significant potential for achieving extreme location accuracy in 6G through innovative beam designs and resource allocation.

\begin{figure}
	\centering
	\includegraphics[width=0.5\linewidth]{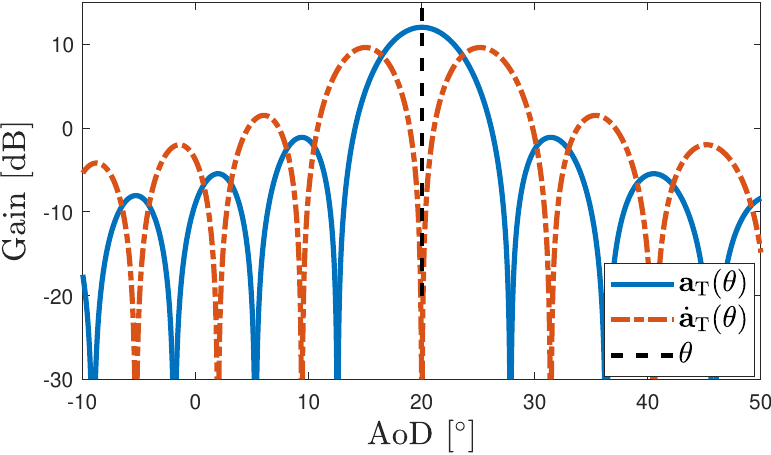}
	\caption{$16$-element ULA beampatterns of directional $\atx(\theta)$ and derivative $\atxdt(\theta)$ beams for the UE located at an AoD $\theta = 20^{\circ}$ with respect to the BS. The directional beam ensures the necessary SNR for AoD estimation, whereas the derivative beam assists the UE in detecting subtle deviations from the targeted direction $\theta$, as indicated by its pronounced curvature around $\theta$. The combined use of directional and derivative beams allows for high-accuracy tracking of the UE in 5G/6G mmWave scenarios.}
	\label{fig_dir_der}
\end{figure}

\begin{figure}
	\centering
	\includegraphics[width=0.5\linewidth]{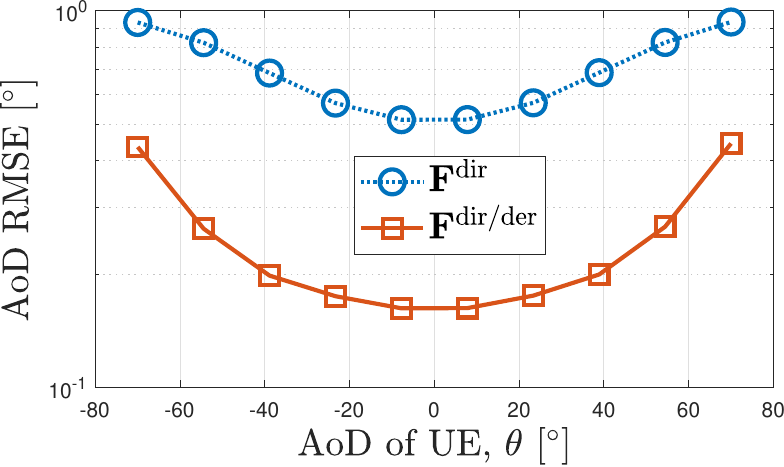}
	\caption{AoD estimation performance with respect to the AoD of the UE, achieved by the considered codebooks in \eqref{eq_ffdir} and \eqref{eq_ffdir_der}.}
	\label{fig_rmse_vs_aod}
\end{figure}

\end{example}

\subsection{Resource allocation} \label{sec:resource-allocation}
\subsubsection{Fundamentals}
In the ISAC architecture, the sensing and communication (S\&C) functions are simultaneously performed based on the unified waveforms to improve the spectrum efficiency as well as reduce the hardware costs, where the radio resources are allocated to achieve the optimal tradeoff between the S\&C performance. Therefore, the design of resource allocation schemes are evaluated and guided by the performance metrics of the dual functions in ISAC systems, which are discussed in detail as follows.

The communication performance of the ISAC system is usually measured by the maximum achievable rate of reliable information transmission over the channel, i.e., the channel capacity, which can be further represented by the maximum mutual information between the unified waveform and the communication symbols, or simply the \ac{SINR} based on the Shannon formula. As for the sensing performance, while it can be evaluated from the detection perspective where the existence of the target is determined based on the received signals, we refer sensing to recovering the target information from the noisy measurements in this section. Then the sensing performance can be measured by the estimation error of target states, which is characterized via the Fisher information analysis in the literature\cite{WinDaiShe:J18}. Namely, let $\hat{\mathbf{s}}$ denote the unbiased estimator of the actual target state vector $\mathbf{s}$, which may include the position $\mathbf{p}$, the orientation $\mathbf{\psi}$, the velocity $v$, and other states of interest. Then according to the information inequality, there exists
\begin{equation}
\label{eq:inforeq}
\mathbb{E}\left\{(\hat{\mathbf{s}}-\mathbf{s})(\hat{\mathbf{s}}-\mathbf{s})^{\mathrm{T}}\right\} \succeq \mathbf{J}^{-1}(\mathbf{s})
\end{equation}
where $\mathbf{J}(\mathbf{s})$ denotes the Fisher information matrix (FIM) given by
\begin{equation}
\label{eq:FIMexp}
\mathbf{J}(\mathbf{s}) = \mathbb{E}\left\{\left[\frac{\partial}{\partial \mathbf{s}} \ln f(\mathbf{y} ; \mathbf{s})\right]\left[\frac{\partial}{\partial \mathbf{s}} \ln f(\mathbf{y};\mathbf{s})\right]^{\mathrm{T}}\right\},
\end{equation}
and the likelihood function $f(\mathbf{y};\mathbf{s})$ is determined by the signal model \eqref{eq_rec}. When the target position $\mathbf{p}$ is concerned, the \ac{SPEB} can be applied to measure the sensing performance in ISAC systems, which is given by
\begin{equation}
 \mathcal{S}(\mathbf{p}) = \mathrm{tr}\big\{\mathbf{J}_\mathrm{e}^{-1}(\mathbf{p})\big\}.
\end{equation}
The notation $\mathbf{J}_\mathrm{e}(\mathbf{p})$ denotes the \ac{EFIM} for $\mathbf{p}$, which is obtained by calculating the Schur complement of submatrix in the original \ac{FIM} $\mathbf{J}(\mathbf{s})$\cite{SheWin:J10}.                           
\subsubsection{Problem formulations}
Note that it remains an open challenge to establish a universal theoretical framework to jointly evaluate the performance of dual-functions in ISAC systems, which can integrate the classical results derived from information theory and estimation theory, respectively. As a consequence, the resource allocation in ISAC no longer follows a unified problem formulation as in localization networks. In particular, the wireless resources including time, frequency, space, power, and code, are allocated to solve the optimization problems with various objective functions, which can be mainly classified into three categories in the literature, i.e., the sensing-oriented formulation, the communication-oriented formulation, and the joint formulation.
\begin{itemize}
    \item 
For the sensing-oriented formulations, the power, space, and other kinds of resources are allocated to optimize the objective functions derived based on sensing metrics, such as the \ac{SPEB} and the detection probability. For example, the sensing-oriented joint waveform, precoding, and combining design in monostatic ISAC systems with signal model \eqref{eq_rec} can be modeled as
\begin{equation}
\label{eq:sen-orien}
\begin{aligned}
&\underset{p(\mathbf{x}),\mathbf{W}_\mathrm{c},\mathbf{W}_\mathrm{s},\mathbf{F}_\mathrm{RF},\mathbf{F}_\mathrm{BB}}{\text{minimize}}  & & \mathbb{E}_{\mathbf{x}}\{\mathcal{S}(\mathbf{p};\mathbf{W}_\mathrm{s},\mathbf{F}_\mathrm{RF},\mathbf{F}_\mathrm{BB})|\mathbf{x}\} \\
&\quad \ \ \text {subject to } & & \mathbf{h}_\mathrm{c}(\mathbf{x},\mathbf{W}_\mathrm{c},\mathbf{F}_\mathrm{RF},\mathbf{F}_\mathrm{BB}) \geq \mathbf{\gamma}_\mathrm{c}\\ 
& & & \mathbf{f}(\mathbf{x},\mathbf{W}_\mathrm{c},\mathbf{W}_\mathrm{s},\mathbf{F}_\mathrm{RF},\mathbf{F}_\mathrm{BB}) \leq \mathbf{0}
\end{aligned} 
\end{equation}
where the objective function denotes the average \ac{SPEB} conditioned on the random transmitted symbols $\mathbf{x}_{n,m}$\cite{MilCho:J78}, the function $\mathbf{h}_\mathrm{c}$ denotes the communication constraints, e.g, the communication \ac{SINR} or the sum-rate, and $p(\mathbf{x})$ denotes the distribution for transmitted symbols. The functions $\mathbf{f}$ still denote the power and structure constraints for the symbols $\mathbf{x}$, the precoding matrices $\mathbf{F}_\mathrm{RF},\mathbf{F}_\mathrm{RB}$ at the transmitter, and the combining matrices $\mathbf{W}_\mathrm{c},\mathbf{W}_\mathrm{s}$ at the communication and sensing receivers. Note that the joint problem \eqref{eq:sen-orien} is hard to solve due to the non-convexity and tight coupling among optimization variables. The \ac{SDR} and \ac{SCA} methods are applied to provide high-quality solutions with acceptable computation costs\cite{LiuYuaChr:J20,LiuChrLi:J18,ZhaZhaLiu:J23}. In addition, the above formulation reduce to the resource allocation in localization and sensing networks with the communication constraints $\mathbf{h}_\mathrm{c}$ removed, where more theoretical insights as well as efficient schemes are provided in this scenario. For example, the sparsity property of power allocation has been revealed in the localization networks, indicating that an optimal power allocation strategy requires only three anchor points to localize an agent\cite{DaiSheWin:J18}. Furthermore, the robust strategies are incorporated in both power allocation and spatial design to account for the uncertainties associated with network parameters essential for the design of resource allocation schemes\cite{LiSheZha:J13,ZhaZhaShe:J20}.
\item 
In the communication-oriented resource allocation for ISAC, the communication metrics such as the mutual information and the network throughput are maximized through efficient resource management strategies including power allocation and beamforming design, i.e., 
\begin{equation}
\label{eq:comm-orien}
\begin{aligned}
&\underset{p(\mathbf{x}),\mathbf{W}_\mathrm{c},\mathbf{W}_\mathrm{s},\mathbf{F}_\mathrm{RF},\mathbf{F}_\mathrm{BB}}{\text{maximize}}  & & I(\mathbf{y}^\mathrm{c};\mathbf{x}) \\
&\quad \ \ \text {subject to } & & \mathbf{h}_\mathrm{s}(\mathbf{x},\mathbf{W}_\mathrm{s},\mathbf{F}_\mathrm{RF},\mathbf{F}_\mathrm{BB}) \leq \mathbf{\gamma}_\mathrm{s}\\ 
& & & \mathbf{f}(\mathbf{x},\mathbf{W}_\mathrm{c},\mathbf{W}_\mathrm{s},\mathbf{F}_\mathrm{RF},\mathbf{F}_\mathrm{BB}) \leq \mathbf{0}
\end{aligned} 
\end{equation}
where the objective function $I(\mathbf{y}^\mathrm{c};\mathbf{x})$ denotes the mutual information between the received communication signals $\mathbf{y}_{n,m}^\mathrm{c}$ and the transmitted symbols $\mathbf{x}_{n,m}$, which can be calculated by the Shannon formula based on the precoding matrices $\mathbf{F}_\mathrm{RF},\mathbf{F}_\mathrm{BB}$ and the combining matrix $\mathbf{W}_\mathrm{c}$. The functions $\mathbf{h}_\mathrm{s}$ denote the sensing constraints, e.g., the sensing accuracy or the deviation  of  the  actual sensing beam from an ideal sensing beam pattern\cite{DinWanZha:J22,XuYuNg:J22}. To solve \eqref{eq:comm-orien}, auxiliary variables are introduced to decompose the original problem, after which the convex relaxation techniques can be applied to provide efficient suboptimal solutions\cite{WanLiGov:J19,MuOuJin:J23}. Furthermore, the two-fold tradeoff consisting of the subspace tradeoff and the deterministic-random tradeoff is revealed in terms of the communication-oriented waveform design with optimal sensing performance constraint, which provides insights for the design and analysis of practical systems\cite{IT_JCAS_Gaussian_2023}.
\item 
In contrast to the above discussions, the S\&C requests hold equal status in the joint formulation of resource allocation for ISAC. For example, the objective function can be designed to involve both S\&C performance measures, in which sense the optimization problem can be modeled as
\begin{equation}
\label{eq:joint}
\begin{aligned}
&\underset{p(\mathbf{x}),\mathbf{W}_\mathrm{c},\mathbf{W}_\mathrm{s},\mathbf{F}_\mathrm{RF},\mathbf{F}_\mathrm{BB}}{\text{maximize}}  & & w_\mathrm{c}R_\mathrm{c}+w_\mathrm{s}R_\mathrm{s} \\
&\quad \ \ \text {subject to } & & \mathbf{f}(\mathbf{x},\mathbf{W}_\mathrm{c},\mathbf{W}_\mathrm{s},\mathbf{F}_\mathrm{RF},\mathbf{F}_\mathrm{BB}) \leq \mathbf{0}\\ 
\end{aligned} 
\end{equation}
where the estimation rate $R_\mathrm{s}$ is introduced as an analog to the communication rate $R_\mathrm{c}$, which measures the reduction in entropy of the target states after estimation\cite{AleBryGar:J16}, and $w_\mathrm{c}, w_\mathrm{s}$ denote the weights for S\&C performance, respectively. The concept termed value of service (VoS) can also be applied to design a proper objective function for \eqref{eq:joint}, where the communication and sensing VoS is defined based on the classical S\&C metrics, respectively, e.g., sensing CRLB and communication \ac{SINR}. Then the radio resources are allocated to optimize the weighted sum of VoS from the S\&C functions\cite{LiWanXin:J23}. Additionally, the objective function can refer to the total resource consumption of the ISAC system in the joint formulation, where the S\&C performance are guaranteed by certain constraints\cite{YanWeiFen:J21}.
\end{itemize}

\section{Technologies for joint bistatic and multistatic sensing and communication}
\label{sec:bistatic-and-multistatic}

\subsection{Introduction}

In this section, we will cover sensing scenarios where the transmitters and receivers are separated. Sensing where transmitter and receiver are co-located will be treated in Section \ref{sec:monostatic}.
When sensing is based on one transmitter and one receiver, it is called bistatic \cite{Puc22}, while sensing based on several transmitters or receivers is called multistatic sensing. Since bistatic sensing can be readily implemented in communication systems, it has been covered extensively in the literature. In contrast, multistatic sensing has received rather limited treatment so-far in communications \cite{fascista2023uplink,vukmirovic2019direct}, but is a classic topic in the radar community \cite{haimovich2007mimo}. 

Bistatic sensing itself is a rich and multi-faceted field. Before we delve into the technical aspects, we first provide a brief overview of the key concepts, summarized in Fig.~\ref{fig:bistatic-overview}. First, the most important use of bistatic sensing is positioning, whereby a \ac{UE} performs bistatic sensing with several \acp{BS}, based on which the UE location can be inferred. 
The positioning topic will be treated in detail in Section \ref{sec:positioning}. Second, bistatic and multistatic sensing, which will be covered in Section  \ref{sec:bistatic-sensing} involves  transmitters and receivers with known locations (e.g., \acp{BS}, but possibly also \acp{UE}), 
to detect and localize objects in the environment, such as vehicles, pedestrians, or buildings. Third, there is the combination of sensing and positioning, known as \ac{SLAM}, which will be covered in Section \ref{sec:SLAM}  involves a \ac{UE} determining its position, while detecting and localizing objects, based on signals to/from \acp{BS}. We note that more traditional \ac{SLAM}, where the \ac{UE} localizes itself and maps the environment based only on sensed backscattered signals is deferred to Section  \ref{sec:monostaticSLAM}.

\subsection{Radio positioning}
\label{sec:positioning}

\subsubsection{Fundamentals of position and orientation estimation}

At its core,  radio positioning aims to estimate the 3D location of the \ac{UE} in a global coordinate system, based on signals to or from one or more  \acp{BS}, each of the form \eqref{eq_rec} \cite{wymeersch2022radio,trevlakis2023localization}. The \acp{BS} are assumed to have known positions and orientations. 
Positioning is thus often a two-stage process, whereby first the channel parameters, i.e.,  the \ac{AoA}, \ac{AoD}, \ac{ToA}, and Doppler of the \ac{LoS} path with respect to each \ac{BS} are estimated, and in a second stage the \ac{UE} location is estimated from the channel parameters. If the \ac{LoS} path is blocked, dedicated \ac{NLoS} detection routines can detect this phenomenon and discard the corresponding measurements \cite{guvencc2007nlos}, or the \ac{NLoS} paths can be used to solve the positioning problem \cite{wymeersch20185g}.

Positioning is generally based on dedicated pilot signals, rather than on unknown data, as this facilitates the channel parameter estimation process and provides more control to improve the resolution and accuracy (see Section \ref{sec:resource-allocation}) \cite{dwivedi2021positioning}. 
Once the \ac{LoS} channel parameter estimates are available, they can be related to the \ac{UE} location. This relation is generally affected by nuisance parameters, e.g., the \ac{LoS} \ac{ToA} in \eqref{eq_rec} say  $\tau_{0} $, assuming this \ac{LoS} path is not blocked, is of the form 
\begin{align}
    \tau_{0} = \Vert \mathbf{p}_{\text{UE}}-\mathbf{p}_{\text{BS}} \Vert /c - \tau_{\text{bias}}, \label{eq:TOA-pos-relation}
\end{align}
where $\mathbf{p}_{\text{UE}}$ is the \ac{UE} position, $\mathbf{p}_{\text{BS}}$ is a \ac{BS} position, $c$ is the speed of light and  $\tau_{\text{bias}}$ is a clock bias of the \ac{UE} with respect to the same \ac{BS} \cite{camajori2023feasibility}. Similarly, the Doppler measurements are affected by the \ac{CFO} between the \ac{UE} and \ac{BS}, and the angles at the \ac{UE} side (i.e., \ac{AoA} in \ac{DL} or \ac{AoD} in \ac{UL}) depend on the unknown user orientation, which is a 3D unknown. This implies that when certain measurements are used for localizing the user, the corresponding nuisance parameters must also be estimated. On the positive side, this means that there are possibilities to jointly estimate the \ac{UE} location while synchronizing it to the network (thanks to the estimation of the clock bias and \ac{CFO}) \cite{koivisto2017joint}, and estimating the complete 6D \ac{UE} pose \cite{nazari2023mmwave,chen2022joint}. 

Mathematically, the \ac{UE} positioning problem is of the form \cite{trevlakis2023localization}
\begin{align}
    \mathbf{y}_{\text{meas}}= \mathbf{f}(\mathbf{x}_{\text{state}}) + \mathbf{n}, \label{eq:positioning-generic}
\end{align}
where $\mathbf{y}_{\text{meas}}$ comprises the estimated angles, delays, and Dopplers, $\mathbf{x}_{\text{state}}$ comprises the 3D \ac{UE} location as well as any nuisance parameters (clock bias, \ac{CFO}, 3D \ac{UE} orientation) and $\mathbf{f}(\cdot)$ is known nonlinear mapping (e.g., containing components of the form \eqref{eq:TOA-pos-relation}, which depends on the known locations and orientations of the \acp{BS}. Recovering $\mathbf{x}_{\text{state}}$ from \eqref{eq:positioning-generic} can be done by, e.g., a least squares or maximum likelihood approach \cite{gustafsson2005mobile}. These problems are generally non-convex, due to the nonlinear relation between the measurements and the \ac{UE} state, so heuristics/approximations/relaxations are employed to to find the global optimum \cite{wang2016robust}. Alternatively, prior information about the \ac{UE} state can be utilized to infer $\mathbf{x}_{\text{state}}$, e.g., when applying tracking filters \cite{lu2023bayesian}.

\begin{figure}
    \centering
    \includegraphics[width=0.6\columnwidth]{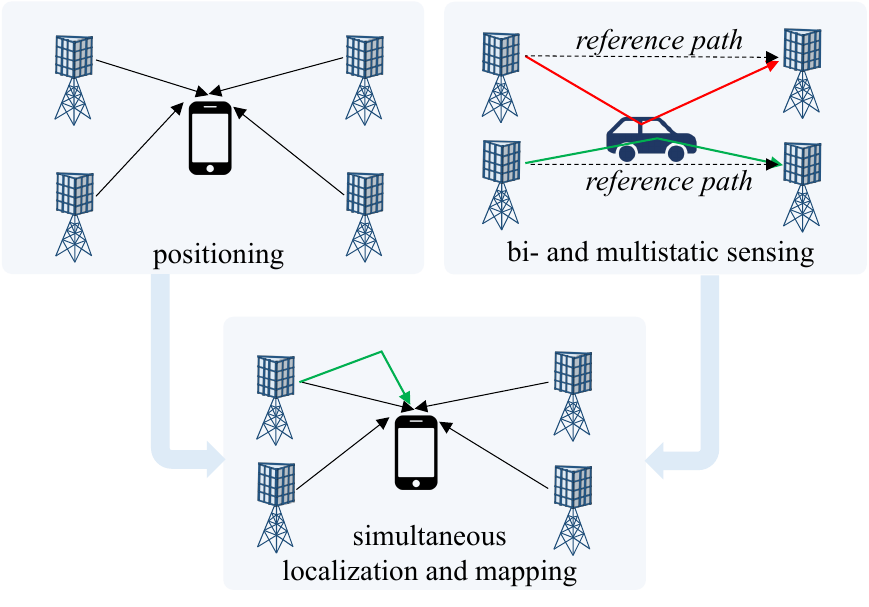}
    \caption{Breakdown of bistatic and multistatic sensing. Operation is shown in downlink, but can equivalently occur in uplink.}
    \label{fig:bistatic-overview}
\end{figure}


\subsubsection{Minimal problems}
\label{sec:mininal-positioning}

Whereas communication to a \ac{UE} in principle requires connection to only a single \ac{BS}, the same does not hold for positioning, which generally needs a much larger number of connected \acp{BS}, especially when some of them may have a blocked \ac{LoS} path to the \ac{UE}. For that reason, understanding the minimal infrastructure needs for positioning as well as the design of new positioning methods or technologies that can reduce the reliance on infrastructure are of great interest. These cases are called minimal problems/minimal solvers \cite{burgess2015toa}, in the sense that if measurements or technologies are removed, the problem can no longer be solved (e.g., in the sense of leading to an infinite number of solutions), which also facilitate outlier detection using RANSAC \cite{li2019massive}.

\begin{example} \label{ex:3BS-positioning}
Consider a scenario with 4 synchronized single-antenna \acp{BS} and a single-antenna \ac{UE}, in \ac{LoS} to all the \acp{BS}. This scenario is visualized in Fig.~\ref{fig:bistatic-overview}. 
Based on \ac{DL} pilots, the \ac{UE} estimated the \ac{ToA} from each \ac{BS}, which provides 4 observations of the form \eqref{eq:TOA-pos-relation}. These delay measurements are sufficient to determine the 3D \ac{UE} location and the 1D clock bias. 
\begin{align}
\hat{\mathbf{p}}_{\text{UE}},\hat{\tau}_{\text{bias}}& =\arg \min_{{\mathbf{p}}_{\text{UE}},{\tau}_{\text{bias}}} f({\mathbf{p}}_{\text{UE}},{\tau}_{\text{bias}})\\
    f({\mathbf{p}}_{\text{UE}},{\tau}_{\text{bias}}) & =\sum_{i=1}^{4}\frac{1}{2\sigma^2_i}\vert \hat{\tau}_{0,i} -\Vert \mathbf{p}_{\text{UE}}-\mathbf{p}_{\text{BS},i} \Vert /c + \tau_{\text{bias}}\vert^2 
\end{align}
where $\sigma_i$ is the standard deviation of the \ac{LoS} \ac{ToA} measurement $ \hat{\tau}_{0,i} $ with respect to \ac{BS} $i$, with location $\mathbf{p}_{\text{BS},i}$.  This problem can be solved iteratively from an initial guess \cite{chan1994simple}.
This clock bias can also be removed by computing 3 (correlated) \ac{TDoA} measurements. 
\end{example}

From this example, the reliance of several \acp{BS} becomes apparent. This reliance can be reduced in a number of ways:
\begin{itemize}
    \item \emph{Additional measurements:} for instance, when augmenting \ac{ToA} measurements with \ac{DL} \ac{AoD} measurements, the number of \ac{BS} can be reduced, but at the cost of more complex multi-antenna \acp{BS} and possibly longer transmission times to support beam sweeping \cite{dwivedi2021positioning}.  Such measurements are further discussed in Sections \ref{sec:FR1-positioning}--\ref{sec:FR2-positioning}. In addition, carrier phase measurements  (i.e., the phase of $\alpha_{\ell}$) can provide extremely precise, but ambiguous location information \cite{dun2020positioning,Talvitie_JSTSP_2023}, as discussed in Section \ref{sec:CPP}.
    \item \emph{Multipath exploitation:} So-far, we have considered the \ac{LoS} path for positioning. While this provides the most direct position information (see again \eqref{eq:TOA-pos-relation}), the \ac{NLoS} paths also provide information, provided they can be resolved. In particular, single-bounce \ac{NLoS} paths are characterized by a single 3D incidence point. While this incidence point is unknown, the cardinality of the measurements provided by the path (e.g., delays and angles) can outweigh the unknowns, and thus improve positioning \cite{wymeersch20185g,chen2022joint}, as will be discussed in Section \ref{sec:FR2-positioning}. Even when the measurements are few, multipath can be leveraged by considering the user at different time instances, see Section \ref{sec:SLAM} \cite{Let:19,gentner2016multipath}.
    \item \emph{New technologies:} Since positioning relies on pilot signals, there is in principle no need for using full-fledged \acp{BS}. Instead, simple beacons may be sufficient \cite{chen2023riss}. Alternatively, low-cost hardware such as \acp{RIS} can be deployed to provide additional controlled multipath components \cite{Elz:21,Wan:22}. More on \acp{RIS} in Section \ref{sec:RIS}. 
    
    \item \emph{New signals:} Conventional \ac{UL} and \ac{DL} signals can be complemented with direct links between \acp{UE}, so-called sidelinks \cite{liuLiaXia:J21}. Such links not only provide additional measurements but also support cooperative, peer-to-peer positioning. These are described in Section \ref{sec:sidelink}.
    
    \item \emph{New methods:} In cases where the channel is complex and the \ac{LoS} cannot easily be extracted, data-driven methods can learn patterns that are beyond the realm of model-based signal processing, see Section \ref{sec:ML-positioning} \cite{li2019machine}. Complementary to data-driven methods, advances in \ac{SLAM} have provided means to perform positioning with reduced infrastructure \cite{kim2022pmbm}, see Section \ref{sec:SLAM}.
\end{itemize}

\begin{figure}
    \centering
    \includegraphics[width=0.7\columnwidth]{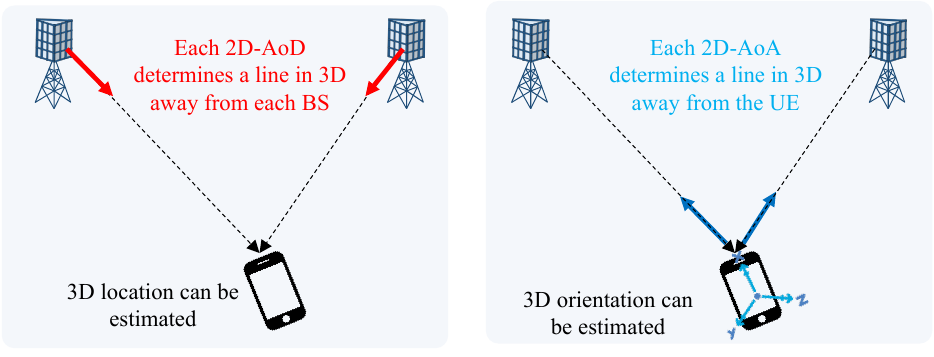}
    \caption{Left: Example of how a \ac{UE} 3D position can be estimated based on the 2D-\ac{AoD} from 2 multi-antenna \acp{BS}. Right: once the 3D \ac{UE} location is known, a multi-antenna UE can use the 2D-\ac{AoA} to determine its 3D orientation. }
    \label{fig:AOAAOD-examples}
\end{figure}

\begin{example} \label{ex:AOD-AOA}
    Consider a case with 2 \acp{BS}, each equipped with a planar array, transmitting downlink pilots to a  \ac{UE}. From the downlink signals, the \ac{UE} can estimate the 2D-\ac{AoD} from each \ac{BS}, i.e., the azimuth and elevation angle. Under noise-free measurements, these angles constrain the \ac{UE} to lie on the intersection of 2 lines in 3D, which is a unique point. See the left side of Fig.~\ref{fig:AOAAOD-examples}.
    If now the \ac{UE} is equipped with a planar array as well, it can estimate the 2D-\ac{AoA} from each \ac{BS}, again in azimuth and elevation. This determines 2 lines in 3D away from the UE. Since the 2 lines are parameterized by 4 parameters and the \ac{UE} orientation has only 3 degrees of freedom, the \ac{UE} orientation can be uniquely determined. See the right side of Fig.~\ref{fig:AOAAOD-examples}.
\end{example}

These examples show the ability of different types of measurement to provide complementary information to the \ac{TDoA} measurements described in Example \ref{ex:3BS-positioning}. A set of additional examples is provided in Table \ref{tab:CoverageAnalysis}. The table provides in particular a more detailed look at the role of uncontrolled and controlled multipath. 

\newcolumntype{P}[1]{>{\centering\arraybackslash}m{#1}}
\begin{table}[]
    \centering
    \resizebox{0.6\columnwidth}{!} {
    \begin{tabular}{|P{2cm}|P{2cm}|P{2cm}|P{2cm}|}
    \hline
    \rowcolor[HTML]{d7d8f9}  \textbf{Approach} & \textbf{3D location} & \textbf{3D orientation} & \textbf{6D pose} \\
     
    \hline
    single-antenna \ac{BS} & 4 BS (TOA) and single-antenna UE & 2 BS and 2D-AOA at multi-antenna UE, known location & 4 BS (TOA) or 3 BS (2D-AOA at multi-antenna UE) \\
    \hline
    multi-antenna \ac{BS} & 2 BS (2D-AOD) & 2 BS and 2D-AOA at multi-antenna UE, known location & 2 BS (2D-AOD) and 2D-AOA at multi-antenna UE\\
    \hline
    multi-antenna \ac{BS} with multipath exploitation & 1 BS (2D-AOD, TOA), 1 IP, multi-antenna UE (2D-AOA)   & 1 BS (2D-AOD, TOA), 1 IP, multi-antenna UE (2D-AOA) & 1 BS (2D-AOD, TOA), 1 IP, multi-antenna UE (2D-AOA) \\
    \hline
     single-antenna \ac{BS} with RIS & 1 BS (TOA) and 1 RIS  (TOA and 2D-AOD) & 1 BS (2D-AOA) and 1 RIS (2D-AOA) & 1 BS (TOA, 2D-AOA) and 1 RIS (TOA, 2D-AOA) \\
     \hline 
    \end{tabular}
    }
     \vspace{0.1mm}
    \caption{Examples of minimal configurations needed to solve the \ac{DL} 3D localization problem without a priori knowledge of \acp{IP} locations. }
    \label{tab:CoverageAnalysis}
\end{table}

\subsubsection{Positioning in sub-6 GHz}
\label{sec:FR1-positioning}
Before the introduction of 5G, cellular positioning was focused exclusively on the sub-6 GHz band, so-called \ac{FR1}. This band has several characteristics of relevance for positioning: 
\begin{itemize}
    \item \emph{Limited bandwidth:} In FR1, bandwidths on the order of 5-20 MHz are available, which limits the distance resolution of delay-based measurements to around 15-60 meters \cite{keating2019overview}. Hence, if \ac{NLoS} paths arrive within 15 meters of the \ac{LoS} path, these paths will merge and appear as one path to standard signal processing methods. This means that delay-based positioning is expected to be poor in cluttered environments (on the order of tens of meters error). 
    \item \emph{Limited array sizes:} At the \ac{UE} side, arrays are generally very small, which means that there is limited angle resolution and thus no possibility to accurately estimate the \ac{UE} orientation. At the \acp{BS}, since the introduction of massive \ac{MIMO} in 5G, larger arrays have been considered with on the order of 64-128 antenna elements. These can provide some amount of angle resolution, provided paths are well separated in the angle domain, as seen from the \ac{BS} \cite{garcia2017direct}. 
    \item \emph{Rich channel:} The challenge of limited resolution in delay and angle is further exacerbated by the richness of the channel \cite{gao2015massive}. This means that the channel matrix $  \HH_{n,m} $ comprises many clusters of paths, coming from many directions, and these clusters may be affected by shadowing, diffraction, multi-bounce reflection, and scattering. From a communication perspective, these effects are combined in statistical models, giving rise to Rayleigh, Rician, or Nakagami fading, which have only a weak relation to the underlying geometry. From a positioning perspective, such models are questionable, not only because they mask the relation to the geometry, but also because they cannot capture the site-specific nature of the channel, which is of direct importance to positioning \cite{TR38.901}. 
\end{itemize}

Due to these characteristics in FR1, conventional methods, e.g., based on FFTs or correlations often perform relatively poorly. To overcome this poor performance, two directions have been pursued: the first is based on machine learning (see Section \ref{sec:ML-positioning}), e.g., in the form of fingerprinting, where the richness of the channel is considered a benefit \cite{li2019machine}. Such methods can bring down location errors below the 10-meter mark, but come at a cost of training complexity, as labeled training data ([fingerprint, location] pairs) must be collected. The second track is based on super-resolution methods \cite{pesavento2023three}. These methods are based on the principle that even if paths differ only to a very small extent, they can be resolved if the \ac{SNR} is sufficiently high.


\subsubsection{Positioning in mmWave}
\label{sec:FR2-positioning}

At mmWave bands, the situation is significantly easier from a positioning perspective. Let us reconsider the characteristics from \ac{FR1} and evaluate them from the \ac{FR2} (24-70 GHz) and sub-THz (100-300 GHz) perspective \cite{shastri2022review,chen2022tutorial}.
\begin{itemize}
    \item \emph{Large bandwidth:} At \ac{FR2} significantly large bandwidths are available, up to 400 MHz, which corresponds to a distance resolution of less than one meter. Hence, delay-based positioning becomes possible in complex and relatively cluttered environments. At sub-THz, the trend is expected to continue, with bandwidths on the order of 1 or more GHz becoming available, with corresponding distance resolution below 30 cm. On the other hand, the ability to provide better accuracy also means that synchronization requirements among \acp{BS} become more strict, and thus more challenging. 
    \item \emph{Large arrays (normalized to the wavelength):} At the \ac{UE} side, even modest arrays of 16-32 elements can provide good angle resolution, not only providing a path towards orientation estimation but also providing an additional dimension to improve resolution. At the \ac{BS}, arrays with 64 or more elements are not exceptions, supporting superior angle resolution at both ends of the link. Note that for a given physical footprint about 25 times more elements can be packed at \ac{FR2} compared to \ac{FR1} \cite{zhang20165g}.     These large arrays come at a cost of shifting from digital arrays at \ac{FR1} to analog or hybrid arrays in \ac{FR2}, and even simpler arrays-of-subarrays in the sub-THz regime \cite{chen2022tutorial}. These arrays not only constrain the signals that can be transmitted (thus affecting \ac{AoD} performance) but also imply that \ac{AoA} estimation should be performed in a lower-dimensional beamspace. In addition, due to hardware impairments and lack of per-device calibration, the generated beams (precoders for transmission and combiners for reception) may deviate significantly from the idealized designs. While this has little or no impact on communication (as long as the beam has a main lobe more or less in the correct direction), this precludes the use of sophisticated methods that rely on knowledge of the complex beam responses \cite{chen2023modeling,tubail2022effect}. 
    \item \emph{Sparse channel:} Further complementing the large bandwidths and large array sizes are the favorable characteristics of the channel. At \ac{FR2}, the channel becomes sparse, with few clusters surviving the propagation between transmitter and receiver, in part because shadowing is so severe \cite{shafi2018microwave}. Multi-bounce reflections become rarer, but due to the reduced wavelength, objects appear more rough, leading to increased diffuse scattering and fewer reflections. Overall, the sparsity of the channel is beneficial since there will be a reduced requirement for multipath resolvability. At sub-THz, these effects are even more pronounced, leading to an even sparser channel, but much more sensitive to blockages, e.g., even due to foliage (at \ac{FR2}) or rainy weather (at sub-THz) \cite{rappaport2019wireless}. 
\end{itemize}

The compound effect of these characteristics makes positioning at mmWave  attractive in support of challenging use cases, such as in the automotive industry. Due to these same factors, relatively low-complexity methods can be employed, e.g., based on FFTs, which facilitates real-time implementation. Another important consideration at mmWave is that due to the sparse channels and the need to form narrow beams to achieve sufficient \ac{SNR}, (i)  communication and positioning are more closely intertwined, and (ii) it is hard for a \ac{UE} to connect to several mmWave \acp{BS} simultaneously. The first consideration has given rise to the concept of location-aided or context-aware communication, the most prolific example of which is location-based beam training. The second consideration is more serious and relates closely to the discussions on minimal problems in Section \ref{sec:mininal-positioning}. This consideration also gives rise to the topic of single-\ac{BS} positioning, which also avoids the need for inter-\ac{BS} synchronization \cite{wymeersch20185g,nazari2023mmwave,chen2022joint}. 

\begin{figure}
    \centering
    \includegraphics[width=0.6\columnwidth]{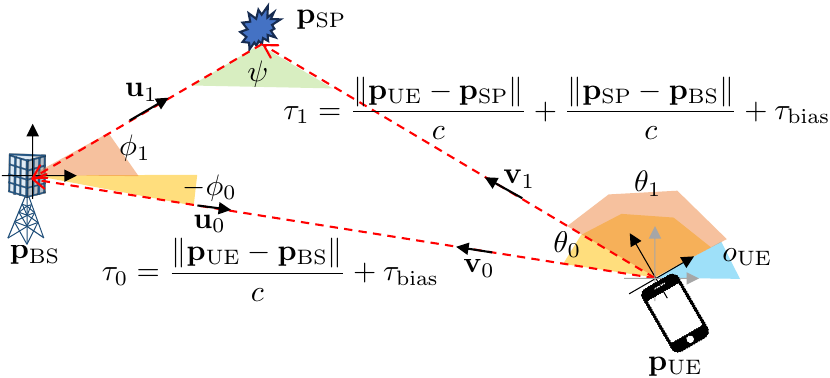}
    \caption{Illustrative example of mmWave positioning  with 1 \ac{BS} by exploiting the uncontrolled multipath, via a scatter point with unknown location. }
    \label{fig:example1BSpos}
\end{figure}

\begin{example}[Single BS Positioning]
Consider a 2D scenario shown in Fig.~\ref{fig:example1BSpos}, with a \ac{BS} that defines the coordinate system,  a \ac{UE} with unknown 2D location $\mathbf{p}_{\text{UE}}$, 1D clock bias $\tau_{\text{bias}}$, and 1D orientation $o_{\text{UE}}$, and a scatter point with unknown 2D location $\mathbf{p}_{\text{SP}}$. From uplink signals, the BS determines the estimates of the \ac{AoA} for the \ac{LoS} path (say $\phi_0$), the \ac{AoA} of the reflected path (say $\phi_1$), the corresponding estimates of the delays ($\tau_0$ and $\tau_1$), as well as the corresponding estimates of the \ac{AoD} ($\theta_0$ and $\theta_1$). The example is visualized in Fig.~\ref{fig:example1BSpos}. 
We see immediately that $\phi_0 = o_{\text{UE}}+\theta_0 + \pi$, from which the \ac{UE} orientation $o_{\text{UE}}$ can be solved. We also immediately find that the angle $\psi=\pi -(\phi_1-\phi_0)-(\theta_1-\theta_0)$. All remaining angles follow from the law of sines. 
{A direct closed-form solution can then be obtained as follows. Let us introduce unit vectors at the BS ($\vu_i$ from the AoA) and at the UE ($\vv_i$  from the AoD) so that the RX position can be defined as
\begin{equation}\label{eq:rx_position}
    \vp_\text{UE} = \vp_\text{BS} + d_i \gamma_i \vu_i - d_i  (1 - \gamma_i) \vv_i
\end{equation}
where $d_i = c (\tau_i - \tau_\text{bias})$ denotes the propagation distance and $\gamma_i \in [0,1]$ represents the fraction of the propagation distance along $\vu_i$. We can re-arrange \eqref{eq:rx_position} as
\begin{equation}\label{eq:rx_position_rearranged} 
\vp_\text{UE} - c \tau_\text{bias} \vv_i = \vmu_i + \gamma_i d_i \vnu_i
\end{equation}
where $\vmu_i = \vp_\text{BS} - c \tau_i \vv_i$ and $\vnu_i = \vu_i  + \vv_i$. Next, we solve for $\gamma_i$ and substitute it back to \eqref{eq:rx_position_rearranged} which yields the following cost function 
 \begin{equation}
     f(\vp_\text{UE}, \tau_\text{bias}) = \sum_{i=0}^{1} \lVert \vH_i \vx_\text{UE} - \vmu_i - \bar{\vnu}_i^\top (\vH_i \vx_\text{UE} - \vmu_i)\bar{\vnu}_i \rVert^2,
 \end{equation}
 where $\vH_i = [\vI, \, -c\vv_i]$, $\vx_\text{UE} = [\vp_\text{UE}^\top, \,  \tau_\text{bias} ]^\top$ and $\bar{\vnu}_i = \bar{\vnu}_i/\lVert \bar{\vnu}_i \rVert$. Now the closed-form solution can be obtained by setting the cost function’s gradient to zero and solving for $\vx_\text{UE}$. It is important to note that for the LOS path ($i=0$), we have $\vu_0 = -\vv_0$ and $\bar{\vnu}_0 = \mathbf{0}$. This estimate will be affected by measurement noise, but can be utilized as a coarse estimate for further refinement, e.g., based on maximum likelihood.}
\end{example}

\begin{figure}
    \centering
    \includegraphics[width=0.6\columnwidth]{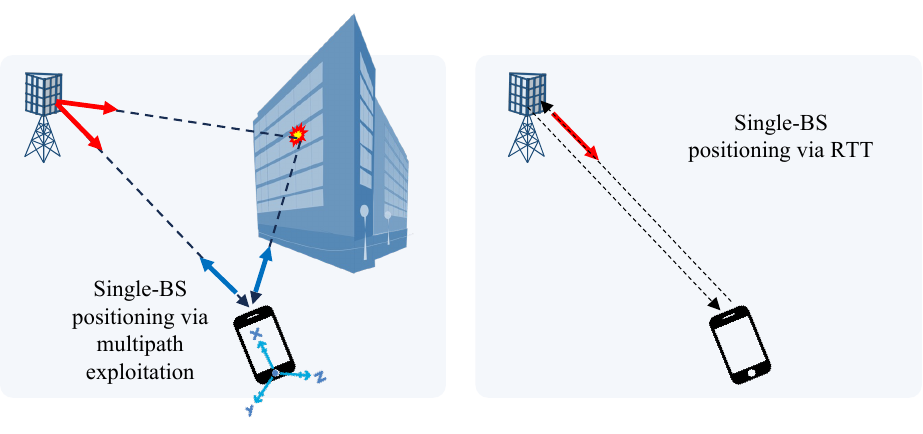}
    \caption{Visualization of single-\ac{BS} positioning approaches. }
    \label{fig:single-BS-positioning}
\end{figure}

A non-exhaustive list of examples of single-BS positioning is provided in Table \ref{tab:single-BS-positioning}, one based on controlled multipath (from a \ac{RIS}), one based on uncontrolled multipath, and one without any multipath. Two of the approaches are also visualized in Fig.~\ref{fig:single-BS-positioning}. Note that methods that rely on multipath (with or without \ac{RIS}) are limited by the strength of the multipath, while methods that rely on angle information suffer from large positioning errors when the \ac{UE} is far away from the multi-antenna \ac{BS} or \ac{RIS}. Note also that the examples in Table \ref{tab:single-BS-positioning} consider far-field propagation. As we will see in Section \ref{sec:near-field}, the reliance on infrastructure can be further reduced when harnessing wavefront curvature. 

\begin{table}[]
    \centering
    \resizebox{0.6\columnwidth}{!} {
    \begin{tabular}{|P{2cm}|P{2cm}|P{2cm}|P{2cm}|}
    \hline
   \rowcolor[HTML]{d7d8f9}  \textbf{Scenario} & \textbf{Measurements} & \textbf{Principle}  & \textbf{Nuisance parameters}\\
     
    \hline
    single-antenna \ac{BS} +\ac{RIS} + single-antenna UE &  \ac{ToA} from \ac{BS}, \ac{ToA} and 2D-\ac{AoD} from \ac{RIS} & 2D-\ac{AoD} determines a line away from the \ac{RIS}, which intersects with hyperboloid from the \ac{TDoA}. & \ac{UE} clock bias can be estimated from the \ac{ToA}.\\
    \hline 
    multi-antenna \ac{BS} + single-bounce \ac{IP} + multi-antenna UE &  \ac{ToA} and 2D-\ac{AoD} from \ac{BS} to \ac{UE} and \ac{IP}, 2D-\ac{AoD}  at UE from \ac{BS} and \ac{IP} & angle measurements determine entire geometry up to scaling. \ac{TDoA} determines scaling & \ac{UE} clock bias, \ac{UE} 3D orientation, and \ac{IP} location can be recovered.\\
    \hline  multi-antenna \ac{BS} +  single-antenna UE &  \ac{RTT} at BS, 2D-\ac{AoA} at \ac{BS} & \ac{UE} on intersection of sphere from \ac{RTT} and line away from \ac{BS} from 2D-\ac{AoA}.  & N/A\\
     \hline 
    \end{tabular}
    }
     \vspace{0.1mm}
    \caption{Examples of minimal configurations for single-BS positioning. }
    \label{tab:single-BS-positioning}
\end{table}



\subsubsection{Positioning in sub-6 GHz vs. mmWave: performance analysis via ray-tracing data}

In this part, we carry out a comparative performance analysis of positioning in sub-6 GHz and mmWave bands using realistic ray-tracing data obtained through the REMCOM Wireless InSite\textregistered  ray-tracer \cite{WirelessInSite}. In the ray-tracer, we consider an urban intersection scenario involving \textit{(i)} a mobile UE (corresponding to a vehicle) crossing the intersection, \textit{(ii)} a BS (corresponding to a road-side unit (RSU) in vehicular settings \cite{tr:37885-3gpp19}) located at the center of the intersection, and \textit{(iii)} four buildings with $30 \, \rm{m}$ in height, located at the corners of the intersection. The UE (with antenna height $1.5 \, \rm{m}$) moves on a straight line starting from $-70 \, \rm{m}$ and ending at $70 \, \rm{m}$, while the BS (with antenna height $10 \, \rm{m}$) is located at $0 \, \rm{m}$ (please see \cite[Sec.~V-B]{v2x_pos_fr1} for further details on the simulation environment). 
At each scenario instance ($101$ in total), the output of the ray-tracer consists of channel gains, delays, AOAs and AODs of the paths between the BS and the UE. To evaluate the positioning performance at sub-6 GHz and mmWave, we utilize the ray-tracer output to generate the channel matrix in \eqref{eq:channel} and the received signal in \eqref{eq_rec}. For fair comparison between sub-6 GHz and mmWave, we keep the same physical aperture size at the BS, resulting in more ULA elements at mmWave.


We consider a single-BS positioning scenario via an RTT approach, as illustrated in Fig.~\ref{fig:single-BS-positioning}, where the goal is to extract the parameters of the LOS path (i.e., the delay $\tau_0$ and the azimuth/elevation AoAs $\phi_{\rm{az},0}$, $\phi_{\rm{el},0}$) using \eqref{eq_rec} and estimate the UE position assuming known UE height. For channel estimation from \eqref{eq_rec}, we employ two algorithms \cite[Sec.~IV-B]{v2x_pos_fr1}: \textit{(i)} a matched filtering (MF) based method that performs correlation processing across frequency and spatial domains, and \textit{(ii)} an ESPRIT-based super-resolution method.  

\begin{figure}
        \begin{center}
        \subfigure[]{
			 \label{fig_pdp_ray_tracer_1}
			 \includegraphics[width=0.4\textwidth]{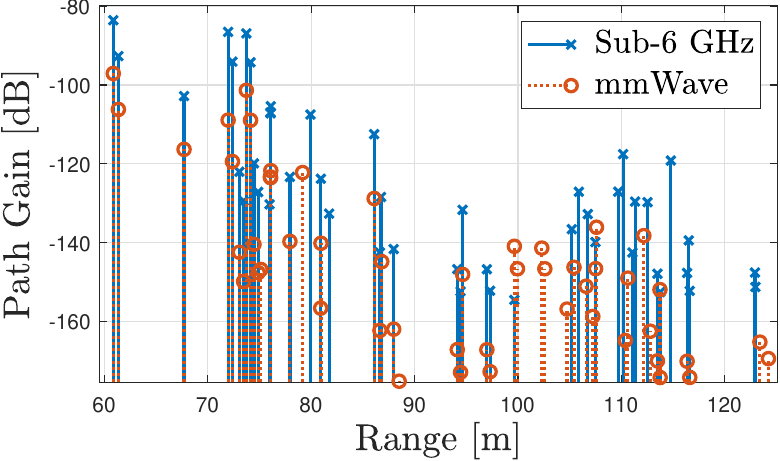}
		}
        \subfigure[]{
			 \label{fig_pdp_ray_tracer_2}
			 \includegraphics[width=0.4\textwidth]{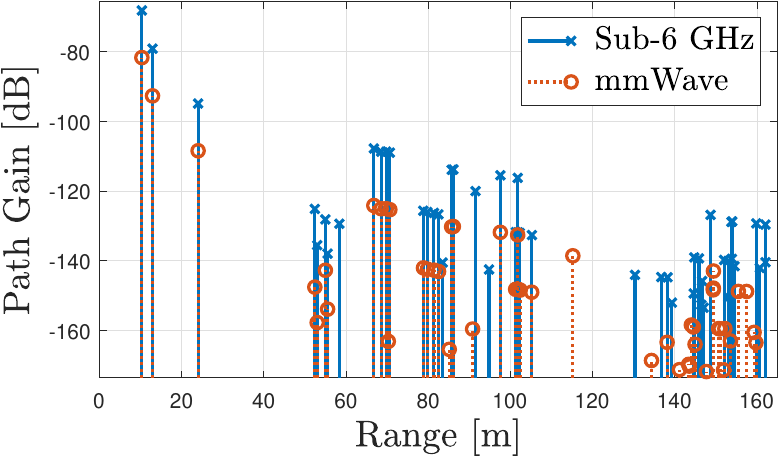}
		}		
		\end{center}
		\vspace{-0.2in}
        \caption{Power delay profiles at sub-6 GHz and mmWave, obtained from the ray-tracing data at two different instances of the single-BS positioning scenario in an urban intersection: \subref{fig_pdp_ray_tracer_1} the UE located at $-60 \, \rm{m}$, leading to a rich multipath environment due to surrounding buildings, \subref{fig_pdp_ray_tracer_2} the UE located at $5 \, \rm{m}$, leading to a small number of dominant paths.}  
        \label{fig_pdp_ray_tracer}
\end{figure}

We first investigate power delay profiles (PDPs) at sub-6 GHz and mmWave to provide an illustration of channel characteristics discussed in Sec.~\ref{sec:FR1-positioning} and Sec.~\ref{sec:FR2-positioning}. Fig.~\ref{fig_pdp_ray_tracer} shows the path gains with respect to range at two distinct scenario instances, representing a rich and a sparse scattering environment (which arises from the presence or absence of large reflectors, such as buildings, between the BS and the UE). As expected, higher path gains are observed at sub-6 GHz. Furthermore, when the UE is located at a considerable distance from the BS, a more diverse multipath channel is formed due to reflections from surrounding buildings.

\begin{figure}
    \centering
    \includegraphics[width=0.5\columnwidth]{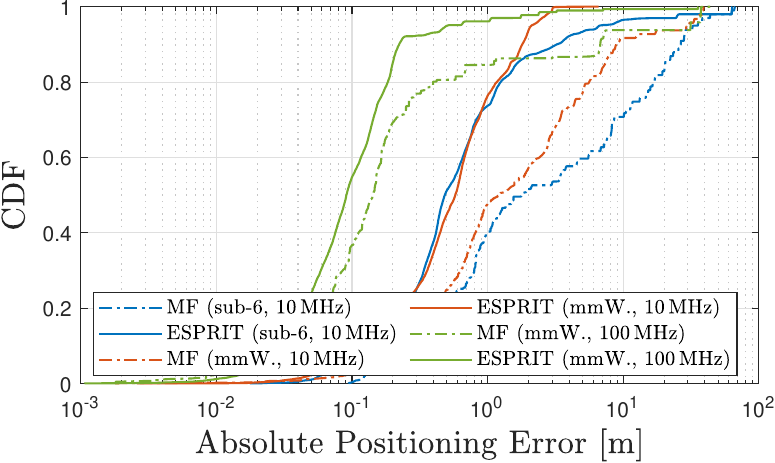}
    \caption{Positioning performances of the MF and ESPRIT algorithms at sub-6 GHz and mmWave under different bandwidths, evaluated using the ray-tracing data. The parameters at sub-6 GHz are $\fc = 5.9 \, \rm{GHz}$, $\deltaf = 60 \, \rm{kHz}$, $M = 12$ and the BS array configuration $1\times 2$ ($2.54 \, \rm{cm}$), while the parameters at mmWave are $\fc = 28 \, \rm{GHz}$, $\deltaf = 120 \, \rm{kHz}$, $M = 24$ and the BS array configuration $1\times 6$ ($2.68 \, \rm{cm}$). The common parameters are set as follows: transmit power $10 \, \rm{dBm}$, the UE array configuration $1\times 1$, the noise PSD $-174 \, \rm{dBm/Hz}$ and the noise figure $8 \, \rm{dB}$.}
    \label{fig_pos_comp_ray_tracer}
\end{figure}

Fig.~\ref{fig_pos_comp_ray_tracer} demonstrates the positioning performances using the considered algorithms at sub-6 GHz and mmWave, considering two different bandwidths at mmWave. It is observed that ESPRIT significantly outperforms MF at both frequency bands and using different bandwidths, through its ability to resolve closely spaced paths, especially in dense multipath environments illustrated in Fig.~\ref{fig_pdp_ray_tracer}. Moreover, under the same bandwidth utilization, both algorithms perform better at mmWave than at sub-6 GHz (despite larger path loss at mmWave). This can be attributed to the use of electrically large arrays at mmWave in the same physical footprint, leading to higher angular resolution and improved path resolvability. Increasing the bandwidth further enhances the performance, which indicates the suitability and attractiveness of mmWave bands for positioning (in the sense of manifestation of the geometric nature of the channel via the use of electrically large arrays and access to large bandwidths). Therefore, in alignment with the explanations in Sec.~\ref{sec:FR1-positioning} and Sec.~\ref{sec:FR2-positioning}, two key take-aways can be deduced from the ray-tracing based simulation results: \textit{(i)} mmWave induces favorable channel characteristics for positioning, \textit{(ii)} MF or correlation based processing suffers from poor resolution, leading to more than an order-of-magnitude degradation in accuracy compared to super-resolution approaches.


\subsubsection{Machine learning for positioning} 
\label{sec:ML-positioning}


Positioning exploiting channel parameters and the geometry of the environment suffers from performance degradation in all types of scenarios and frequency bands. At sub-6 GHz frequencies, non-line of sight (NLOS) multipath acts as an interference that degrades the geometric localization performance in both indoor and outdoor urban environments. At mmWave frequencies, strategies like trilateration and triangulation are often not practical, since links to several \acp{BS} are required. Although at mmWave it is possible to estimate the position from the CSI of a single BS-\ac{UE} link by exploiting the sparsity of the channel and the large arrays and bandwidths, high-accuracy localization requires very accurate channel estimates, which come at the cost of high pilot transmission overhead and high complexity. In addition, the sensitivity of geometric localization to channel estimation errors and impairments such as phase noise, residual clock offsets, array calibration errors, or beam squint may reduce the accuracy that can be achieved in practice. 

CSI-fingerprinting positioning is an alternative to geometric approaches that can provide enhanced performance in LOS and NLOS scenarios \cite{He2016, Klus_SPAWC_2022, Wang2020Globecom}. Although it was initially proposed for indoor environments and WLANs \cite{Bahl2000INFOCOM}, it has been successfully extended to outdoor scenarios and cellular networks \cite{Vo2016CST}.  It requires of an offline phase to create a database of fingerprints (one or more parameters associated to the propagation channel such as RSS or full CSI) and their associated locations. In conventional fingerprinting, the  online phase is used to compare the fingerprint obtained in real time with the stored ones and infer the location exploiting algorithms such as K nearest neighbors \cite{Hu2019ITJ}, Horus \cite{Youssef2005MobiSys} or RADAR \cite{Bahl2000INFOCOM} for example. Mathematically, this approach can be written as 
\begin{equation}
    g:\mathbf{F} \rightarrow \mathbf{p}_{\text{UE}}
    \label{eq:fingerprinting}
\end{equation}
where $g$ is a mapping operation from the fingerprint $\mathbf{F}$ to the user position $\mathbf{p}_{\text{UE}}$.
The drawbacks of conventional fingerprinting come from (i) the requirement of permanently storing a large database and updating it periodically as the environment changes \cite{mager2015fingerprint} (ii) the complexity of searching the whole database for every new position to be predicted.

ML-based fingerprinting can overcome these limitations. A large database will be collected for training the network offline, but it does not need to be stored or searched during online operation. During training, the network will learn the mapping $g$ in \eqref{eq:fingerprinting}, while in the online phase it will perform a regression operation to compute the position given the fingerprint. 
RSS is the most commonly used fingerprint in conventional designs, but it can only capture coarse channel information, it is highly dependent on the device and it suffers from a high variability due to multipath. Recent works on ML-based fingerprinting for massive MIMO at sub-6GHz and mmWave MIMO exploit richer fingerprints based on the CSI which provide enhanced performance: the angle-delay domain channel power matrix \cite{Sun2018TVT,Wu2021TWC}, a weighted average of CSI values over multiple antennas \cite{Xiao2012ICCCN}, the CSI per subcarrier \cite{Wang2017TVT,Zhang2023TWC} or a decimated delay-domain CSI representation followed by autocorrelation to extract features invariant to the system impairments \cite{Ferrand2020Globecom} to name a few. Alternatively, other recent works avoid the design of specific features by introducing the full CSI in time or frequency as the input to the deep network. This way, the first stage of the network itself extracts a suitable feature using the attention mechanisms in Transformer networks \cite{chen2023learning,Chen2023SPAWC,Salihu2022SPAWC}.

Most of the deep networks that have been designed for fingerprinting-based positioning approach the problem as a regression task. Designs based on \ac{CNN} architectures leverage image-like inputs and exploit convolutional layers' ability to extract features and relationships among adjacent data points \cite{Chen2017Access,butt2021ml, gante2020deep, kia2022cnn, wang2021deep, sadr2021uncertainty,Wu2021TWC,Zhang2023TWC,Wang2017TVT}. Long short-term memory (LSTM) networks have been also proposed  to explore the correlation of CSI at different subcarriers \cite{Chen2023PIMRC}. To tackle the problem of 
outdated network weights  in a dynamic environment, it is possible to use transfer learning \cite{Gao2021ITJ}, which enables the reconstruction of the fingerprinting database using the outdated fingerprints and a small number of new measurements. Hybrid approaches \cite{chen2023learning,chen2022joint} that combine model-based geometric localization and deep networks exploiting site-specific data can provide very high accuracy even with small training datasets, as shown in Example 4. 

Position tracking can also be implemented exploiting ML. An obvious extension to the previously described approaches is to include a tracking stage that uses some kind of Bayesian filter (Kalman or extended Kalman for example) on the results of a ML-based fingerprinting approach \cite{Shi2018TVT}. However, recent designs replace the Bayesian filtering stage by a deep network specialized in tracking, so there is no need to build a mathematical evolution model---which may not hold in practice---to be exploited by the Bayesian filter. 
LSTM networks are common choices to implement tracking \cite{Zhang2021Sensors}, but 
newer designs exploiting transformers and their attention mechanisms exhibit enhanced performance. For example,  \cite{Chen2023SPAWC} proposes V-ChATNet for position tracking, an attention network that exploits the series of previous
channel and position estimates to build the estimation fir the new position. This design keeps the location error below 20cm for 95\% of the time when evaluated with realistic vehicular channels generated by ray tracing.

\begin{example}[ML-based positioning]
Consider a mmWave vehicular communication system operating with a hybrid MIMO architecture and UPAs at both ends. During initial access, the frequency selective mmWave channel between a vehicle and a single BS is estimated using MOMP \cite{Palacios2022EUSIPCO} and some training symbols.  The hybrid data/model-driven positioning system proposed in \cite{chen2023learning,chen2022joint} localizes the vehicles on the road exploiting three stages: 1) PathNet, a fully connected network that classifies the estimated channel paths as LOS, first NLOS or higher order; 2) a geometric localization algorithm that accounts for the clock offset and exploits the parameters of the LOS and first order paths extracted by PathNet to obtain an initial position estimation; and 3) ChanFormer, a Transformer network that exploits the concept of “attention” to evaluate which estimated paths are more credible and formulates the problem of position refinement as a classification task, by computing the probability of corresponding to the true location for a grid of points built around the initial position estimation. As illustrated in Fig.~\ref{fig:ML-based_positioning_ini}, the system achieves sub-meter accuracy localization for 95\% of the users in channels with a LoS path when evaluated with ray tracing channels. In addition, the system provides sub-meter accuracy localization for 50\% of the users in NLoS channels. During the tracking stage, a different attention network, V-ChATNet \cite{Chen2023SPAWC}, further reduces the localization error, achieving 20cm accuracy for 95\% of the users in a combination of LOS and NLOS channels, significantly outperforming Kalman Filtering based tracking, as shown in Fig.~\ref{fig:ML-based_positioning_track}. 
\begin{figure}
    \centering
      \vspace{-0.2in}
     \subfigure[]{
			 \label{fig:ML-based_positioning_ini}
			 \includegraphics[width=0.48\textwidth]{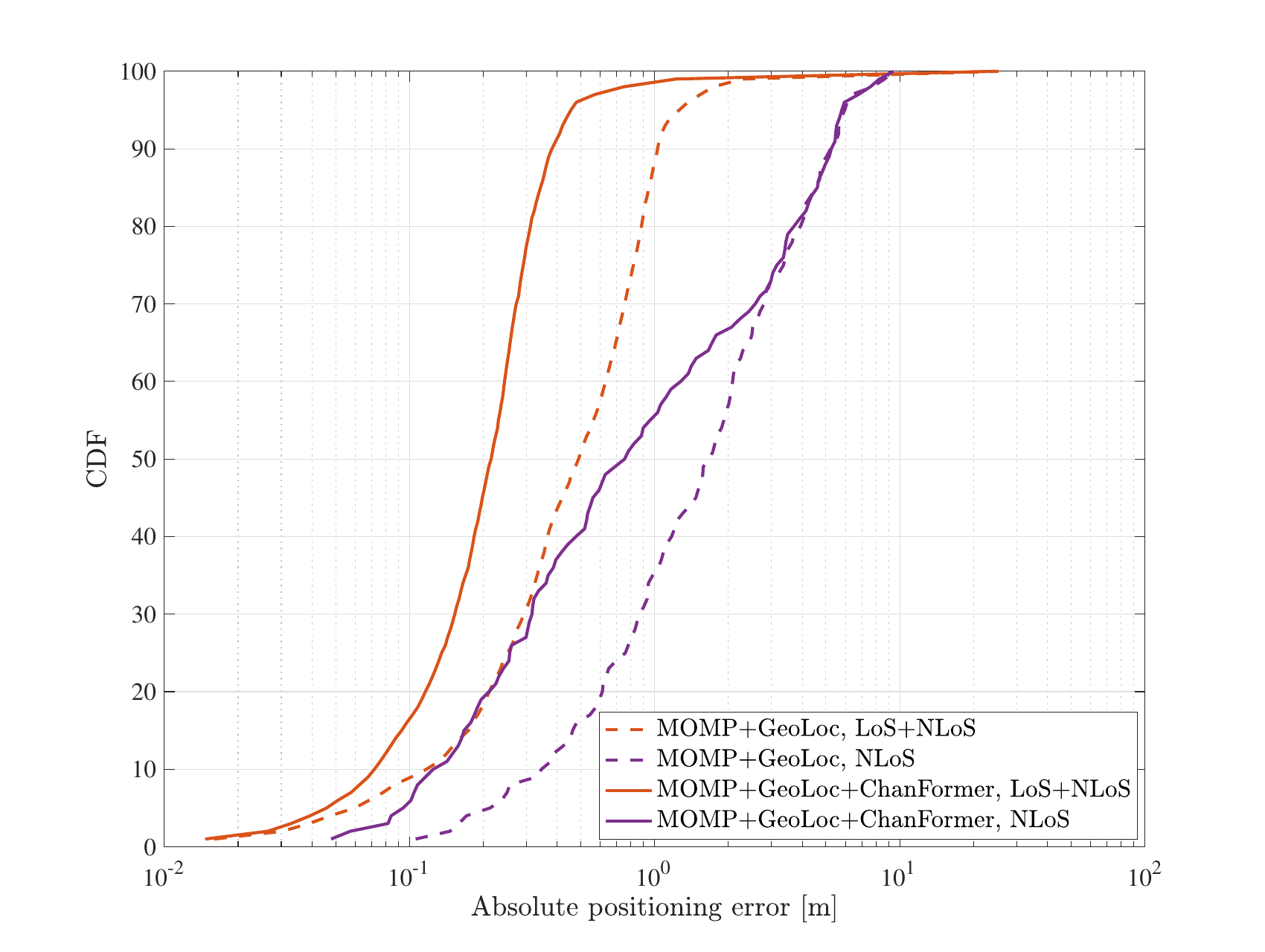}}
   \subfigure[]{
			 \label{fig:ML-based_positioning_track}
			 \includegraphics[width=0.48\textwidth]{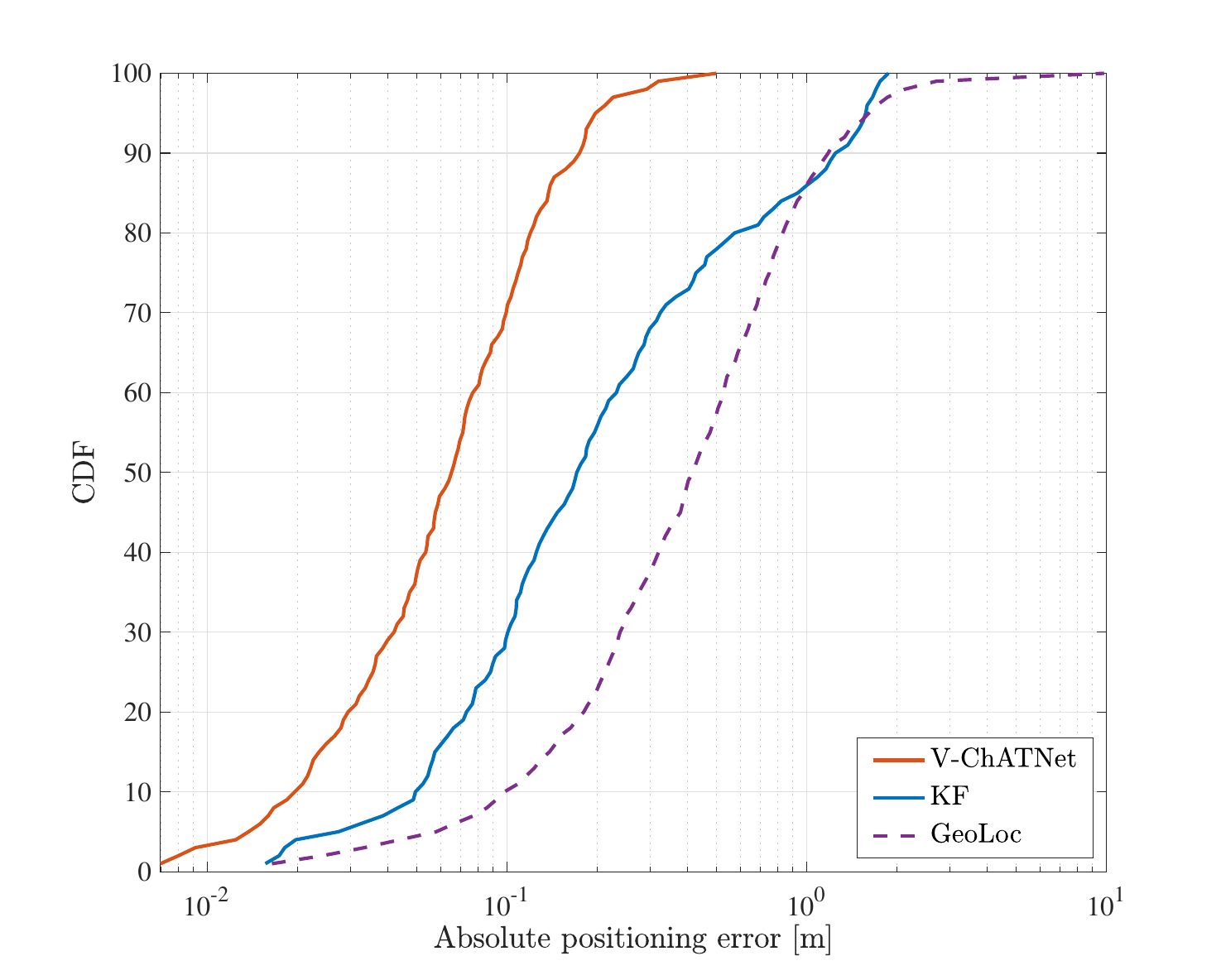}}
    \caption{Performance of the positioning strategies based on attention networks--- ChanFormer for the initial access scenario and V-ChATNet for the tracking case---compared to positioning based only on channel parameter estimation and  exploitation of the geometry of the environment (MOMP + GeoLoc) evaluated using the ray tracing setup described in \cite{chen2023learning,Chen2023SPAWC}: (a) Results for the initial access and position estimation scenario, where only one snapshot of the channel is estimated and leveraged for positioning; (b) Results for the tracking scenario, where series of channel and position estimates are exploited by both a Kalman Filter (KF) and V-ChATNet, while MOMP+GeoLoc only exploits previous channel estimates to reduce computational complexity in the channel parameter estimation stage.}
\end{figure}
\end{example}

\subsubsection{Sidelink positioning} \label{sec:sidelink}
Sidelink positioning is a technique which exploits sidelinks between \ac{UE}s to transmit signals and perform measurements to enhance the positioning performance for \ac{V2X}. 

During the sidelink positioning process, the target \ac{UE} aims to determine its own position under the assistance of the anchor \ac{UE}s, which is further divided into three stages: the configuration stage, the signal transmission and measurement stage, and the position calculation stage\cite{Tho:J21}. First in the configuration stage, the target \ac{UE} and anchor \ac{UE}s are scheduled to form the sidelink positioning group for the following operations. Then in the signal transmission and measurement stage, the measurements are acquired by sending the positioning reference signals over the sidelinks between \ac{UE}s. Finally in the position calculation stage, the \ac{ToA} and \ac{AoA} information obtained from link-level measurements are fused to calculate the absolute or relative position of the target \ac{UE}, which can be performed either at the network location server or at the \ac{UE} itself.

As sidelink positioning emerges as a promising technique to complement the traditional methods, e.g., \ac{GNSS}, in V2X, there have been extensive research on this area. In \cite{liuLiaXia:J21}, the authors describe basic system architectures and key technologies for high-accuracy sidelink positioning. In \cite{GeMaxFur:C23}, the authors present analysis of V2X sidelink positioning in sub-6 GHz, where a novel performance bound is derived to predict the positioning performance in the presence of severe multipath. In addition, the sidelinks in V2X are exploited for near-field localization in \cite{NicAnnCat:J23}, where the fundamental positioning limits are determined in order to assess the possibility of the proposed scheme.

Furthermore, sidelink positioning can be viewed as the special case of the general cooperative localization paradigm, in which the unknown positions of multiple \ac{UE}s are jointly inferred from the network measurements, and the performance gain comes from the exploitation of the relative position information provided by the \ac{UE}-\ac{UE} measurement links. Compared with conventional localization techniques, cooperative localization demonstrates significant advantages in harsh propagation environments such as indoors and urban canyons, which attracts great attention from the research society in both theoretical and algorithmic aspects. As for the theoretical basis, the fundamental limits of localization accuracy in cooperative networks have been derived in \cite{SheHenWim:J10,WinSheDai:J18}, where the structure of \ac{EFIM} is revealed to characterize the information gain brought by \ac{UE} cooperation. Then the spatio-temporal information coupling is investigated to quantify the cooperation efficiency, where the asymptotic error propagation laws are determined to provide guidelines for large-scale networks\cite{SanSheWin:J18,SheLiuShe:C21,XioWuShe:J22}. As for the operation strategies, efficient resource allocation schemes are  developed  from the  perspectives  of  convex  programming  and  game  theory, which  can  provide  practical  solutions  for  high-accuracy cooperative localization in both centralized and distributed schemes\cite{DaiSheWin:J15,CheDaiShe:J16}.

\subsubsection{RIS-aided positioning}
\label{sec:RIS}

A \ac{RIS} is a programmable surface that can be used to control the reflection of radio waves by changing the electric and magnetic properties of the surface, in either a continuous or discrete way \cite{basar2019wireless}. While there are different \ac{RIS} technologies, their distinction is irrelevant for our purposes.
\Acp{RIS} were originally devised to overcome \ac{LoS} blockages, especially at \ac{FR2}, by creating an additional path with high \ac{SNR}. The most common kind of a \ac{RIS} is a so-called passive reflective RIS, which creates an additional channel  (in its more simple form)
\begin{align}
    \HH^{\text{RIS}}_{n,m} = \alpha_m^{\text{RIS}}  
    e^{-j 2 \pi n \deltaf \tau^{\text{RIS}} } 
    e^{j 2 \pi m \Tsym \nu^{\text{RIS}} } 
    \arx(\phi^{\text{RIS}}) \atx^T(\theta^{\text{RIS}}) ~,\label{eq:channel-RIS}
\end{align}
where the only difference lies in the $\alpha_m^{\text{RIS}}$, which is of the form \cite{keykhosravi2022ris}
\begin{align}
    \alpha_m^{\text{RIS}} = \alpha^{\text{T-RIS}}\alpha^{\text{RIS-R}} \mathbf{a}^T_{\text{RIS}}(\boldsymbol{\vartheta})\boldsymbol{\Omega}_m \mathbf{a}_{\text{RIS}}(\boldsymbol{\varphi}) \label{eq:channel-RIS-detail}
\end{align}
in which $\alpha^{\text{T-RIS}}$
 denotes the channel from transmitter to \ac{RIS}, $\alpha^{\text{RIS-R}} $ is the channel from \ac{RIS} to receiver, $\mathbf{a}_{\text{RIS}}(\boldsymbol{\varphi})$ is the RIS steering vector as a function of the \ac{AoA} $\boldsymbol{\varphi}$ (from the receiver) and \ac{AoD} $\boldsymbol{\vartheta}$ (to the transmitter), and $\boldsymbol{\Omega}_m $ denotes the diagonal \ac{RIS} control matrix, which represents the time-varying state of the \ac{RIS}. The entries of $\boldsymbol{\Omega}_m $ have at most unit amplitude (since a passive RIS cannot amplify) and controllable phase. In communications $\boldsymbol{\Omega}_m $ can be set so that $\mathbf{a}^T_{\text{RIS}}(\boldsymbol{\vartheta})\boldsymbol{\Omega}_m \mathbf{a}_{\text{RIS}}(\boldsymbol{\varphi}) \to M$, where $M$ is the number of \ac{RIS} elements, thereby providing an \ac{SNR} scaling of up to $M^2$. The value of $M$ can be increased by making the RIS larger (leading to more power illuminated on the \ac{RIS}, but also wavefront curvature effects appear (see Section \ref{sec:JCAS-large-RIS}) or by making the RIS denser (though the power illuminated on the RIS is constant and mutual coupling will play an important role). 
 In addition to the considered passive reflective RIS, there are many other RIS variants, such as amplifying RIS (sometimes called active RIS), sensing RIS (sometimes called hybrid RIS), RISs that can simultaneously transmit (in the sense of passing through) and reflect (called STAR-RIS) \cite{he2023star}, and RISs that allow non-diagonal control $\boldsymbol{\Omega}_m$ \cite{Cle:23}. 

The role of \ac{RIS} for positioning lies in the ability to provide (i) an additional position reference, similar to a BS, (ii) additional measurements as any other path (e.g., the \ac{AoA}, \ac{AoD}, delay, and Doppler in \eqref{eq:channel-RIS}), (iii) unique measurements of the \ac{AoA} or \ac{AoD} in \eqref{eq:channel-RIS-detail} \cite{bjornson2022reconfigurable}. While the \ac{AoA} and \ac{AoD} in \eqref{eq:channel-RIS-detail} are not jointly identifiable, typically one of the two angles is known since either the transmitter or the receiver is a \ac{BS}. This leads to two practical issues: how can we know the \ac{RIS} location and orientation \cite{mizmizi2023target} and how can the receiver separate the signal from the RIS with respect to all the uncontrolled multipath, including the \ac{LoS} path? To address the first issue, \ac{RIS} calibration methods have been devised, based either on transmissions between \acp{BS} or while simultaneously localizing the user. To address the second issue, which is especially relevant in a multi-RIS scenario, dedicated RIS control sequences have been designed that allow separation, e.g., such that $\sum_m \boldsymbol{\Omega}_m=\mathbf{0}$, so that under zero Doppler, the uncontrolled multipath can be recovered by adding the signals over time, for each transmit beam. 

\begin{figure}
    \centering
    \includegraphics[width=0.5\columnwidth]{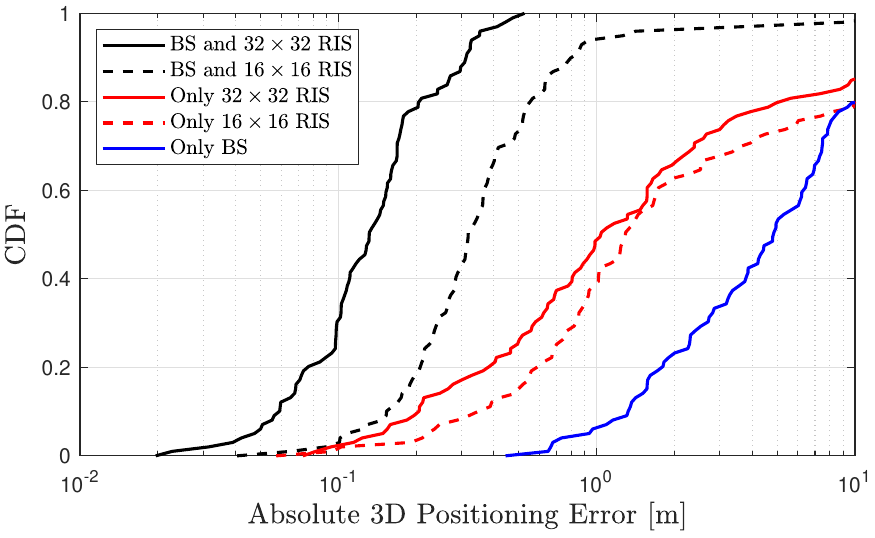}
    \caption{\ac{RIS}-aided 3D positioning performance with an $8 \times 8$ \ac{BS}. Evaluated using the ray-tracing data from a smart factory environment with transmit power $20 \, \rm{dBm}$, center frequency $\fc  = 60 \, \rm{GHz}$, and bandwidth $B = 100 \, \rm{MHz}$.}
    \label{fig_RIS_localization}
\end{figure}

\begin{example} \label{ex:RIS_localization}
Consider a point-to-point RIS-aided mmWave communication system where both the \ac{BS} and the \ac{UE} are equipped with planar arrays. In this case, the received signal at the \ac{UE} contains contributions of the paths from two sources, i.e., from the environment and through the \ac{RIS}, which can be expressed as
\begin{equation}\label{eq_rec_RIS}
     \yy_{n,m}^{\emph{RIS}} = \WW\herm \left(\HH_{n,m} + \HH^{\emph{RIS}}_{n,m}\right) \FFrf \FFbb[n,m] \xx_{n,m} + \zz_{n,m}.
\end{equation}
In \cite{Bayraktar2022SPAWC}, it is shown that the sparse channel parameters from both sources can be simultaneously estimated by using a slightly modified version of the low-complexity multidimensional orthogonal pursuit matching (MOMP) algorithm proposed in \cite{Palacios2022EUSIPCO}, which estimates the parameters by using their own individual dictionaries instead of a Kronecker product of them. Note that the work in \cite{Bayraktar2022SPAWC} considers the pulse shaping effects and clock offset, and estimates the time-domain channel while assuming that the \ac{LoS} path between the \ac{BS} and the \ac{RIS} has been pre-estimated in the RIS calibration stage. The positioning stage contains two different cases: (1) \ac{LoS} path from the \ac{BS} and \ac{RIS} exist, (2) \ac{LoS} path from either \ac{BS} or \ac{RIS} exists. The first case can be solved with a simple linear equation whereas the second case can be solved via a linear system of equations constructed with \ac{LoS} and \ac{NLoS} paths of one source (i.e., \ac{BS} or \ac{RIS}) by leveraging the reflection properties of the indoor environment \cite{Bayraktar2022SPAWC,Palacios2022EUSIPCO}. Positioning error results with the described method obtained via ray tracing based simulations for an indoor environment are given in Fig~\ref{fig_RIS_localization}. It can be seen that highly accurate positioning can be achieved when the \ac{LoS} path exists for both \ac{BS} and \ac{RIS}. Specifically, $\rm{\%}80$ of the users have positioning error below $20 \, \rm{cm}$ with a $32 \times 32 $ \ac{RIS}.
 \end{example}

\subsubsection{Carrier phase positioning} 
\label{sec:CPP}

\begin{figure}
    \centering
    \includegraphics[width=0.6\columnwidth]{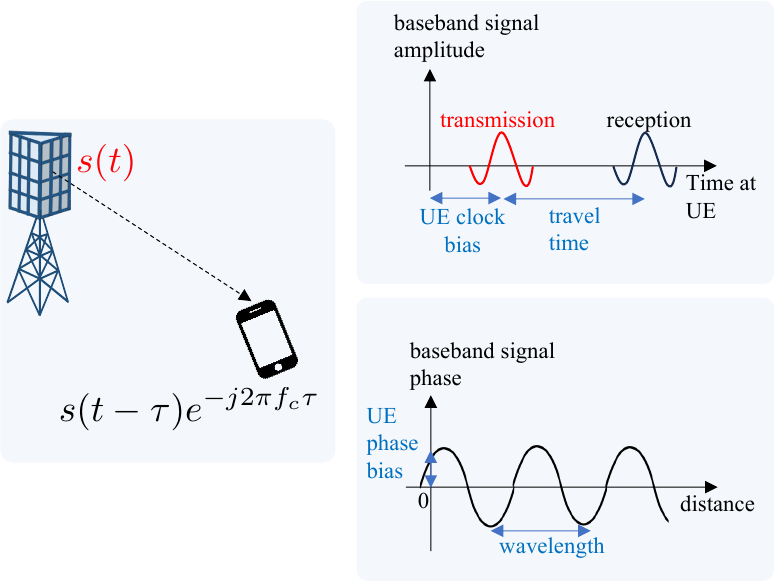}
    \caption{Carrier phase positioning provides accurate distance information, but is subject to ambiguities equal to the wavelength. }
    \label{fig:CPP}
\end{figure}

In general, the carrier phase of the incoming received signal depends explicitly on the propagation delay, as shown in Fig.~\ref{fig:CPP}. Hence, measuring carrier phase accurately, relative to a reference, allows carrying out delay estimation and thereon ranging with accuracies that are fractions of a wavelength. Especially in millimeter-wave networks, using carrier phase measurements can thus enable ranging accuracies in the order of few millimeters while in C-band networks a centimeter-level ranging accuracy is feasible.

In the context of localization and positioning, carrier phase based methods are conceptually known and broadly utilized in GNSS systems, particularly the GNSS-RTK approach, see for example \cite{Ding_RM_2021} and the references therein. Additionally, and importantly, the utilization of carrier phase measurements is recently also considered in 3GPP cellular network standardization---particularly in the context of 5G-Advanced and corresponding enhanced positioning capabilities \cite{3GPP_38859,3GPP_RP-223549}.

One of the most prominent technical challenges related to the use of carrier phase measurements is the so-called integer ambiguity problem~\cite{Talvitie_JSTSP_2023,Teunissen_lambda,Cox_lambda}. This refers to the fact that the carrier phase is immune to any integer multiple of the wavelength, i.e., the amount of full wavelengths in the distance between the TX and RX entities cannot be directly measured. Practical ways to solve or relax the integer ambiguity problem stem from differential and/or double-differential measurements, that build on the concept of a reference device. In the cellular network context, a UE with a known and fixed reference position could serve for such purpose.

Example recent works in this area in the cellular system context covers, e.g., \cite{Chen_IoTJ_2022,Fan_TCOM_2022} focusing on estimation, synchronization and positiong algorithms for different use cases. Additionally, \cite{Talvitie_JSTSP_2023} describes methods how cellular carrier phase measurements combined with appropriate Bayesian filtering can facilitate super-resolution and low-latency 6DoF tracking of 5G-empowered XR headsets without any additional sensors.

\subsection{Bistatic and multistatic sensing of non-connected objects}
\label{sec:bistatic-sensing}

In contrast to positioning, which involves determining the state of connected users, in this section, we will focus on sensing (detecting and localizing) non-connected objects (sometimes also called landmarks or targets, depending on the context). We refer back to Fig.~\ref{fig:bistatic-overview}.  

\subsubsection{Fundamentals of mapping and target tracking, multi-site processing}
First, it is necessary to clarify the terminology: when the objects of interest are static, they are called landmarks. The corresponding sensing problem is called \emph{mapping}. When the objects of interest are moving, they are called targets and the corresponding sensing problem is called \emph{\ac{MTT}}. The reference to 'objects of interest' is due to there being other objects (e.g., ground reflections when tracking a moving vehicle), which generate measurements, but are not of interest to the application. 

\begin{figure}
    \centering
    \includegraphics[width=0.7\columnwidth]{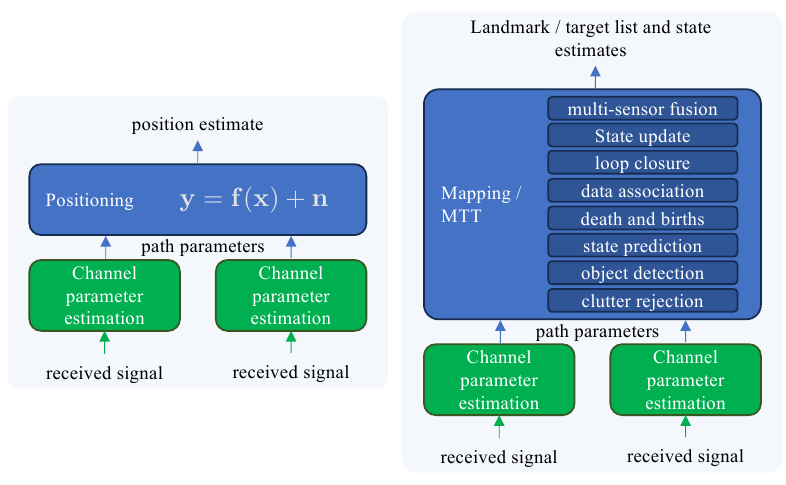}
    \caption{Both positioning and sensing require channel parameter estimation. The subsequent processing for mapping and \ac{MTT} is much more involved than for positioning.  }
    \label{fig:two-step-processing}
\end{figure}

To separate positioning from mapping and \ac{MTT}, let us first describe the commonalities (see Fig.~\ref{fig:two-step-processing}): in both cases, processing involves two stages, wherein the first stage radio signals are sent and received, based on which \ac{LoS} and \ac{NLoS} paths are extracted, with associated channel parameters (delays, angles, Dopplers). In the second stage, those parameters are converted to a geometric state of the user/object. In terms of the differences, the first stage for positioning is generally focused on extracting the \ac{LoS} path parameters, while in mapping and \ac{MTT}, the \ac{LoS} path only serves as a reference path, while the \ac{NLoS} paths relate to the non-connected objects. The main difference lies in the subsequent processing, as visualized in Fig.~\ref{fig:two-step-processing}. 
The reasons are as follows \cite{vo2015multitarget}:
\begin{itemize}
    \item \emph{Unknown number of objects:} The sensing system does not know a priori how many objects there are. This means that objects need to be first detected before they can be localized. This problem is also affected by background clutter, which is present but not of interest. Hence, sensing involves both a detection (involving false alarms and missed detections) and an estimation problem. 
    \item \emph{Complicated object states:} In contrast to positioning, where the goal is to estimate the \ac{UE} 3D position, objects are characterized by a much more complicated state definition, which may include the velocity, the extent/shape of the object, and even the material type. Objects may also give rise to more than one measurement \cite{granstrom2016extended}. 
    \item \emph{Unknown data association:} Another fundamental difference compared to positioning, is that the measurements carry no information about the object. Hence, measurements taken at some time $t$ should be associated with objects detected at some earlier time $t'<t$. This type of combinatorial problem lies at the heart of all mapping and \ac{MTT} methods. As time progresses the number of data associations grows quickly, which requires dedicated routines to mitigate computational complexity \cite{williams2010data}. 
    
    \item \emph{Multi-sensor fusion:} When several receivers are processing the signals from one or more transmitters, some form of fusion is needed to provide a consistent view of the landmarks or targets. The type of fusion depends on the level of coherence between the different \acp{BS}. When the receiving \acp{BS} are phase coherent, then can be interpreted as one large distributed sensor and processing can be performed on the aggregated received waveforms, as in distributed MIMO radar. Such processing involves sharing raw I/Q data with a central processing unit.  In contrast, when the receiving \acp{BS} are not phase-synchronized, they act as independent observers, and fusion should occur after local channel parameter estimation. The different field of view of the different receivers makes this challenging \cite{yi2020distributed}. 
\end{itemize}

\begin{figure}
    \centering
    \includegraphics[width=0.7\columnwidth]{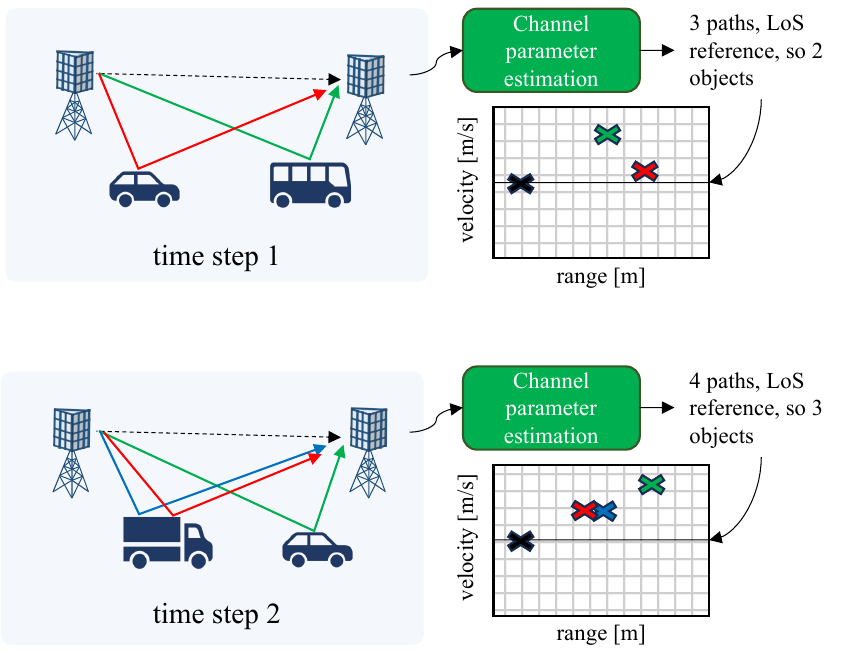}
    \caption{Each measurement gives rise to a potential object. Objects may generate 0, 1, or several measurements at each time. Objects detected at time step 1 should be associated to measurements at time step 2. }
    \label{fig:data-association}
\end{figure}

The problems of an unknown number of objects, complicated object states, and unknown data association are visualized in Fig.~\ref{fig:data-association} for a bistatic scenario.      

\subsection{Radio SLAM in bistatic scenarios} 
\label{sec:SLAM}


In this section, we will focus on bistatic Radio \ac{SLAM} which combines aspects of positioning (see Section \ref{sec:positioning}) and sensing (see Section \ref{sec:bistatic-sensing}) and it involves an \ac{UE} determining its position, while simultaneously detecting and localizing landmarks, based on signals to/from the \acp{BS}. The Radio \ac{SLAM} problem has gained widespread attention in the research field over the past years (see e.g., \cite{shahmansoori2018,wymeersch20185g,hyowon2020,kaltiokallio2022spawc}) since it is closely related to the minimal problems discussed in Section \ref{sec:mininal-positioning} and solving it provides three important benefits. First, the \ac{NLoS} paths provide additional information and can enhance positioning accuracy; second, reliance on infrastructure can be reduced, enabling for example single-\ac{BS} positioning in \ac{mmWave}; third, localization is possible in mixed \ac{LoS}/\ac{NLoS} conditions, even in the absence of \ac{LoS}. 


The processing pipeline of Radio \ac{SLAM} is similar to that of positioning/sensing (see Fig.~\ref{fig:two-step-processing}). The key difference is that in the second stage, the channel parameter estimates are converted to a geometric state of both the \ac{UE} and landmarks. Since Radio \ac{SLAM} has many commonalities to positioning and sensing, we will first introduce the mapping problem which is closely related to the sensing problem and thereafter, we will present the overall Radio \ac{SLAM} problem. In the following, mapping is used to refer to positioning $M$ landmarks, $\vm_{1:M}$, and a map is defined as $\vm = [\vm_1^\top, \; \vm_2^\top ,\; \ldots, \; \vm_M^\top]^\top$.

The mapping problem, commonly referred to as mapping with known poses in \ac{SLAM} literature \cite{thrun2005}, aims to estimate the map posterior, $p(\vm \mid \vx_{1:t}, \vy_{1:t})$, using the sequence of \ac{UE} poses $\vx_{1:t}$ and measurements $\vy_{1:t}$. Since the \ac{UE} trajectory is known, the landmarks are independent and the map posterior can be factorized as \cite{thrun2005}
\begin{equation}\label{eq:map_posterior}
    p(\vm \mid \vx_{1:t}, \vy_{1:t}) = \prod_{i=1}^M p(\vm_i \mid \vx_{1:t}, \vy_{1:t}).
\end{equation}
The map posterior can be computed for example using Bayesian filtering and if the landmark density is approximated using a Gaussian, \eqref{eq:map_posterior} can be efficiently estimated using $M$ \acp{EKF} in parallel, one for each landmark. Landmark-based mapping approaches commonly decompose the physical environmental landmarks such as reflecting surfaces and scattering objects into parametric representations such as a point \cite{hyowon2020}.

The objective of  \ac{SLAM} is to compute the joint posterior density of the \ac{UE} trajectory and map, $p(\vx_{1:t},\vm \mid \vy_{1:t})$, given the measurements up to time $t$. An important characteristic of the \ac{SLAM} problem is that by conditioning the map to the \ac{UE} trajectory renders the landmark estimates conditionally independent. Exploiting this feature, the joint \ac{SLAM} density can be factorized as \cite{montemerlo2002}
\begin{equation}\label{eq:slam_posterior}
    p(\vx_{1:t},\vm \mid \vy_{1:t}) = p(\vx_{1:t} \mid \vy_{1:t}) p(\vm \mid \vx_{1:t}, \vy_{1:t})
\end{equation}
in which $p(\vx_{1:t} \mid \vy_{1:t})$ and $p(\vm \mid \vx_{1:t}, \vy_{1:t})$ are posterior of the \ac{UE} and map, respectively. This factorization makes it natural to apply \ac{RBPF} solutions, in which a \ac{PF} is used to approximate $p(\vx_{1:t} \mid \vy_{1:t})$ and computing $p(\vm \mid \vx_{1:t}, \vy_{1:t})$ is equivalent to solving \eqref{eq:map_posterior}. It is important to note that each particle represents a single \ac{UE} trajectory and a unique map is associated to every particle. In Radio \ac{SLAM}, the \ac{UE} state consists of the pose and clock. The pose represents the position and orientation of the \ac{UE}, whereas the clock represents the required parameters needed to synchronize the local clock of the \ac{UE} to the network clock. 

In literature, a large variety of solutions to \ac{SLAM} are available and these can be classiﬁed as snapshot, ﬁltering and smoothing approaches. The concrete difference between the approaches is the time horizon. Snapshot \ac{SLAM} only considers observations at time $t'$ for estimating $p(\vx_{t'},\vm \mid \vy_{t'})$, filtering approaches utilize measurements up to time $t$ for approximating $p(\vx_{t},\vm \mid \vy_{1:t})$, while smoothing approaches perform batch processing to estimate the full \ac{SLAM} posterior $p(\vx_{1:T},\vm \mid \vy_{1:T})$ in which $T \geq t$. The early works in conventional \ac{SLAM} mainly considered the filtering problem but due to inconsistency issues, majority of the works nowadays consider the smoothing problem \cite{CadEtAl:J16}. In bistatic Radio \ac{SLAM}, the smoothing problem is yet unexplored and the other two approaches have been considered instead. The main differentiator of bistatic Radio \ac{SLAM} is that the \ac{UE} state is estimated in a global reference system that is defined with respect to the \ac{BS}, whereas in conventional \ac{SLAM}, estimation is typically performed in the local frame of the sensor (see also Section  \ref{sec:monostaticSLAM}). This has significant ramifications to the bistatic Radio \ac{SLAM} problem and two prominent examples include: i) it is possible to estimate the \ac{UE} state with respect to a global reference system just from single snapshot observation as presented in Section \ref{sec:positioning}; ii) filter divergence can be identified by comparing the filtering solution to the snapshot solution. It is important to note that both snapshot and filtering approaches are relevant in bistatic Radio \ac{SLAM} and the preferred choice depends on the overall system and application scenario. Snapshot \ac{SLAM} is fundamentally important as it serves as a baseline for what can be done with radio signals alone, while filtering based \ac{SLAM} methods are expected to improve the accuracy. The snapshot approach can be solved for example using numerical optimization methods \cite{wen2021,shahmansoori2018}, whereas the filtering problem is commonly solved using Bayesian filtering \cite{hyowon2020,kaltiokallio2022spawc} or belief propagation on factor graphs \cite{wymeersch20185g,Let:19}. 

\begin{example}

\begin{figure}[!t]
\centering
\includegraphics[width=8cm,trim={1.2cm 0 1.6cm 0.8cm},clip]{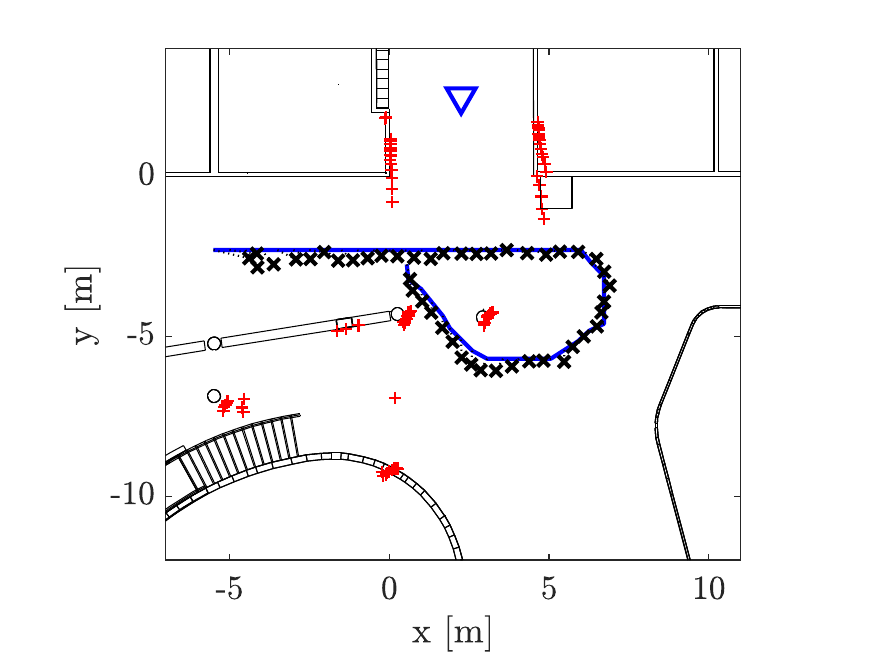}
\caption{Example bistatic Radio SLAM performance obtained using an RBPF-PHD filtering solution. The TX position is illustrated with (\protect\bluetriangle), UE trajectory using (\protect\blueline), and estimated UE path and landmark locations with (\protect\blackcross) and (\protect\redplus), respectively.}
\label{fig:kampusarena_depleted_BM2}
\end{figure}

In this example, we consider an indoor bistatic \ac{SLAM} scenario composing of a single \ac{BS} that transmits \acp{PRS} which are received by the \ac{UE} to jointly estimate its pose and map the surrounding environment. The experiment was conducted at 60 GHz carrier frequency and using 400 MHz \ac{PRS} bandwidth. The \ac{BS} and \ac{UE} were equipped with 4×16 planar antenna arrays and Sivers Semiconductors Evaluation Kits EVK06002 were used as the \ac{TX} and \ac{RX} entities. The \ac{TX} and \ac{RX} used $126$ and $252$ beams, respectively, which corresponds to a $180^\circ$ \ac{FOV} for the \ac{BS} and $360^\circ$ \ac{FOV} for the \ac{UE}. Overall, the UE trajectory consisted of $45$ measurements positions. The experimental scenario is  illustrated in Fig.~\ref{fig:kampusarena_depleted_BM2} and for further details, please see \cite{rastorguevafoi2023}.

The considered \ac{SLAM} problem is solved using an \ac{RBPF}-\ac{PHD} filter \cite{hyowon2020} that utilizes an \ac{OID} approximation \cite{kaltiokallio2022spawc} to decrease the number of required particles. The performance of the algorithm is visualized in Fig.~\ref{fig:kampusarena_depleted_BM2} and as illustrated, the estimated path closely follows the \ac{UE} trajectory and the estimated landmarks coincide with the actual floor plan. In the experiment, the \ac{RMSE} position, heading and synchronization error are: $0.55 \text{ m}$, $2.43^\circ$ and $1.53 \text{ ns}$, respectively. The benefit of using a filtering approach is three fold. First of all, the filter can operate in mixed \ac{LoS}/\ac{NLoS} conditions and the posterior from the previous time step can be viewed as a regularization term which constrains the posterior update so that the system state is identifiable at every measurement position. Second, sequential processing of the measurements improves the accuracy. Third, the filter can inherently deal with the challenges of bistatic Radio \ac{SLAM} which include: i) the landmark can be misdetected due to limitations in the \ac{RX} and channel estimation routine; ii) clutter measurements and multi-bounce signals can generate false detections that are not inline with the measurement model; and iii) measurement ambiguities can lead to situations where a wrong measurement is associated to the wrong landmark.

\end{example}

\section{Technologies for  joint monostatic sensing and communication}
\label{sec:monostatic}


\subsection{Introduction}
In this section, we address the mono-static sensing and SLAM paradigms in the spirit of cellular ISAC. We first shortly review the fundamentals and challenges of monostatic system scenarios. Then, in Section V.B, we present the basics of the related \ac{SI} waveform modeling and discuss shortly the potential \ac{TX}-\ac{RX} isolation solutions, while Section V.C provides examples how precoding/beamforming optimization can contribute to \ac{SI} suppression. Section V.D then looks into relevant impairments, most notably oscillator phase noise and Doppler-induced \ac{ICI}, and shows how these can be turned from foe to friend in monostatic sensing. Lastly, section V.E presents monostatic \ac{SLAM}, which is one potential application of monostatic sensing.

\begin{figure*}[!t]
    \centering
    \includegraphics[width=\textwidth]{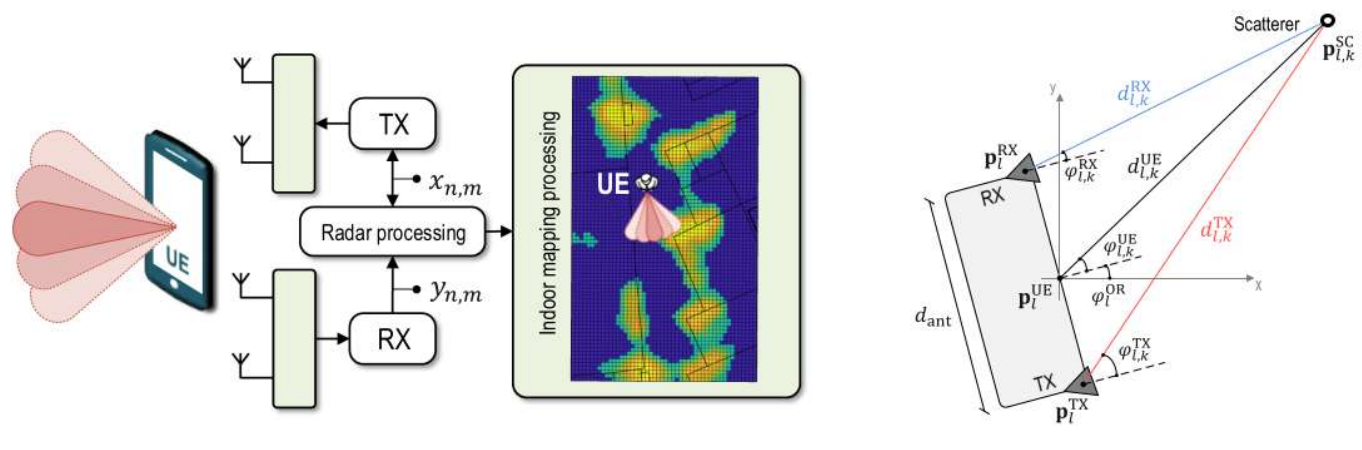}
    \caption{Illustration of the monostatic sensing paradigm at UE end, together with the corresponding problem geometry.}
    \label{fig:monostatic_sensing_mobile_UE}
\end{figure*}

In general, different from the previously discussed bi-static and multi-static transmitter-receiver arrangements, the \emph{mono-static} case refers to a scenario where the transmitting and receiving entities are essentially collocated \cite{richards2005fundamentals,Barneto2021WC,Baquero2019TMTT}. A principal illustration of such mono-static sensing and mapping scenario is shown in {Fig.~\ref{fig:monostatic_sensing_mobile_UE}}, indicating also the opportunity for separate antenna systems at the TX and RX ends \cite{Puc22}. Such mono-static approach allows to turn the individual \acp{UE} or \acp{gNB} essentially into cellular radars, providing standalone situational awareness related to the surrounding environment without relying on other network entities. Example representative use cases could be, e.g., in different vehicular or industrial systems where connected moving machines can extract and harness situational awareness in terms of the environment landmarks while also tracking their own coordinates relative to a reference point, by using their own transmit waveform as the illumination signal. Compared to the bi-static or multi-static counterparts, the mono-static approach is appealing as the complete I/Q transmit waveform is directly known also to the receiver. Additionally, arranging for very accurate time- and frequency-synchronization between the TX and RX chains is clearly much more straight-forward and can be resolved even at hardware level.

The mono-static operation is, as such, well-known from ordinary radars. Pulsed radars \cite{richards2005fundamentals} operate on the basis of time-multiplexing between the exact transmit and receive time periods. In the context of cellular systems, specifically in TDD networks, the individual network nodes (\acp{UE} and \acp{gNB} and the corresponding \acp{TRP}) also operate on the basis of dividing the TX and RX active periods in time. However, the individual active transmit periods in both uplink and downlink are commonly in the order of a millisecond – a period that is very long compared to pulsed radars, and does not facilitate measuring transmit waveform interaction with targets or landmarks at any meaningful distance. Hence, any cellular monostatic sensor must be able to execute the receiver simultaneous to transmitting, in order to measure the reflecting and scattering waves. This, in turn, means that such sensor is essentially operating as an \ac{IBFD} transceiver, commonly also referred to as a \ac{STAR} system \cite{Kolodziej2019TMTT,Barneto2021WC,Baquero2019TMTT}.

One fundamental technical challenge in \ac{STAR} systems is, in general, the \emph{self-interference} (SI) or direct transmit-receive coupling \cite{Kolodziej2019TMTT,Khaledian2018TMTT,Liu2017TMTT}. As a concrete example, the \ac{EIRP} of a macro gNB can be in the order of +70\,dBm, or more, while the receiver thermal noise floor with e.g. 100\,MHz channel bandwidth is around --90\,dBm. Thus, from the sensing or radar function point of view, the direct TX-RX coupling acts as an extremely powerful target at a very short distance, and will mask everything else if not properly suppressed through antenna processing, active RF cancellation and baseband digital cancellation \cite{Kolodziej2019TMTT,Khaledian2018TMTT,Liu2017TMTT,Pascual2021TSP,Kiayani2018TMTT}. Importantly, sufficient amount of TX-RX isolation must be obtained already in the antenna or RF circuit domain prior to the receiver \ac{LNA} – otherwise, the powerful SI will totally desensitize and block the \ac{LNA} and thereon the sensing receiver. In the following Section V.B, we provide further modeling of the \ac{SI} waveform while also discussing the basic \ac{TX}-\ac{RX} isolation approaches.

\subsection{Basic SI modeling and isolation solutions}
We next shortly discuss the different potential solutions to arrange sufficient TX-RX isolation, as well as the related topic of SI channel and waveform modeling. There are, generally speaking, two alternative ways to arrange the antenna interface for the simultaneously operating transmitter and receiver, namely, sharing a set of antennas through circulators for both TX and RX, or then utilizing different sets of physical antennas for TX and RX. One may consider the latter more feasible approach, as the circulators are commonly providing only isolation levels in the order of 20\,dB while the different sets of antennas allow for larger isolation levels through separating the antenna systems in space---especially in gNB type of entities as well as in UE with larger form factors such as vehicles. Additionally, an importantly separate antenna systems allow harnessing beamforming optimization for TX-RX isolation in a more efficient manner. Hence, we also primarily assume in the following that separate antenna systems are utilized, though the RF and baseband digital canceller principles apply to both scenarios.

Considering first the basic SI channel and waveform modeling, let us denote the baseband precoded transmit waveform at transmit path $l$ by $x_l(n)$, and further assume that subarray based hybrid beamforming is adopted at TX where each precoded signal is transmitted through a subarray of size $M$. Then, the complex baseband equivalent SI waveform at receiver antenna $j$, contributed by the TX subarray $l$, reads
\begin{align}
    x^{\text{RF},\text{SI}}_{j,l}(n)= \sum_{i=1}^{M}h^{\text{SI},l}_{i,j}(n)*(f_{i,l}\phi^\text{TX}_l(x_l(n))), \label{eq:SI_RF}
\end{align}
where $h^{\text{SI},l}_{i,j}(n)$ denotes the physical SI channel from the antenna element $i$ of the TX subarray $l$ to RX antenna element $j$, $*$ denotes convolution, while the function $\phi^\text{TX}_i(\cdot)$ refers to a transmitter hardware model that can accommodate different impairments such as power amplifier (PA) nonlinearity and oscillator phase noise. Additionally, the beamforming weights of the TX subarray $l$ are denoted by $f_{i,l}$.  

Next, taking into account the contribution of all involved TX subarrays, say $l=1,2,...,L$, while incorporating also RX beamforming as well as RX hardware path impairments, the observable baseband SI waveform reads eventually
\begin{align}
    x^{\text{BB},\text{SI}}(n) &=  \phi^\text{RX}(\,\sum_{j=1}^{K}w_{j}\sum_{l=1}^{L}x^{\text{RF},\text{SI}}_{j,l}(n)), \label{eq:SI_BB}
\end{align}
where $K$ denotes the array size at RX, $w_{j}$ denotes the corresponding beamforming weight, and $\phi^\text{RX}$ refers to the RX path hardware model.
This model is valid for an arbitrary RX (sub)array.

Importantly, the models in \eqref{eq:SI_RF}-\eqref{eq:SI_BB} show that both TX and RX beamforming contribute to the observable SI at RX. Thus, beamforming is one fundamental processing tool to facilitate or improve TX-RX isolation---and particularly the TX beamforming as it allows to control the SI waveform already at the RX LNA input as shown by \eqref{eq:SI_RF}. Beamforming optimization solutions that seek for controlled and favorable tradeoffs between SI suppression, target illumination capability and achievable communication rate are described, e.g., in \cite{Barneto2022TCOM,Islam2022ICC,Liyanaarachchi2021JCS,Baquero2020ICC}, and will also be discussed in the following Section V.C. Furthermore, the accommodation of the prominent hardware imperfections in \eqref{eq:SI_RF}-\eqref{eq:SI_BB} is of great importance. For one, as shown e.g. in \cite{Pascual2021TSP, Pascual2018GC,KorpiAsilomar13}, appropriate HW modeling allows to craft advanced digital SI cancellation solutions that can suppress also the nonlinear effects and distortion products imposed by TX PA, RX LNA and other analog components. Such is important when optimizing the receiver dynamic range. And what is more, as discussed and demonstrated concretely later in this Chapter, properly modeling of hardware imperfections can turn them from a foe to a friend, when interpreted from the sensing point of view---this is particularly so for the oscillator phase noise as discussed further in the following in Section V.D.

\subsection{Triple-function precoding/combining in full duplex transceivers}
In this section, we discuss the precoder and combiner design that enable STAR operation for monostatic sensing with communication-centric FD transceivers. We focus on designs suitable for initial access and also tracking scenarios. 

\subsubsection{Precoding/combining for tracking}
We focus on the tracking stage of the joint \ac{DL} communication and sensing with an FD \ac{BS} in a single \ac{UE} setting. The three functions of the spatial filters at both the TX and the RX side can be stated as follows: (i) enabling \ac{DL} communication, (ii) providing high gain in the target direction, (iii) and contributing to SI suppression.
 To provide simultaneous communication and sensing the system should operate with a multibeam precoder that illuminates the target angle and the multiple paths of the communication channel \cite{Barneto2021WC}. Considering a system operating in FR2 with a hybrid analog/digital architecture at both ends of the FD \ac{BS} and OFDM signaling, the received signal at the FD \ac{BS} can be expressed as
\begin{multline}\label{eq_rec_FD}
    \yy_{n,m} = \WW\herm \HH_{n,m} \FFrf \FFbb[n] \xx_{n,m} \\ + \underbrace{\WW\herm \HH^{\rm{SI}}_{n} \FFrf \FFbb[n] \xx_{n,m}}_{\text{SI}} + \zz_{n,m},
\end{multline}
where $\HH_{n,m},\,\HH^{\rm{SI}}_{n} \in \complexset{\Nrx}{\Ntx}$ are the radar and SI channels, respectively. The entries of the SI channel are the coupling between each TX-RX antenna pair, as described previously. We assume that the FD \ac{BS} is stationary, which is why the SI channel remains constant at different symbols. The second term in \eqref{eq_rec_FD} should be suppressed with precoding/combining so that the LNAs are not saturated and the targets can be detected. 

The precoders should also provide reliable \ac{DL} communication and target gain for the radar operation. We can quantify the communication performance with the subcarrier dependent rate $\mathcal{R}_{n}$, which is a function of the precoders, the \ac{DL} channel, and the combiners at the \ac{UE}. The precoders and the combiners at the \ac{UE} are designed separately, which is the common approach in communications. Moreover, we can quantify the sensing performance with the TX target gain per stream for a given target angle $\theta_{\rm r}$, which is expressed as
\begin{equation}\label{eq_transmit_target_gain}
     G_{\mathrm{T}, n_s, n}(\theta_{\rm r}) = \left|\atx\herm (\theta_{\rm r}) \left[\FFrf \FFbb[n,m]\right]_{:,n_s} \right|^2,
\end{equation}
for $n_s = 1,\dots,\Ns$. The final objective of the precoders is to mitigate the SI that is equivalent to nulling the SI term in \eqref{eq_rec_FD} as
\begin{equation}\label{eq_SI}
     \WW\herm \HH^{\rm{SI}}_{n} \FFrf \FFbb[n] = \boldzero_{\Nrfr \times \Ns},
\end{equation}
which is a joint task for the combiner. Aside from SI suppression, the goal of the combiner is to provide a high target gain at the $n_{\rm RF}$-th RX RF chain that is expressed as
\begin{equation}\label{eq_receive_target_gain}
     G_{\mathrm{R}, n_{\rm RF}}(\theta_{\rm r}) = \left|[\WW]_{:,n_{\rm RF}}\herm \arx (\theta_{\rm r}) \right|^2,
\end{equation}
for $n_{\rm RF}=1,\dots,\Nrfr$. The joint precoding/combining problem can be formulated as the maximization of the sum DL rate over all subcarriers under the target gain, SI suppression and hardware constraints \cite{Barneto2022TCOM,Islam2022ICC}. We can define TX and RX gain thresholds denoted by $\tau_{\rm T}$ and $\tau_{\rm R}$, respectively. With the described formulation, we can write the overall optimization problem as
\begin{equation}\label{eq_problem_overall}
    \begin{aligned}
    & \underset{ \substack{\WW,\,\FFrf, \\ \FFbb[n] }}{\text{maximize}} & & \sum_{n=0}^{N-1} \mathcal{R}_{n} \\
    & \text{subject to} & &  G_{\mathrm{T}, n_s, n}(\theta_{\rm r}) \geq \tau_{\rm T}, \:\: G_{\mathrm{R}, n_{\rm RF}}(\theta_{\rm r}) \geq \tau_{\rm R}, \\
    & & & \eqref{eq_SI}, \:\: \FFrf \in \mathcal{A}_{\rm T}, \:\: \WW \in \mathcal{A}_{\rm R}, \\
    & & & \|\FFrf\FFbb[n]\|_{\rm F}^{2} = \Ns,
\end{aligned}
\end{equation}
where $\mathcal{A}_{\rm T}$ and $\mathcal{A}_{\rm R}$ are the sets of feasible structures for the analog precoder and the analog combiner. For example, unit-modulus entries, which correspond to a phase shifter network, and the subarray architecture can be imposed on the analog precoder/combiner by using the defined sets. Finally, normalization is imposed on the precoders to satisfy the power constraint.

The problem in \eqref{eq_problem_overall} is hard to solve due to the coupling between the precoders and the combiner, and the hardware constraints related to the analog precoder/combiner. Hence, alternating optimization can be adopted to decompose the problem into several subproblems, each of them responsible for one of the variables while the others are fixed. Furthermore, relaxation techniques such as convex approximations and semidefinite problems can be used to solve each subproblem. The work in \cite{Barneto2022TCOM} solves this problem by relaxing the cost function from rate to gain in the \ac{LoS} angle of the users. Moreover, SI cancellation is satisfied with only the combiner by using the null-space projection (NSP) method which requires fine grained attenuators in addition to phase shifters. In \cite{Islam2022ICC}, the authors maximize the data rate by assuming that a subset of the targets in the environment contribute to downlink communication. They utilize an analog SI cancellation method which reduces the SI. Then, the analog precoder and combiner are designed by using a DFT codebook. Finally, the residual SI is suppressed by the digital precoders, which also maximize the \ac{DL} communication rate. In \cite{Bayraktar2023ArXiV}, a more general system model is considered where the \ac{DL} channel can have multiple paths which do not necessarily coincide with the targets. This work utilizes generalized eigenvalue-based precoders that suppress the SI while maximizing the data rate. The precoders are coherently combined with the precoders that maximize the TX target gain. Then, a hybrid decomposition algorithm is used to obtain analog and digital precoders, which reduce the SI suppression capability. Finally, a convex relaxation-based algorithm is employed to design the combiner that minimize the residual SI while keeping the RX target gain above a threshold.

\subsubsection{Precoding/combining for initial access}
In this setting, the \ac{BS} and \ac{UE} have to establish communication for the first time. Thus, the precoders at the \ac{BS} and the combiners at the \ac{UE} should be designed such that they cover a wide angular region. This stage, in nature, is similar to the target discovery step of radar operation. Hence, the precoders and combiners at the \ac{BS} that are used for initial access and channel estimation for communication are suitable for radar sensing. If we consider the triple-function precoding/combining for tracking, we could see that the spatial resources are shared between the communication and sensing, which would not be necessary for the initial access stage. In other words, the triple-function requirement for precoding/combining would be reduced to two-function as the communication and sensing requirement would be unified.

\begin{figure}
        \begin{center}
        \subfigure[]{
			 \label{fig_initial_codebook_SINR}
			 \includegraphics[width=0.6\linewidth]{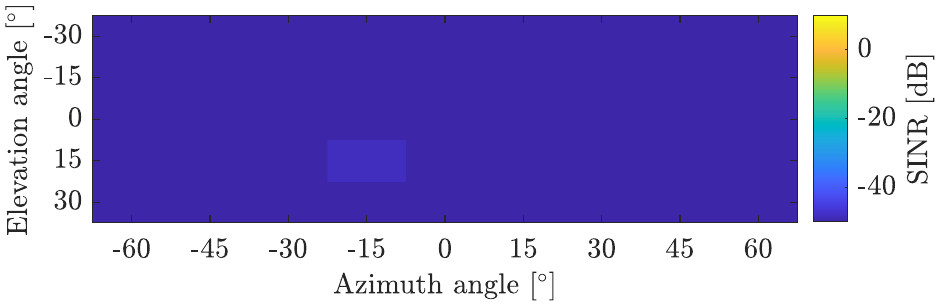}
		}
        \subfigure[]{
			 \label{fig_proposed_codebook_SINR}
			 \includegraphics[width=0.6\linewidth]{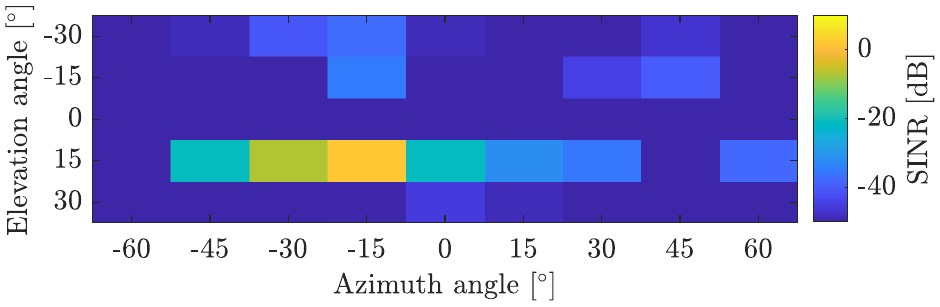}
		}		
		\end{center}
		\vspace{-0.2in}
        \caption{Radar SINR at the FD \ac{BS} equipped with $8\times8$ arrays that are separated by $10\lambda$ for: (a) the initial codebook, (b) the SI-aware codebook. The center frequency is $28\,\rm{GHz}$, the bandwidth is $300\,\rm{GHz}$, the transmit power is $20\,\rm{dBm}$, and the SI-to-noise ratio is $80\,\rm{dBm}$. The target is located at $80\,\rm{m}$ from the FD \ac{BS} with azimuth angle $-20^\circ$ and elevation angle $15^\circ$.}  
        \label{fig_radar_SINR}
\end{figure}

Initial access can be established with beam training for \ac{FR2} in 5G NR. The beams are selected from a codebook that comprise directional beams. Let us denote the directional codebooks for TX and RX by $\bA_{\rmtx} = [ \atx(\theta_1) \, \ldots \, \atx(\theta_{\Mtx})] \in \complexset{\Ntx}{\Mtx}$ and $\bA_{\rmrx} = [ \arx(\phi_1) \, \ldots \, \atx(\phi_{\Mrx})] \in \complexset{\Nrx}{\Mrx}$, where $\Mtx$ and $\Mrx$ are the sizes of AoA set $\{\theta_i\}_{1}^{\Mtx}$ and AoD set $\{\phi_j\}_{1}^{\Mrx}$, respectively. If these codebooks are used for sensing at the FD \ac{BS}, we would observe high residual SI. Let us denote the SI-aware beam codebooks that we would like design as $ \bar{\bA}_{\rmtx} \in \complexset{\Ntx}{\Mtx}$ and $\bar{\bA}_{\rmrx} \in \complexset{\Nrx}{\Mrx}$. The coupling matrix that shows the residual SI between beam pairs can be defined as $\bC = \bar{\bA}_{\rmrx}\herm \HH^{\rm{SI}} \bar{\bA}_{\rmtx}$, where $\HH^{\rm{SI}} \in \complexset{\Nrx}{\Ntx}$ is the frequency-flat SI channel. For a wideband SI channel, $\HH^{\rm{SI}}$ can be selected as the the SI channel at the center frequency. The goal of the codebook design is to minimize the total coupling power, i.e., $\|\bC\|_{\rm F}^{2}$, while the beams maintain a high gain at the AoA and AoD grids. One other constraint is to have unit-modulus entries so that the beams can be realized with a purely analog architecture that comprises phase shifters, which makes the problem non-convex. The TX and RX codebooks are designed with alternating minimization, and convex relaxation combined with block-coordinate descent method by using $\bar{\bA}_{\rmtx}$ and $\bar{\bA}_{\rmrx}$ as initial codebooks in \cite{Bayraktar2023CAMSAP}. The beam training procedure for sensing is designed such that a TX-RX beam pair that correspond to the same angle on the AoD and AoA grids is employed for each symbol at the FD \ac{BS}. We show the radar SINR obtained at the angular grid in a single point target scenario in Fig.~\ref{fig_radar_SINR}. The FD \ac{BS} is equipped with two $8\times8$ planar arrays that are separated by $10\lambda$. The angular grid for both TX and RX codebooks is selected as $[-60^\circ,-45^\circ,\ldots,60^\circ]\times[-30^\circ,-15^\circ,\ldots,30^\circ]$. The results show that the SI-aware codebook efficiently suppresses SI and yields high radar SINR at target angle, while the initial codebook is corrupted by the SI, which makes the target detection impossible. The beams from the SI-aware codebook can also be used for the tracking stage by utilizing the beams from the codebook at the analog precoder/combiner design.

The described monostatic sensing scenarios for the tracking and initial access stages include only \ac{DL} users for communication. Thus, the precoder at the FD \ac{BS} is designed such that it serves for both sensing and communication while considering SI suppression. From a communication perspective, only half-duplex communication is considered. One of the most important challenges is to integrate \ac{UL} users to enable FD communication which would significantly improve the spectrum efficiency of the overall system. 

\subsection{Joint monostatic sensing and communication under non-idealities: repurposing challenges into benefits}
In this part, we address the problem of joint monostatic sensing and communication with OFDM waveform in the presence of non-idealities, and demonstrate how such imperfections can be turned into an advantage for sensing. We begin by presenting the OFDM radar signal model and basic operations for range-Doppler detection/estimation. Then, we focus on two specific non-idealities, namely inter-carrier interference (ICI) and phase noise (PN), elaborating on their impacts on sensing as well as on how they can be exploited to improve sensing performance. For ease of exposition, we consider a single-input single-output (SISO) system at both radar and communication receivers. 

\begin{figure*}
	\centering
	\includegraphics[width=1\linewidth]{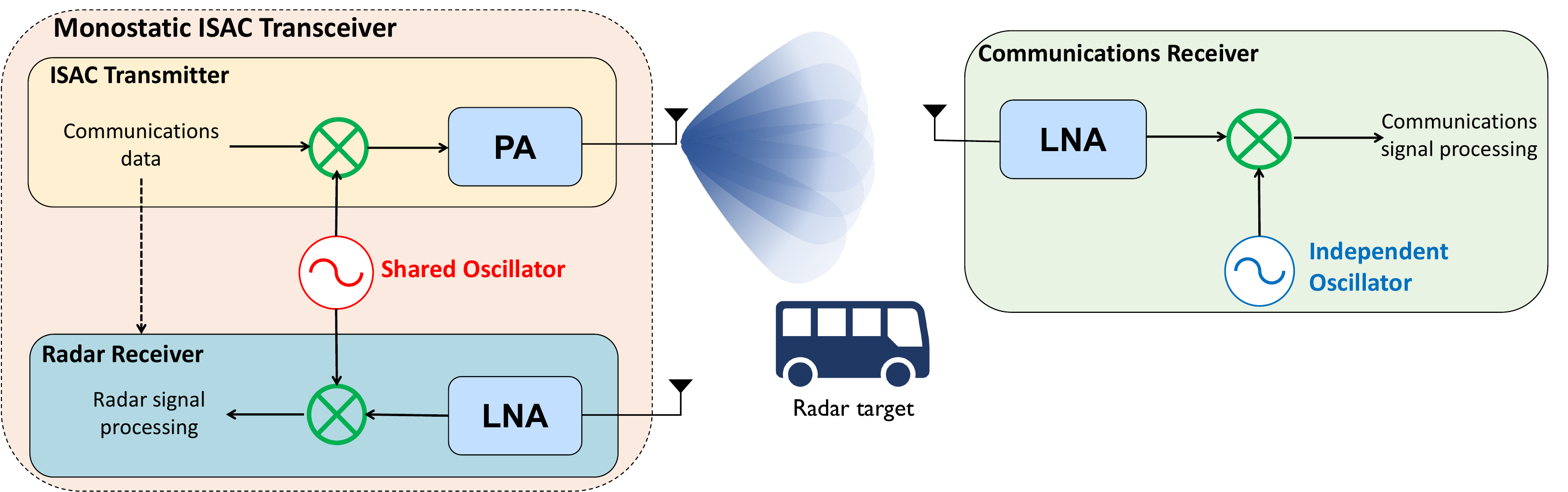}
	\caption{Monostatic ISAC setup with a monostatic ISAC transceiver containing an ISAC transmitter and a radar receiver co-located on a shared hardware, and a communications receiver located on a remote device. In contrast to the independent oscillator employed at the communications receiver, the use of a shared oscillator in monostatic sensing implies that impairments, such as phase noise (PN) and intercarrier interference (ICI), carry geometric information on the parameters of radar targets (i.e., delay and Doppler, respectively) and thus can be turned into benefits for sensing. Please see Table~\ref{tab_expl_isac} for further details.}
	\label{fig_monostatic_isac}
\end{figure*}

\subsubsection{OFDM monostatic radar sensing without non-idealities}
Consider a monostatic ISAC setup in Fig.~\ref{fig_monostatic_isac}, where a monostatic ISAC transceiver sends data symbols to a remote communications receiver and collects the backscattered signals at the co-located radar receiver for sensing the objects in the environment. Under a SISO setup, the monostatic OFDM radar observations can be expressed, using \eqref{eq_rec} and \eqref{eq:channel}, as
\begin{equation}\label{eq_rec_siso}
    y_{n,m} =  \sum_{\ell=0}^{L-1} \alpha_{\ell}  e^{-j 2 \pi n \deltaf \tau_{\ell} } e^{j 2 \pi m \Tsym \nu_{\ell} }  x_{n,m} + z_{n,m}  ~.
\end{equation}
Stacking \eqref{eq_rec_siso} over $N$ subcarriers and $M$ symbols, we obtain the frequency-time domain observations in compact matrix form as \cite{OFDM_DFRC_TSP_2021}
\begin{align} \label{eq_y_radar_matrix}
    \YY =  \XX \odot \sum_{\ell=0}^{L-1} \alpha_{\ell} \,    \bb(\tau_{\ell}) \cc\trp(\nu_{\ell})   + \ZZ \in \complexset{N}{M} ~,
\end{align}
where $\XX \in \complexset{N}{M}$ with $\left[ \XX \right]_{n,m} = x_{n,m}$, $\YY \in \complexset{N}{M}$ with $\left[ \YY \right]_{n,m} = y_{n,m} $, and, $\bb(\tau) \in \complexset{N}{1}$ and $\cc(\nu) \in \complexset{M}{1}$ denote, respectively, the frequency-domain and time-domain steering vectors, i.e., $[\bb(\tau)]_n = e^{-j 2 \pi n \deltaf \tau} $ and $[\cc(\nu)]_m = e^{j 2 \pi m \Tsym \nu} $. To detect targets from \eqref{eq_y_radar_matrix} and estimate their delay-Doppler parameters, we first remove the impact of data symbols $\XX$ either via reciprocal filtering/zero-forcing (i.e., dividing $\YY$ by $\XX$ element-wise) or matched filtering (i.e., multiplying $\YY$ by the conjugate of $\XX$ element-wise) \cite{RadCom_Proc_IEEE_2011,OFDM_Radar_Phd_2014,OFDM_Radar_Corr_TAES_2020,Fan:22}. Since $\bb(\tau)$ and $\cc(\nu)$ correspond to DFT/IDFT matrix columns on a uniformly sampled delay-Doppler grid, taking IDFT over the columns and DFT over the rows of the resulting observation matrix provides the delay-Doppler/range-velocity spectrum, from which target detection and parameter estimation can be performed \cite{RadCom_Proc_IEEE_2011,MIMO_OFDM_ICI_JSTSP_2021,Fan:22,ISAC_survey_IoT_2023}, e.g., via constant false alarm rate (CFAR) processing \cite[Ch.~6.2.4]{richards2005fundamentals}. Alternatively, super-resolution algorithms, such as MUSIC and ESPRIT, can be applied by harnessing the structure in $\bb(\tau)$ and $\cc(\nu)$ (e.g., shift-invariance property) \cite{OFDM_ISAC_ESPRIT_TVT_2020,OFDM_ISAC_MUSIC_ESPRIT_TVT_2020}.

\subsubsection{OFDM monostatic radar sensing under ICI}
\begin{figure}
	\centering
	\includegraphics[width=0.5\linewidth]{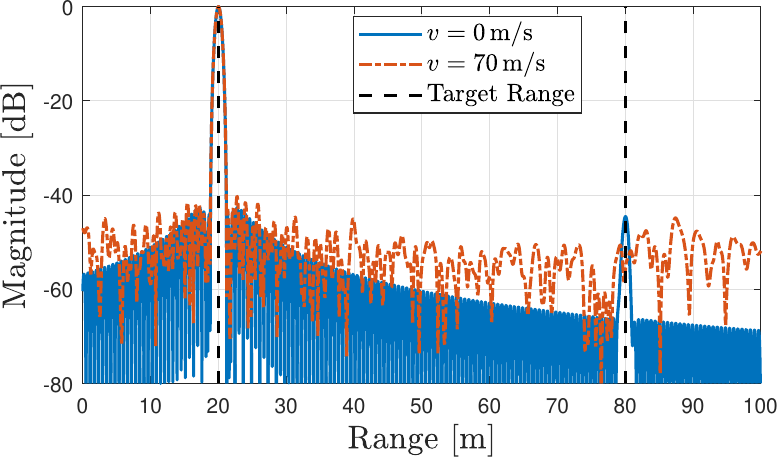}
	\caption{Range spectrum in OFDM-based sensing under the impact of ICI. The OFDM parameters are $\fc = 28 \, \rm{GHz}$, $\deltaf = 60 \, \rm{kHz}$, $N = 4096$ and $M = 16$. The scenario includes $2$ targets with the same velocity $v$, the ranges $(20, 80) \, \rm{m}$ and the SNRs $(25, -20) \, \rm{dB}$, respectively. $v = 70 \, \rm{m/s}$ represents the case where two cars are approaching one another on a highway.}
	\label{fig_ici_range_profile}
	\vspace{-0.1in}
\end{figure}

When dealing with high-speed targets and/or small $\deltaf$, the validity of the model \eqref{eq_y_radar_matrix} diminishes since intra-symbol Doppler-induced phase shifts become non-negligible, destroying subcarrier orthogonality and leading to ICI. Under scenarios with high-mobility and/or small $\deltaf$, \eqref{eq_y_radar_matrix} can be generalized to  \cite{MIMO_OFDM_ICI_JSTSP_2021}
\begin{align} \label{eq_y_ici}
    \YY = \sum_{\ell=0}^{L-1} \alpha_{\ell}  \FF_N \underbrace{\DD(\nu_{\ell})}_{\substack{\rm{ICI} } } \FF_N\herm \Big(\XX \odot \bb(\tau_{\ell}) \cc\trp(\nu_{\ell}) \Big)  + \ZZ ~,
\end{align}
where $\DD(\nu) = \diag{1, e^{j 2 \pi \frac{T}{N} \nu}, \ldots, e^{j 2 \pi \frac{T(N-1)}{N} \nu} } \in \complexset{N}{N}$ encodes intra-symbol (\textit{fast-time} in radar nomenclature) phase shifts as a function of Doppler $\nu$. We note that inter-symbol (\textit{slow-time}) phase shifts are captured by $\cc(\nu)$. For low velocities and/or large subcarrier spacing, the maximum phase progression in $\DD(\nu)$ satisfies $2 \pi T \nu \ll 2 \pi$. In this case, $\DD(\nu) \approx \Imatrix$ and \eqref{eq_y_ici} boils down to \eqref{eq_y_radar_matrix}. To illustrate the impact of ICI on sensing, Fig.~\ref{fig_ici_range_profile} shows the range spectrum of OFDM radar obtained via standard DFT/IDFT based processing, using typical 5G NR FR2 parameters in a high-mobility scenario. We observe a noticeable increase in side-lobe levels induced by ICI, leading to a masking effect on targets. This phenomenon poses a significant challenge for sensing as it could severely impede the detection of weaker targets \cite{OFDM_ICI_TVT_2017,MIMO_OFDM_ICI_JSTSP_2021}.


\subsubsection{OFDM monostatic radar sensing under PN}
\begin{figure}
	\centering
	\includegraphics[width=0.5\linewidth]{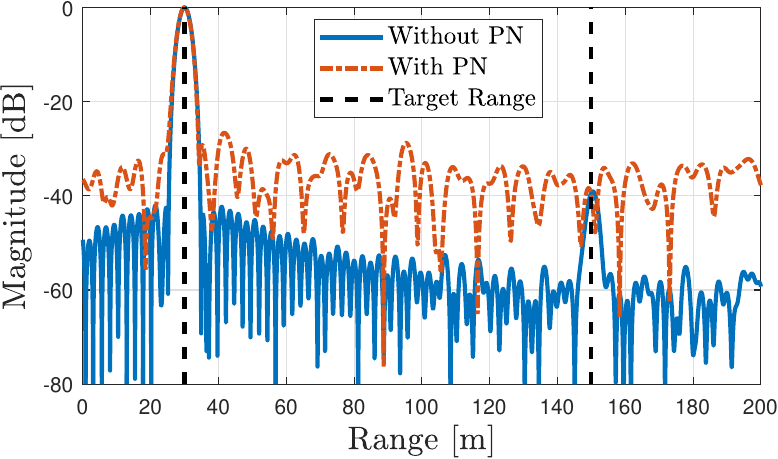}
	\caption{Range spectrum in OFDM-based sensing under the impact of PN. The OFDM parameters are $\fc = 28 \, \rm{GHz}$, $\deltaf = 120 \, \rm{kHz}$, $N = 512$ and $M = 10$, while the oscillator is a free-running oscillator with the $3 \,\rm{dB}$ bandwidth $200 \, \rm{kHz}$. The scenario contains $2$ targets with the ranges $(30, 150) \, \rm{m}$, the velocities $(10, 10) \, \rm{m/s}$, and the SNRs $(30, -10) \, \rm{dB}$, respectively.}
	\label{fig_pn_range_profile}
	\vspace{-0.1in}
\end{figure}

For practical, non-ideal oscillators, PN should be taken into account in sensing. In this case, \eqref{eq_y_radar_matrix} becomes
\cite{PN_Exploitation_TSP_2023}
\begin{align} \label{eq_y_pn}
    \YY = \sum_{\ell=0}^{L-1} \alpha_{\ell} \FF_N \Big[  \underbrace{\WW(\tau_{\ell})}_{\substack{\rm{PN} } } \odot\, \FF_N\herm \Big(\XX \odot \bb(\tau_{\ell}) \cc\trp(\nu_{\ell}) \Big) \Big]  + \ZZ ~,
\end{align}
where $\WW(\tau_{\ell}) \in \complexset{N}{M}$ represents the multiplicative PN matrix in the fast-time/slow-time domain, belonging to the $\thn{\ell}$ target. Due to the use of a shared oscillator at the ISAC transmitter and the radar receiver, as shown in Fig.~\ref{fig_monostatic_isac}, $\WW(\tau_{\ell})$ corresponds to a realization of a differential (self-correlated) PN process \cite{Demir_PN_2006}, which has \textit{delay-dependent statistics} \cite{PN_Exploitation_TSP_2023} (the so-called \textit{range correlation} effect \cite{Range_Correlation_93,SPM_PN_2019,Canan_SPM_2020}). For the special case of an ideal oscillator, $\WW(\tau_{\ell})$ degenerates to an all-ones matrix and \eqref{eq_y_pn} reverts to \eqref{eq_y_radar_matrix}. Fig.~\ref{fig_pn_range_profile} illustrates the impact of PN on the range spectrum of OFDM radar. Similar to ICI, PN distorts subcarrier orthogonality and reduces the dynamic range of the radar, which, in turn, degrades detection performance, particularly for weak targets.

\subsubsection{Exploitation of ICI and PN in monostatic sensing}
As illustrated in Fig.~\ref{fig_monostatic_isac}, a critical distinction emerges between monostatic radar sensing and communications in ISAC systems when impairments, such as ICI and PN, are present. Specifically, in a monostatic sensing configuration, the radar receiver shares the same oscillator with the ISAC transmitter. In contrast, the communications receiver uses an independent oscillator. This implies that \textit{(i)} the ICI effect in \eqref{eq_y_ici} involves only target Doppler without any carrier frequency offset (CFO) \cite{MIMO_OFDM_ICI_JSTSP_2021}, and \textit{(ii)} the statistics of the PN in \eqref{eq_y_pn} are delay-dependent since this self-correlated PN process represents the difference between the original PN process and a time-shifted version, where the shift corresponds to the round-trip delay of the target \cite{Bliss_PN_2016,PN_Exploitation_TSP_2023}. Table~\ref{tab_expl_isac} summarizes the differences between monostatic sensing and communications in the face of impairments.

\begin{table}
\caption{Exploitation of Impairments in Joint Monostatic Sensing and Communications}
\centering
    \begin{tabular}{|l|l|l|}
        \hline
         & \textbf{Monostatic Sensing} & \textbf{Communications} \\ \hline
        \textbf{Oscillator} & Shared & Independent \\ \hline      
        \textbf{ICI} & Doppler & Doppler + CFO \\ \hline 
        \textbf{PN} & Delay-dependent statistics & Independent statistics \\ \hline 
    \end{tabular}
    \label{tab_expl_isac}
\end{table}

\begin{figure}
	\centering
	\includegraphics[width=0.8\linewidth]{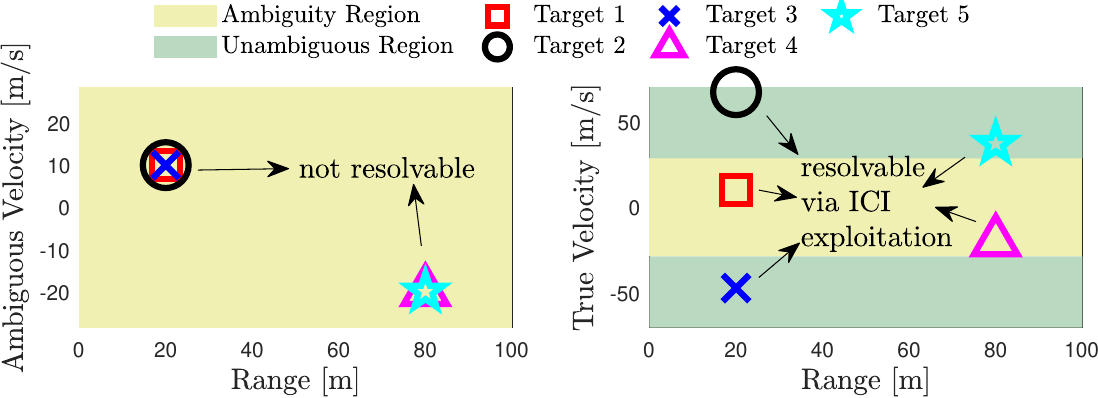}
	\caption{Exploitation of ICI in a multi-target scenario to resolve Doppler ambiguity of high-mobility targets and introduce additional dimension for target resolvability, where $\fc = 60 \, \rm{GHz}$, $B = 50 \, \rm{MHz}$ and $N = 2048$.}
	\label{fig_ici_exploitation}
\end{figure}

\begin{figure}
	\centering
	\includegraphics[width=0.6\linewidth]{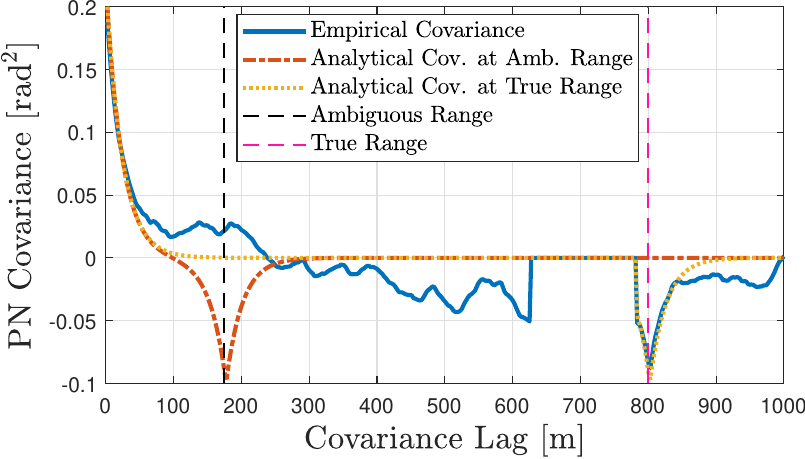}
	\caption{Exploitation of PN via covariance matching to resolve range ambiguity of a target at $800 \, \rm{m}$, which appears at $175 \, \rm{m}$ due to the maximum unambiguous range of $625 \, \rm{m}$. The OFDM parameters are $\fc = 28 \, \rm{GHz}$, $\deltaf = 240 \, \rm{kHz}$, $N = 256$ and $M = 10$, while the oscillator is a phase-locked loop (PLL) synthesizer with the loop bandwidth $1 \, \rm{MHz}$ and the $3 \,\rm{dB}$ bandwidth $100 \, \rm{kHz}$.}
	\label{fig_pn_exploitation}
\end{figure}

Such distinctive properties of ICI and PN in monostatic sensing present opportunities to turn these traditionally detrimental effects into beneficial elements to improve sensing performance. As seen from \eqref{eq_y_ici}, ICI brings additional Doppler information through $\DD(\nu)$ on top of slow-time Doppler information carried by $\cc(\nu)$. Due to the $N$ times higher frequency of time-domain sampling in $\DD(\nu)$ compared to $\cc(\nu)$, we can derive an unambiguous velocity from $\DD(\nu)$ that is $N$ times greater. Hence, ICI can be exploited in two different ways: \textit{(i)} resolving Doppler ambiguity of high-speed targets, and \textit{(ii)} enhancing target resolvability by introducing an additional unambiguous Doppler dimension, allowing us to distinguish targets located in the same delay-Doppler-angle bin \cite{MIMO_OFDM_ICI_JSTSP_2021}. Fig.~\ref{fig_ici_exploitation} shows an example range-velocity scenario illustrating how ICI can be turned into a benefit for sensing. Analogous to ICI, the delay-dependency of the PN statistics in \eqref{eq_y_pn} can be exploited to resolve range ambiguity of far-away targets since the range information conveyed through the statistics of $\WW(\tau)$ is not subject to any ambiguity, unlike the one carried by $\bb(\tau)$, which has a range ambiguity of $c/(2\deltaf)$ \cite{PN_Exploitation_TSP_2023}. Fig.~\ref{fig_pn_exploitation} provides an illustrative example of PN exploitation for resolving range ambiguity of a distant target by a covariance matching approach \cite[Alg.~2]{PN_Exploitation_TSP_2023}. We note that employing this exploitation strategy elevates the performance of a radar with PN even beyond that of an ideal radar without PN, effectively repurposing challenges into benefits for sensing.

\subsection{Radio SLAM in monostatic scenarios} \label{sec:monostaticSLAM}

Classical monostatic \ac{SLAM} originated in the field of robotics, where a mobile user (e.g., a robot) continuously scans its surrounding environment using a laser device (lidar) or a camera. The primary goal is to detect specific features (\emph{landmarks}) of the scenario and simultaneously estimate the mutual positions of the user and the landmarks.
At first glance, this process may seem like a "chicken and egg" problem. 
Mapping, i.e., determining the position of landmarks, requires knowledge of the user's position. Localization, i.e., determining the user's position, requires knowledge of the map. However, this apparent dilemma can be generally resolved through appropriate algorithms. A crucial requirement for a monostatic \ac{SLAM} algorithm to function effectively is that the user must be in motion to collect measurements from different perspectives. It is even more advantageous if the relative movements of the user can be measured through odometry or inertial devices.
Surveys on the general  \ac{SLAM} problem can be found in \cite{CadEtAl:J16}, providing an overview of various techniques, including FastSLAM, GraphSLAM, and belief propagation \ac{SLAM}.

In the classical setup, the (unknown) state of the system at the discrete time instant $n$ consists of the state $\mathbf{x}_n$ of the mobile user, encompassing factors such as its position, orientation, and speed, and the position of the landmarks $\mathbf{m}$ representing the map. 
The objective is to determine the probability density function $p(\mathbf{x}_n , \mathbf{m} | \mathbf{y}_{1:n})$ (referred to as the \emph{belief}) and deduce both $\mathbf{x}_n$ and $\mathbf{m}$, where $\mathbf{y}_{1:n}$ denotes the set of measurements collected up to time instant $n$. To achieve real-time processing, Bayesian filtering approaches can be employed. In each iteration, $p(\mathbf{x}_n , \mathbf{m} | \mathbf{y}_{1:n})$ is computed, starting from the belief at time $n-1$ incorporating the new measurement $\mathbf{y}_n$ and utilizing the statistical mobility model of the user, characterized by the probability density function $p(\mathbf{x}_n|\mathbf{x}_{n-1})$ \cite{DarCloDju:J15}.

The incorporation of ISAC functionalities into smartphones opens up the possibility of utilizing them as monostatic SLAM devices, employing radio signals for scenario scanning (Radio SLAM). This allows for infrastructure-less localization and the automatic generation of digital maps, all while safeguarding user privacy and minimizing energy consumption compared to lidar-based solutions that demand perfect visibility and manual operation \cite{PasEtAl:J20}.
Early investigations into Radio SLAM on handheld devices were previously referred to as \emph{personal radar} and were introduced in \cite{GuiGueDar:J16}. In Radio SLAM, landmarks are identified through specular reflections of signals emitted by the mobile user, which is assumed to be equipped with an antenna array.

\begin{figure}
\centering
\includegraphics[width= 0.6\linewidth]{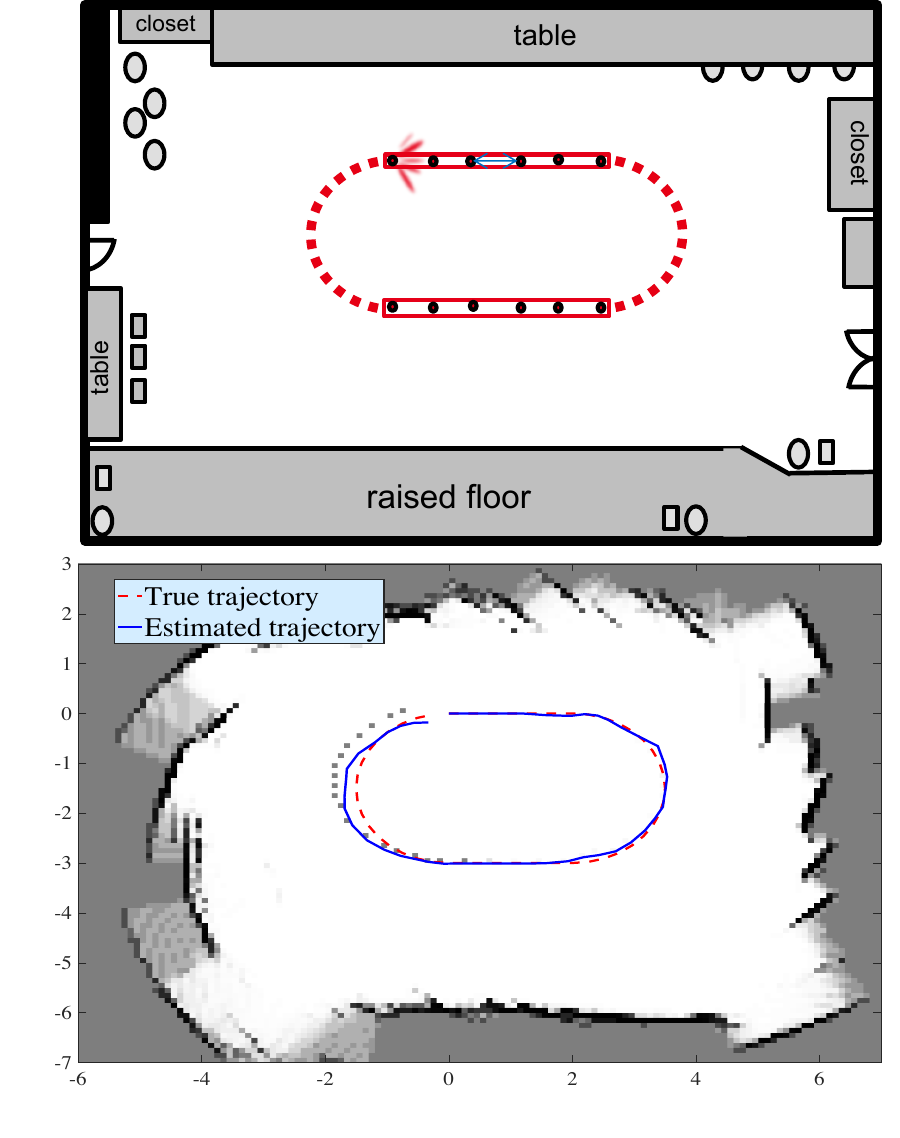}
\caption{Radio SLAM using measurements in the THz band. (top) Sketch of the measured indoor scenario; (bottom) Trajectory and map reconstruction using the MFM algorithm and an occupancy grid mapping method, respectively \cite{LotPasGueGuiDErDar:J23}.   
}
\label{Fig:R-SLAM}
\end{figure}

From the perspective of signal processing algorithms, two main approaches are typically followed: i) Classical SLAM algorithms that rely on the joint estimation of landmarks and user positions, as described earlier; ii) Algorithms based on relative pose estimation by comparing measurements (scenes) collected in the last two time instants, $\mathbf{y}_n$ and $\mathbf{y}_{n-1}$, respectively. In the latter approach, the goal is to estimate the position and orientation displacement of the mobile user, denoted as $\mathbf{d}_n=\mathbf{x}_n-\mathbf{x}_{n-1}$, at time $n$ with respect to time instant $n-1$.
A wide range of algorithms has been developed for this purpose, including the widely-used scan matching algorithm \cite{HeKoRaAn:C16} and approaches borrowed from image processing, such as those utilizing the Fourier-Mellin transform \cite{CheEtAl:J10, LotPasGueGuiDErDar:J23}.

Nevertheless, the presence of multipath, diffuse reflections, and the side lobes' impact from the antenna array can introduce artifacts in the backscattered received signal, leading to the emergence of "ghost" landmarks. This complication renders the application of algorithms in Radio SLAM significantly more challenging compared to lidar-based or vision-based SLAM. Furthermore, at frequencies up to millimeter waves, specular reflections tend to dominate over diffuse reflections. Consequently, in a monostatic setup, the signal emitted by the user terminal is only reflected back to the user if it impinges the obstacle almost perpendicularly to its surface. Otherwise, the obstacle might remain invisible, posing a critical challenge to the SLAM process.

In this context, the capability to operate in the THz band is expected to play a crucial role in 6G systems. The wide bandwidth available in this range will result in high spatial resolution, and the feasibility of large antenna arrays will enable unprecedented angular resolution \cite{SarEtAl:J20}. Moreover, at THz frequencies, the wavelength becomes comparable to the typical roughness of objects, potentially causing the diffuse component to dominate over the specular component. This aspect enhances object visibility even when the impinging angle of the signal is not perpendicular.
Experimental investigations of Radio SLAM at THz frequencies are in their early stages. Initial results are reported in \cite{LotPasGueGuiDErDar:J23}, where the performance of scan matching algorithms and a modified version of the Fourier-Mellin transform (MFM) is assessed using real-world THz radar measurements in an indoor environment. An example is provided in Fig. \ref{Fig:R-SLAM}, where the performance of the MFM algorithm in terms of mapping and trajectory reconstruction is shown. The results were obtained starting from measurements taken at 300 GHz (lower THz frequency band) in a typical large office scenario by moving the measurement set up along a circular shape (red curve) and scanning the scene with a span of $\pm 90^{\circ}$ with respect to the direction of movement and a beamwidth of $18^{\circ}$. The obtained localization \ac{RMSE} is $12\,$cm.  More details on the algorithms and the measurement campaign can be found in \cite{LotPasGueGuiDErDar:J23}. 

\section{Sensing and communicating with wide apertures}
\label{sec:near-field}

\begin{figure}
\centering
\includegraphics[width= 0.5\linewidth]{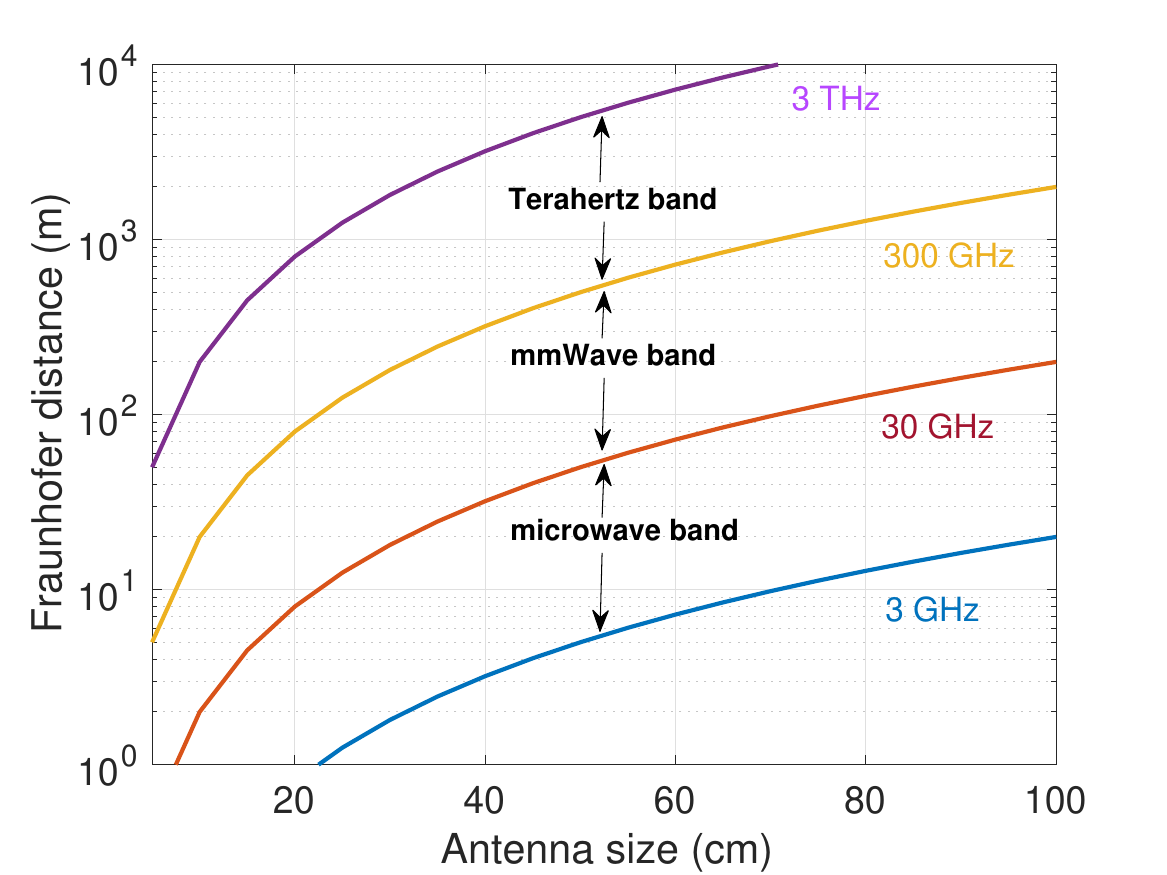}
\caption{Fraunhofer region boundaries as a function of the antenna size at different frequency bands. The area above each curve corresponds to the far-field region (Fraunhofer region), whereas the area below each curve corresponds to the near-field region. 
}
\label{Fig:Fraunhofer}
\end{figure}

\subsection{Introduction}

In contemporary wireless systems, antenna arrays are typically adopted to obtain beamforming and spatial multiplexing. The former involves directing the transmitted signal precisely toward its designated receiver in space, whereas the latter consists in simultaneously transmitting multiple data streams to the same user or users located at different positions. These are the core functionalities of (massive) \ac{MIMO} systems.  As a natural evolution of the massive \ac{MIMO} technology, \ac{XL-MIMO} further boosts the number of antennas by at least an order of magnitude, e.g., several hundreds or even thousands of antennas, thus obtaining \acp{ELAA} and unprecedentedly improving the spectral efficiency and spatial resolution for wireless communication and sensing \cite{Bjo:19,Wan:23,Bjo:23}. 
In the following, we will refer to an \ac{ELAA} as an antenna whose dimension $D$ is much larger than the wavelength $\lambda$ (electrically large antenna). 
Depending on the technology adopted and the way it is modeled, an \ac{ELAA} may be dubbed as Holographic \ac{MIMO} and \ac{LIS} \cite{Hua:20,Hu:18,Dar:J20}. Typically, these terms are applied when arrays consist of electrically small and densely packed elements.

\subsection{Near-field vs. far-field communication and sensing}

The exploitation of higher frequency bands, from millimeter waves to THz, foreseen in 6G networks and the availability of new antenna technologies for \acp{ELAA}   opens new opportunities for communication and sensing \cite{Hu:18,Dar:J20,Shl:21,Bjo:23,ElzGueGuiDarAlo:23}.    
In fact, traditional wireless systems typically work at distances beyond the Fraunhofer distance, defined as $d_{\text{F}}=2D^2/\lambda$ (\emph{far-field boundary}), where the \ac{EM} wavefront can be well approximated as being planar. Instead, using an \ac{ELAA}, the wireless links are likely to operate at distances below  $d_{\text{F}}$, corresponding to the radiating \emph{near-field region}, where such an approximation no longer holds and, consequently, the wavefront impinging on the antenna is spherical \cite{Kri:17}. 
While this aspect requires a revisitation of the classical channel models that in many cases are based on the far-field assumption and then may fail in the near field, at the same time, it provides the chance to improve the communication and sensing capabilities of the system \cite{DarDec:J21,Wan:23}. 
In Fig. \ref{Fig:Fraunhofer}, the Fraunhofer distance is plotted as a function of the antenna aperture size $D$ for different frequency bands ranging from microwave to THz. As it can be noticed, at millimeter waves and beyond, the radiating near-field region might correspond to practical operating distances of several meters or hundred meters even with relatively physically small (but electrically large) antennas.

Regarding communication, a spherical \ac{EM} waveform is more informative than a plane wave, as it will be discussed in Sec. \ref{Sec:ELAAComm}. 
For sensing/positioning, incident spherical wavefronts embed not only angular information, as in the far-field regime, but also distance information. This property can be exploited to determine the position of a transmitting source by analyzing the phase profile of the received signal along the antenna aperture, as it will be detailed later \cite{ElzGueGuiDarAlo:23}.

\begin{figure}
\centering
\includegraphics[width= 0.6\linewidth]{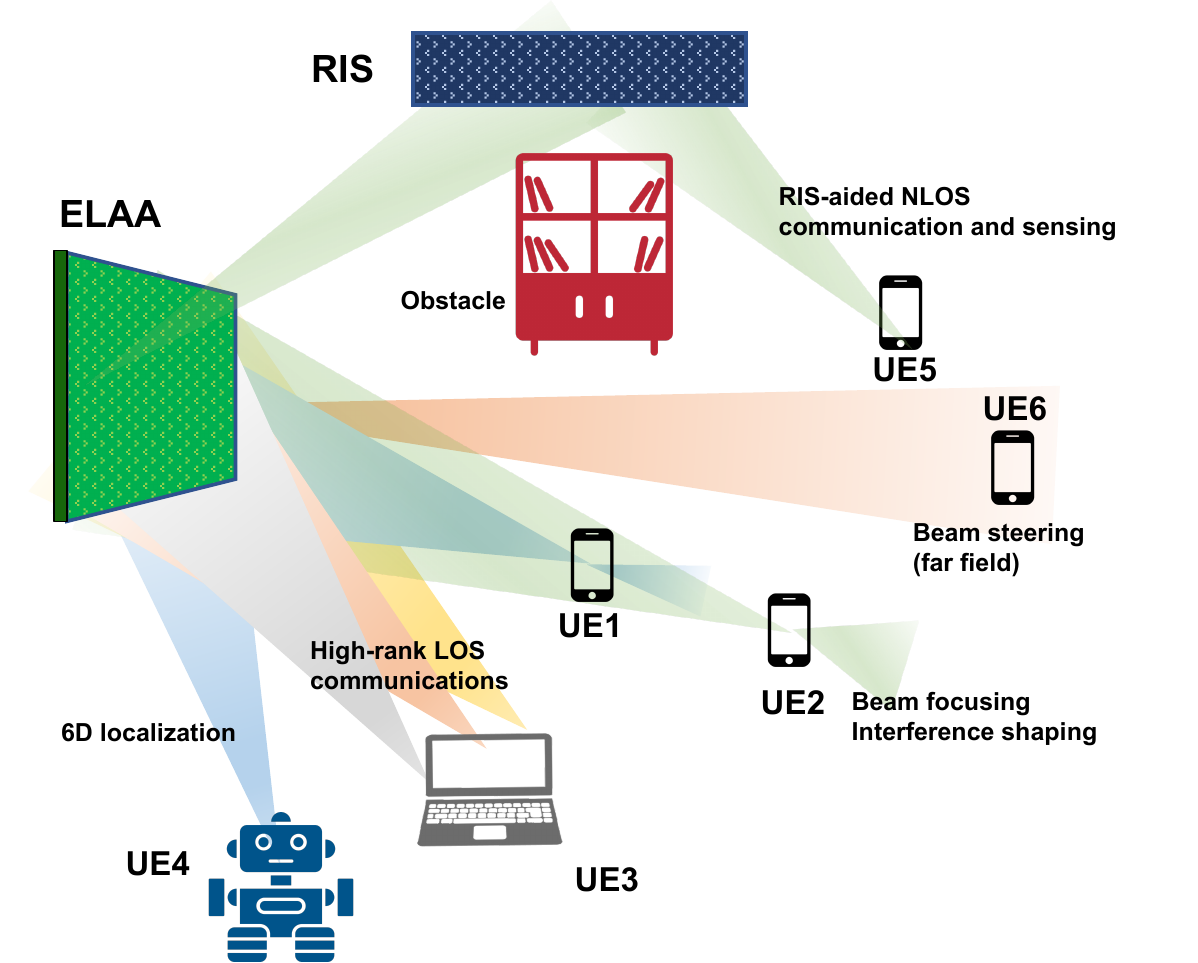}
\caption{Near-field communication and sensing: UE1 and UE2 can be discriminated even if they are at the same angle of view thanks to focusing (interference shaping); UE3 can establish a high-rank communication link (spatial multiplexing in \ac{LoS}); the position and orientation of UE4 can be estimated (single-antenna 6D localization); UE5 can be localized/sensed in NLOS thanks to the \ac{RIS}; UE6 is in the far-field region, then only rank 1 communication and beam steering can be realized in strong \ac{LoS} conditions.
}
\label{Fig:ScenarioELAA}
\end{figure}

\subsection{Communication and sensing with \acp{ELAA}}
\label{Sec:ELAAComm}

The adoption of an \ac{ELAA} provides improved flexibility in forming the beam and sensing the \ac{EM} wave (see Fig. \ref{Fig:ScenarioELAA}).
While in the far field, only differently steered beams can formed, thus providing user discrimination only at the angular level, in the near-field region the beam can be focused on a specific location similarly to an optical lens \cite{Nep:17}. It not only allows for the control of multi-user interference (interference shaping) in terms of angular direction, similar to traditional beam steering but also provides control over interference in terms of distance thus allowing the discrimination of users seen with the same angle of view  \cite{ZhaShlGuiDarEld:J23}.  

\begin{figure}
\centering
\includegraphics[width= 0.4\linewidth]{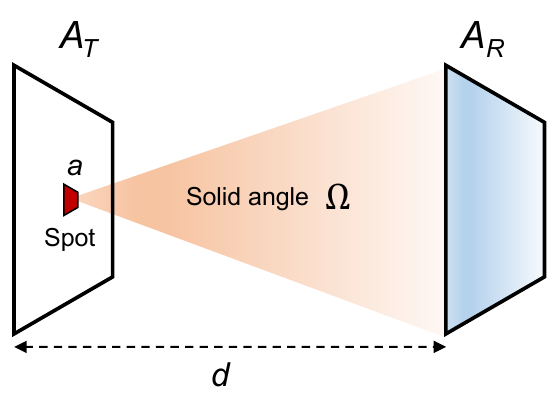}
\caption{Diffraction effect at distance $d$ of a small spot of area $a$ composing the transmitting surface.
}
\label{Fig:DOF}
\end{figure}

The near-field effect can be exploited to combat the multiplexing gain degradation caused by sparse multipath channels in \ac{MIMO} links when operating in strong \ac{LoS} conditions at high frequencies. 
In principle, it is possible to establish orthogonal channels (\emph{communication modes}) by generating non-overlapping beams, each one focusing on different properly chosen locations of the receiving antenna (see UE3 in Fig. \ref{Fig:ScenarioELAA}). 
A rough estimate of the number of communication modes, i.e., communication \ac{DoF}, can be calculated starting from simple arguments of diffraction theory \cite{Miller:00}.  Consider two antennas, modeled as continuous surfaces for convenience, in parallax configuration with areas $A_{\text{T}}$ and $A_{\text{R}}$, respectively, at distance $d$, as shown in Fig. \ref{Fig:DOF}.  The smallest spot of area $a$ we can use at the transmitting antenna will be the one from which the diffracted \ac{EM} field approximately fills the aperture of the receiving antenna. In particular, the diffraction solid angle from the spot is $\Omega \approx \frac{\lambda^2}{a}$, where it must be $\Omega \, d^2 \approx A_{\text{R}}$ (full illumination). The number of distinct spots on $A_{\text{R}}$, i.e., communication modes, is
\begin{equation}
\mathsf{DoF} \approx \frac{A_{\text{R}}}{a}=\frac{A_{\text{T}}  A_{\text{R}}}{d^2 \lambda^2} \, .
\end{equation}
The previous result confirms that with \ac{ELAA}, i.e., $A_{\text{T}}/\lambda^2 \gg 1$, it is possible to obtain high-rank communications even in strong \ac{LoS} conditions.  
A more accurate theoretical bound on \ac{DoF} valid also for asymptotically large \ac{ELAA} modeled as a \ac{LIS} can be found in  
\cite{Dar:J20}. Achieving capacity-approaching \ac{MIMO} communications in the near field results in intricate array phase profiles. As indicated in \cite{DecDar:J21}, these profiles can be effectively approximated by using a combination of simpler multiple-focused beams whose configuration depends on the geometry.

Regarding the localization task, as anticipated it is possible to localize a source or a target using only one \ac{ELAA} if it is in the antenna's near-field region through the analysis of the phase profile of the impinging \ac{EM} wave \cite{GueGuiDar:J18,GuiDar:J21}.

\begin{example}
    Consider a simple 2D scenario where a receiving linear \ac{ELAA} located 
in the origin is  composed of $N$ elements deployed along the $x$-axis spaced apart of $\Delta=\lambda/2$ with coordinates $\mathbf{r}_n=[(n-1) \, \Delta,0]$, for $n=1,2, \ldots , N$.  The size of the ELAA is $D=N\, \Delta$, with $N\gg 1$. A punctiform monochromatic RF source located at position $\mathbf{p}=[p_x,p_y]$ in the near field of the \ac{ELAA} is present (e.g., an active node to be localized or a reflecting target in case of sensing). Denote with $d=\| \mathbf{p}- \mathbf{r}_1 \|$ the distance between the first element of the \ac{ELAA}  and the source, and by $\theta$ the steering angle so that $\sin (\theta)=p_x/d$ and $\mathbf{p}=[d \sin(\theta), d \cos(\theta)]$.
The complex baseband channel between the source and the \ac{ELAA} can be modeled as
\begin{equation} \label{eq:hn}
h_n=\beta_n e^{-\jmath \frac{2\pi}{\lambda} d_n} 
\end{equation}
where $\beta_n$ represents the channel amplitude and 
\begin{equation} \label{eq:dn1}
d_n=\|  \mathbf{p}- \mathbf{r}_n \|=d \sqrt{1+\frac{(n-1)^2 \lambda^2}{4\, d^2} - \frac{2\, p_x\, (n-1) \lambda}{ 2\, d^2}} 
\end{equation}
is the Euclidean distance between the source and the $n$th receiving antenna element. 
if $D<d$, then $\beta_n \simeq \beta_1$ $\forall n$ and we can use in \eqref{eq:dn1} the following Taylor expansion $\sqrt{1+x}=1+\frac{x}{2}- \frac{x^2}{8}+o(x^2)$ leading to
\begin{equation} \label{eq:dn}
d_n \simeq  d - \frac{(n-1) \lambda\, \sin (\theta)}{ 2} +\frac{(n-1)^2 \lambda^2}{8\, d} \, .
\end{equation}
The first term of \eqref{eq:dn} contributes in \eqref{eq:hn} with a constant phase shift which is not informative if $d>\lambda$ because of the $2\pi$ phase ambiguity and the need for the source to be perfectly synchronized with the antenna. The second and third terms indicate a parabolic behavior of the phase profile observed along the array which is a function of the steering angle and the distance, i.e., the position $\mathbf{p}$ of the source, which can be therefore estimated.  
If  $d>d_{\text{F}}$ (far-field condition), the third term contributes with a phase shift smaller than $\pi/8$ at the edge of the antenna and hence becomes negligible. As a consequence, $d_n \simeq  d - (n-1) \lambda\, \sin (\theta)/2$ and only the angle of arrival $\theta$ can be estimated. 
\end{example}

The case where the source employs an antenna array opens the door to 6D positioning, i.e., the estimation of the source's position and orientation if located in the near-field region of the \ac{ELAA}. (see UE4 in Fig. \ref{Fig:ScenarioELAA}).
The theoretical performance bound of 6D positioning has been derived in \cite{GueGuiDar:J18} for generic shapes of the antenna arrays and wideband signals. 
 
It has to be remarked that, single-antenna positioning is possible also in the far field but that would require, in addition to the signal's \ac{AoA} measurement,  the exchange of wideband signals to estimate the distance between the source and the antenna through \ac{ToA} measurements and a ranging protocol to cope with the lack of a common clock.  On the contrary, with an  \ac{ELAA}, localization, and sensing can be obtained with a narrowband signal and lower latency because no ad-hoc synchronization procedures and time-base ranging protocols are needed. As a consequence, more resources become available for communication which allows for more efficient joint communication and sensing schemes.

A deep investigation of near-field integrated sensing and communication is still missing in the literature. In this direction,  \cite{Wan:23_nf} derives the minimization of the Cramér-Rao bound for the near-field joint distance and angle sensing subject to the minimum communication rate requirement of each user. In the same work, both fully digital antennas and hybrid digital and analog antennas are investigated. The authors in \cite{dehkordi2023multistatic} consider a multi-static \ac{OFDM}-based \ac{ISAC} scenario by proposing a low-complexity two-stage estimation process where first a rough estimate of the target is obtained assuming a far-field condition, then refined accounting for the correct near field model. 


The primary challenges in modeling and designing joint communication and sensing functionalities using \acp{ELAA} can be summarized as follows:
i) \emph{Near-field channel modeling}: The extremely large array aperture introduces spatial non-wide sense stationary properties. Different regions of the array observe the propagation environment from varying perspectives, with diverse polarizations. This implies that regions may perceive signals transmitted along a specific propagation path with differing powers and polarizations, or signals from distinct propagation paths \cite{Yua:23}.
ii) \emph{Channel estimation}: Achieving optimal transmission demands highly accurate channel estimation. Unfortunately, the vast number of elements constituting the \ac{ELAA} generally entails estimating a considerable number of channel parameters, particularly in challenging propagation environments, leading to increased signaling overhead. In traditional \ac{MIMO} systems, exploiting the sparsity of the angular-domain channel (e.g., through compressed sensing) simplifies this task. However, near-field propagation exhibits sparsity in the location domain due to different electromagnetic characteristics. Consequently, existing angular-domain-based algorithms are not directly applicable. For instance, in \cite{Cui:22}, the authors proposed a polar-domain representation of the channel that comprehensively captures near-field spherical wave characteristics. Differently, in cases of strong \ac{LoS}, the channel is closely linked to geometry, specifically the source's position. In principle, only three parameters are sufficient for estimating the channel. Consequently, channel estimation aligns with the localization task, further intertwining sensing and communication.
iii) \emph{\ac{ELAA} Modeling}: Depending on the technology adopted, if the \ac{ELAA} comprises dense elements, mutual coupling effects between them cannot be overlooked and must be accurately modeled \cite{Wan:23}.

\subsection{Distributed MIMO/cell-free massive MIMO for multi-perspective localization and sensing}

\begin{figure}
    \centering
    \includegraphics[width=0.6\columnwidth]{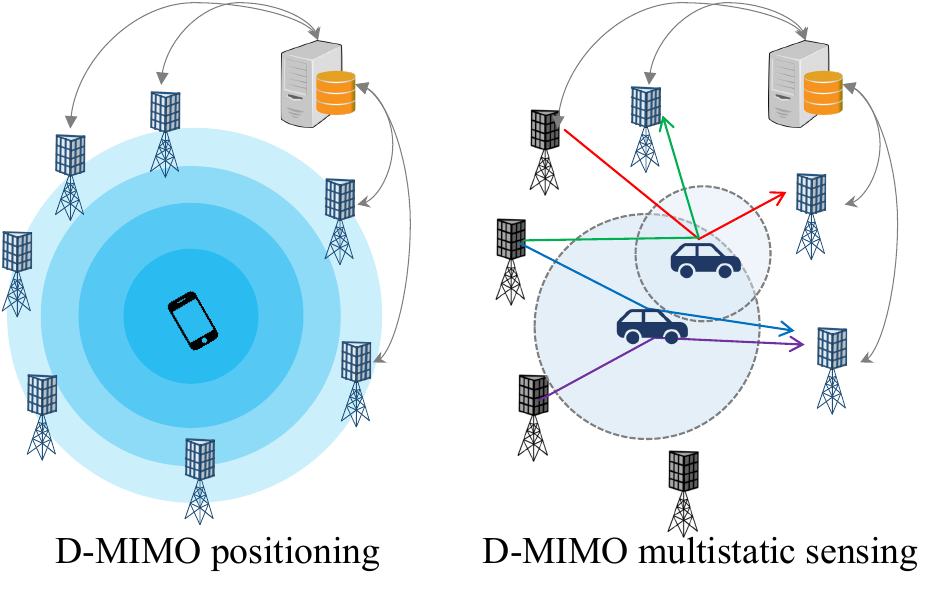}
    \caption{D-MIMO for positioning (left) and multistatic sensing (right). Phase-coherent processing leads to wavefront curvature, considering all \acp{RU} as part of a distributed array. Black \acp{RU} are transmitters, while blue \acp{RU} are receivers. }
    \label{fig:DMIMO}
\end{figure}

A \ac{D-MIMO} system can be viewed as a massive MIMO system, where a large number of antennas are deployed over a wide area with irregular inter-element spacing (see Fig.~\ref{fig:DMIMO}) \cite{demir2021foundations}. Each antenna (or small array) is called a \ac{RU}, several of which may be connected to a \ac{DU}, which takes care of most of the baseband processing \cite{cf_mimo_measurement_2022}. The \acp{DU} are connected to \ac{CU}, which manages higher layers and coordinates the \acp{DU}. Various \ac{D-MIMO} architectures correspond to different node topologies, different levels of local versus central processing, and different levels of synchronization \cite{seqRS_TCOM_2021}. The operation is otherwise similar to classical massive MIMO, involving uplink pilot transmission, downlink multi-user beamforming, and uplink data transmission \cite{cf_survey_2022}. 


Typical \ac{D-MIMO} assumes phases synchronization among the \acp{RU}, enabling spatial focusing towards \acp{UE} \cite{demir2021foundations,phaseCalib_2023}.
While for communication purposes, the locations of \acp{RU} are irrelevant, and phase stability during each coherence interval suffices, more stringent requirements exist for localization and sensing functions. In these cases, considering all \ac{D-MIMO} \acp{RU} as a large array, the overall array steering vector needs to be known, which requires not only phase stability, but also precisely known locations of the \acp{RU} (within a fraction of the wavelength) \cite{fascista2023uplink,nearfieldSense_TWC_2022,chen20236g}. Combined, this places \acp{UE} and objects in the near-field of a massive distributed array, while remaining in the far-field of individual elements \cite{fascista2023uplink}. Meeting these stringent requirements is likely only feasible at FR1 \cite[Sec.~6.2.1]{hexax_d22}, \cite{6g_testbeds_2023}, whereas at higher frequencies (FR2 and above), classical time-coherent processing can be performed. At FR1, the operation of \ac{D-MIMO} bears resemblance to carrier phase positioning (see Section \ref{sec:CPP}), with multistatic sensing (see Section \ref{sec:bistatic-and-multistatic}) and with \ac{ELAA} (see Section \ref{Sec:ELAAComm}). The main differences lie in the large, dense, and distributed architecture of \ac{D-MIMO}.


The main benefits of \ac{D-MIMO} include more uniform coverage for communication, localization, and sensing \cite{demir2021foundations,chen20236g}. In addition, the large aperture provides enhanced multipath resolution and improved accuracy, though at the expense of high computational complexity \cite{manzoni2023wavefield}. 





\begin{figure}[t]
\centering
\includegraphics[width=0.6\columnwidth]{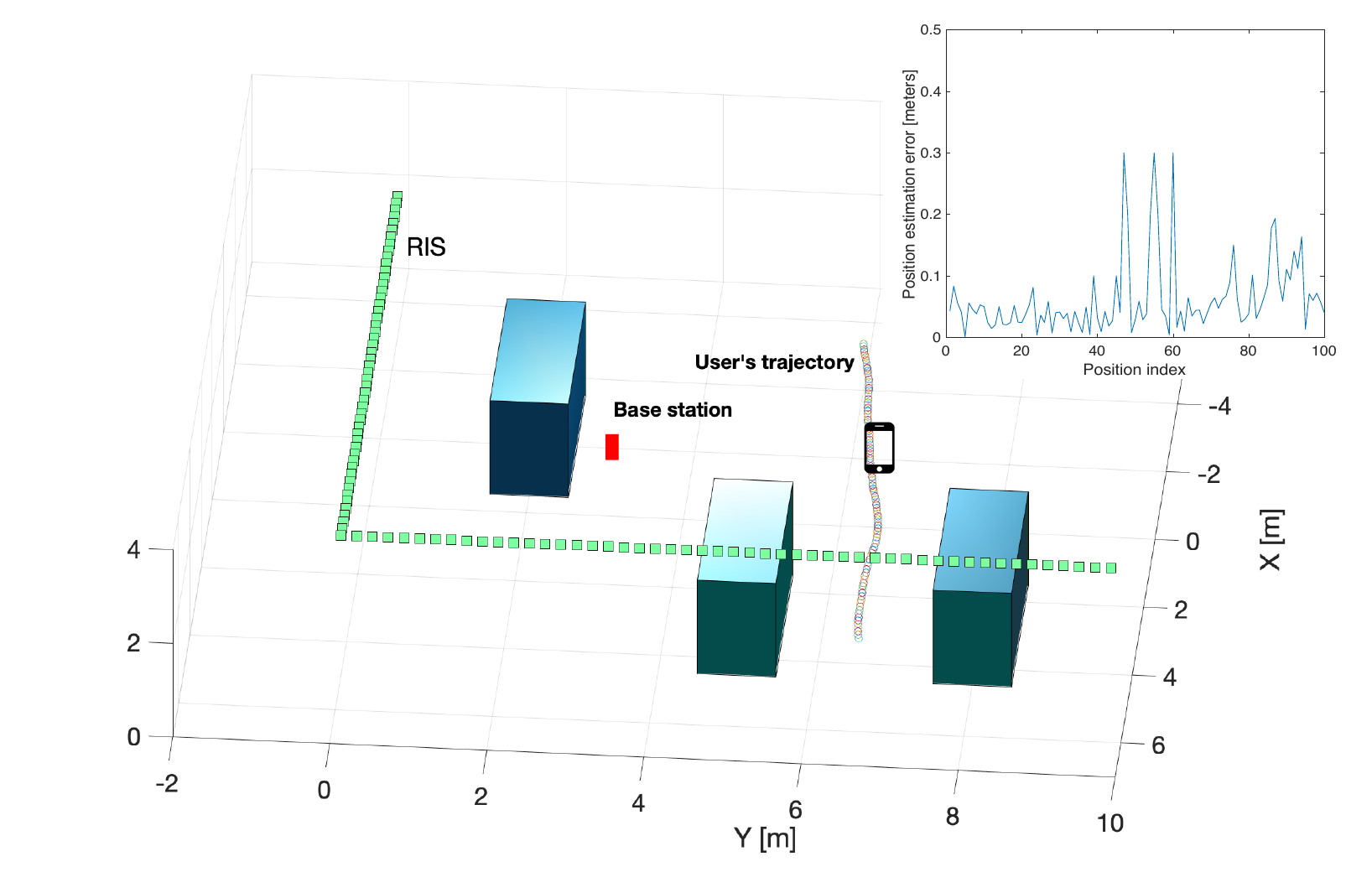}
\caption{RIS-assisted localization in a near-field \ac{NLoS} scenario. $28\,$GHz \ac{OFDM} system with $250\,$MHz bandwidth.  
The error evolution along the trajectory is reported \cite{DarDecGueGui:J22}.}
\label{Fig:LocRIS}
\end{figure}

\subsection{Sensing and localization aided by large RIS}
\label{sec:JCAS-large-RIS}

 The use of \ac{RIS} for localization was discussed in Section \ref{sec:RIS}, where the \ac{UE} was in the far field of the \ac{RIS}, providing additional delay and angle information. When the \ac{RIS} becomes large and the \ac{UE} is in the near-field region of the \ac{RIS}, the model \eqref{eq:channel-RIS-detail} changes to \cite{Shaban2021,DarDecGueGui:J22,rinchi2022compressive,NF_RIS_cuneyd_2023}
\begin{align}
    \alpha_m^{\text{RIS}} = \alpha^{\text{T-RIS}}\alpha^{\text{RIS-R}} \mathbf{a}^T_{\text{RIS}}(\mathbf{p})\boldsymbol{\Omega}_m \mathbf{a}_{\text{RIS}}(\boldsymbol{\varphi}). \label{eq:channel-RIS-detail-near-field}
\end{align}
This means that a single-antenna \ac{UE}, observing ${y}_m=\alpha_m^{\text{RIS}} x_m + z_m$ over a single subcarrier across different times $m$, can in principle estimate its 3D location $\mathbf{p}$, even when the \ac{LoS} path between the \ac{UE} and \ac{BS} is blocked \cite{ris_nf_cont_2021,rinchi2022compressive,NF_RIS_cuneyd_2023,ris_opt_2022}, or when some of the \ac{RIS} elements are failing \cite{pixel_RIS_2023}. In \eqref{eq:channel-RIS-detail-near-field}, $\mathbf{a}_{\text{RIS}}(\mathbf{p})$ represents the near-field response vector as a function of the \ac{UE} position. Fig. \ref{Fig:LocRIS} illustrates a practical example of \ac{NLoS} localization facilitated by \ac{RIS}. In this scenario, a single-antenna \ac{UE} navigates within an indoor setting in the presence of several obstacles, with a \ac{BS} emitting a reference \ac{OFDM} signal. A long linear \ac{RIS} is deployed (green patches in the figure), allowing the \ac{UE} to determine its location by analyzing only the signals reflected by the \ac{RIS}.
In this specific setup, as proposed in \cite{DarDecGueGui:J22}, the reflection coefficient of each element in the \ac{RIS} matrix $\boldsymbol{\Omega}_m$ dynamically varies during the transmission of the reference signal. This variation follows a predefined pattern designed to enable the \ac{UE} to estimate the delay of the signal component reflected by each visible element of the \ac{RIS}. Utilizing these \ac{BS}-\ac{RIS}-\ac{UE} delay measurements, the \ac{UE} can then compute its own position.
The inset plot demonstrates that, despite numerous obstacles partially shadowing the \ac{RIS}, the localization error along the trajectory of the \ac{UE} remains confined to $20-30$ cm. Additionally, a narrowband version of the algorithm, leveraging the phase profile of the received signal, is proposed in \cite{DarDecGueGui:J22}. 
For sensing, the model \eqref{eq:channel-RIS-detail-near-field} can be used, for example, in a monostatic setup with a single-antenna \ac{BS} monitoring an area and making use of the high spatial resolution of the large \ac{RIS} \cite{foundations_RIS_radar_TSP_2022,grossi_radar_2023,Buzzi_Radar_2021}. By generating \ac{RIS} configurations $\boldsymbol{\Omega}_m $ that scan an area, targets can be detected and localized, even with limited bandwidth, from the observations at the BS, harnessing the paths from BS to RIS to target, back to RIS and back to the BS.
 The near-field imaging problem utilizing \ac{XL-MIMO} antennas and \ac{RIS}  has been investigated in \cite{torcolacci2023holographic} where the design of the optimal illumination waveform and \ac{RIS} configuration is addressed.   


\noindent

\section{Technologies for sensing assisted communication}\label{sec:sensing-assisted-comm}

MIMO communication operating at mmWave with large arrays and bandwidths provides the angular and delay resolution required for high accuracy localization and sensing, as we have discussed  in previous sections. Resilient communication becomes more challenging, however, as cellular networks advance to higher carrier frequencies. On the other hand, fast adaptation of communication strategies is more difficult due to large MIMO arrays, higher bandwidths and challenging circuit designs. For example, initial access to configure the mmWave beams in 5G \cite{Giordani2019CST} can be up to 5s with a simple analog beamforming architecture, with more time expected at sub-THz frequencies \cite{Polese2020ComMag}.  On the other hand, there are fewer opportunities for high-rate communication due to the larger performance differential between LOS and NLOS links, more frequent blockage (smaller first-order Fresnel zone), higher penetration losses, increased scattering and less reflection \cite{mmWavetutorial2016}. For example, commercial 5G mmWave throughput measurements with a blocked LOS path show a $2\times$ throughput reduction when reconfiguring the beamforming to use an available NLOS path and a $4\times$  throughput reduction when the device switches to a lower frequency in 4G because no NLOS path is available \cite{Narayanan2020ACM}. 

Sensing-assisted network operation can mitigate the impact of  these issues. For example, ISAC-maps can incorporate localization information about the environment and about the success of past communication configurations to design position dependent beam codebooks with a reduced number of beams, as explained in Section~\ref{sec:intro-sensing-assisted-communication} and illustrated in Fig.~\ref{fig:ISLACmap-uses-cases}. In addition, Fig.~\ref{fig:ISLACmap-uses-cases} shows how ISAC maps can also provide the position and velocity of potential blockers of the communication signal, so the network can proactively react to mitigate the impact of blockage.   
In this section, we describe specific approaches for adapting communication operation based on the exploitation of the ISAC map. In particular, we address the problems of sensing-aided array configuration and blockage prediction and management.

\subsection{Sensor-aided array configuration}

The earliest prior work on enhancing network operation in an ISAC setting considers the problem of mmWave MIMO beam training for initial access aided by some type of sensing information and past communication. 
The problem is to find the transmit beam $\ff$ in a codebook $\cal{F}$ and the receive beam $\ww$ in a codebook  $\cal{W}$ to maximize a given performance metric, for example signal-to-noise ratio (SNR). The standard approaches are a bruteforce search over pairs \cite{Hur2013TWC} or a hierarchical search \cite{Wang2009Globecom} over increasingly refined codebooks. In both cases, search time generally grows with antenna size but it can be reduced by searching (intelligently designed) smaller codebooks. Achieving this reduction is especially important in vehicular settings, where the channel is highly dynamic and frequent antenna array reconfiguration is required. In the next paragraphs we will show examples of how sensing and learning can be used together to make communication more efficient.

\subsubsection{Position-aided beam training}

Initial work exploited position information obtained with a GPS \cite{Garcia2016SPAWC} or a radar mounted at the BS  \cite{AliRadarConf2019} to reduce the size of the beam codebook. The basic idea consists of using the estimated direction of the user $\phi_\text{UE}$ suggested by its position to define a reduced set of beams to be tested    \cite{AliRadarConf2019}. This reduced beam codebook $\cal{W}_{\cal{L}}$ can be computed to account for the estimation error in the  angular direction of the user, denoted as $\Delta\phi_\text{UE}$. This way, the reduced codebook will include the beams from an initial grid indexed by the index set $\cal{L}$  such that $n \in \cal{L}$ if
\begin{align}
\sin(\hat\phi-\Delta\phi)+1 \leq \frac{2n}{\Ntx} \leq \sin(\hat\phi+\Delta\phi)+1+2/\Ntx.
\label{eq:reduction}
\end{align}
With this type of strategy, it is possible to achieve an overhead reduction of $(1-|\cal{L}|/\Ntx)$  \cite{AliRadarConf2019}. Newer strategies that exploit location information obtained via radio positioning have also been proposed \cite{Lu2020Access}. However, all these solutions are only feasible in LOS settings. 

More elaborated approaches exploiting position and also ML can achieve a significant overhead reduction in both LOS and NLOS channels. For example, inverse fingerprinting learns a subset of location-dependent beam-pairs based on past measurements in similar locations, such that with high probability at least one of the vectors in the subset works well \cite{Va2018TVT,Va2019Access,Satyanarayana2019TVT}. The recommendation algorithm is trained based on past measurements of the strength of different beam pairs as a function of location, made under different snapshots of the environment \cite{Va2018TVT}. The recommendations may be refined using online learning, by further exploring the angular space to achieve better beam pointing \cite{Va2019Access}. Additional information may also be used, like the traffic density, to make more accurate predictions \cite{Satyanarayana2019TVT}.

\begin{figure}[htbp]
\begin{center}
			 \includegraphics[width=0.5\textwidth]{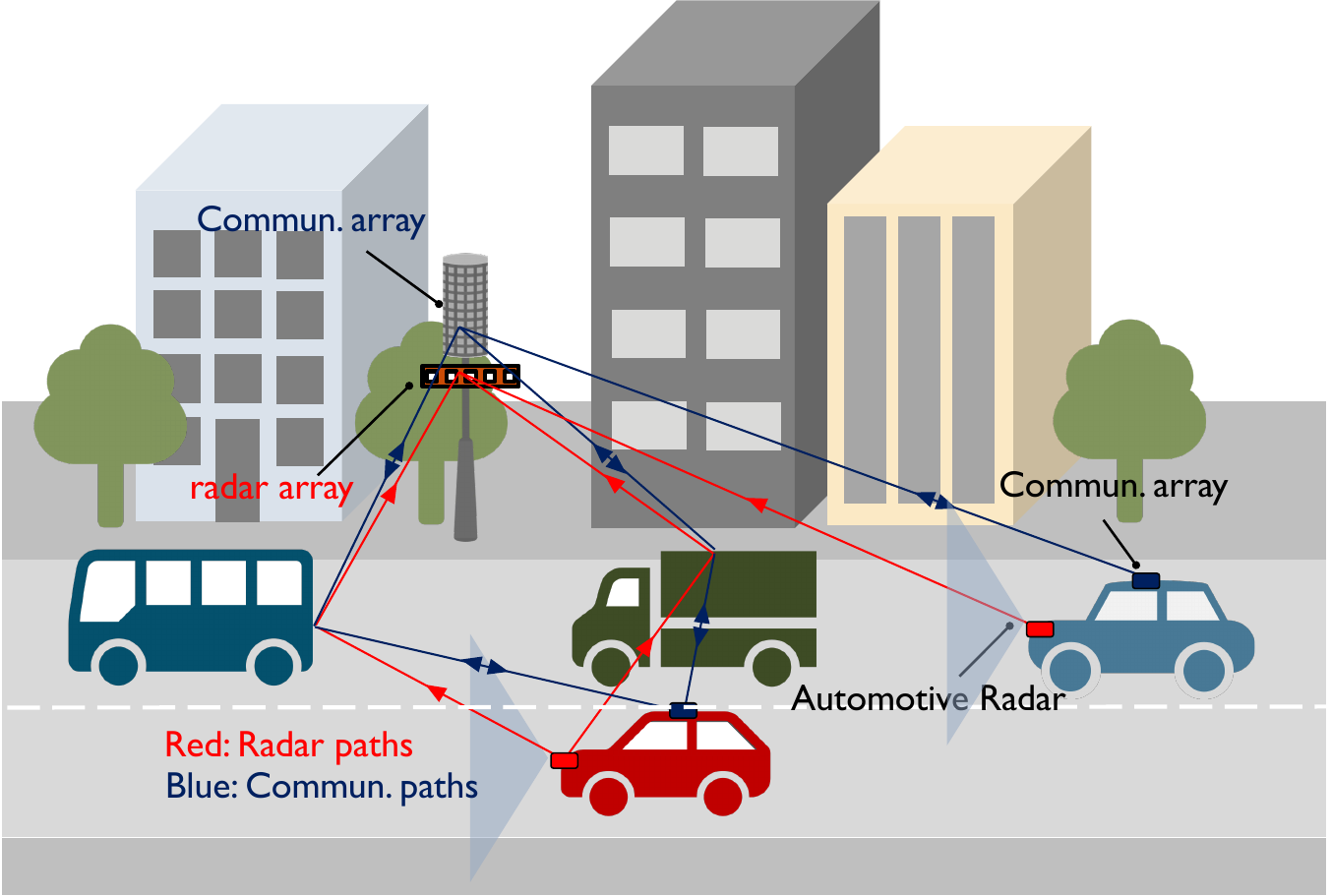}
\caption{Illustration of a vehicular communication system with radar-aided beam training. There is similarity between the radar and communication main channel directions, which can be exploited to reduce beam training overhead in NLOS and multiuser scenarios.}
\label{fig:radar-aided-beamtraining}
\end{center}
\end{figure}

\subsubsection{Radar-aided beam configuration}

For NLOS settings and a hybrid architecture, it is possible to design the precoders and combiners from the full instantaneous channel state information (CSI) or from the  spatial channel covariance information (also known as statistical CSI, partial CSI, or imperfect CSI). However, estimating either the channel or the spatial covariance introduces a significant overhead \cite{Park2020TWC,Venugopal2017,Javier2018TWC}. The spatial similarity between the radar and communication channels have been studied in \cite{Gonzalez-Prelcic2016ITA,Graff2019,Ali2020TVT} via ray tracing simulation and also experimental measurements. 
These studies show that the similarity in the main channel directions---in other words similarity in spatial covariance---between radar and communication channels operating at close but different mmWave frequencies is high, so that the radar covariance can be used as a prior for the communication covariance.

The first work on radar-aided spatial mmWave link configuration proposes the estimation of the spatial covariance in a V2I link with an active radar mounted at the BS \cite{Gonzalez-Prelcic2016ITA}. An alternative design in \cite{Ali2020TVT} considers a passive radar receiver at the BS, which is listening to the automotive radar signals coming from the different vehicles on the road, as illustrated in Fig~\ref{fig:radar-aided-beamtraining}. Some of the vehicles are already connected to the BS, while others need to join the network. Even in the absence of the chirp reference signal used by the automotive radar in the vehicles, it is possible to estimate the radar covariance at the BS radar receiver.  Using this information, \cite{Ali2020TVT} shows that communication overheads can be reduced by 77\% in a realistic vehicular environment simulated by ray tracing. Though the results are promising, there is still a mismatch between radar and communication channels due to using different frequencies or different locations of radar modules and transceivers in the vehicle, as illustrated in Fig~\ref{fig:radar-aided-beamtraining}. Some subsequent work \cite{Chen2021Globecom,Graff2023TVT} introduces multiple users into the environment and designs a deep learning strategy that translates the radar covariance into a communication covariance to compensate for these mismatches, achieving further overhead reductions which result in higher effective rates as shown in Example~\ref{ex:radar-aided-beamtraining}. 

\begin{table}[h!]
\begin{center}
\caption{Simulation of a radar-aided mmWave vehicular  communication system: parameters for the communication signals and the automotive radar signals. \label{tab:radar-aided-beamtraining}}{%
 \begin{tabular}{|p{1.8in}|p{0.3in}|p{0.3in}|p{0.3in}|}
 \hline \hline
 {\bf Parameter} & {\bf Symbol} & {\bf Value} & {\bf Units} \\ \hline \hline 
\multicolumn{4}{|c|}{ Communication system} \\ \hline 
Transmit power & $P_c$ & $30$ & dBm \\
Carrier frequency & $\fc$ & $73$  & GHz \\
Bandwidth & $B$ & $1$ & GHz \\
BS height &  & $5$ & m \\
Vertical separation of arrays at the BS &  & $10$ & cm \\ 
Distance BS to closest point on the road & $d$ & $10$ & m \\
Number of antennas at the BS & & 128 & \\
Number of RF chains at the BS &  & 1 & \\
Number of antennas at the vehicle &  & 16 & \\
Number of arrays at the vehicle & & 4 & \\
Number of RF chains at the vehicle &  & 1 & \\
Height of the communication arrays at the vehicle & & 1.6 & m \\
Number of phase shifter bits & $D$ &   2 & bits \\
Number of subcarriers & $N$ & 2048 & \\
Subcarrier spacing & $\deltaf$ & $240$ & kHz \\
Cyclic prefix length & $L_\text{c}$ & 511 & samples  \\ \hline
\multicolumn{4}{|c|}{Radar system}  \\ \hline
Center frequency & $f_\text{r}$ & $76$  & GHz \\
Bandwidth & $B_\text{r}$ & $1$ &  GHz \\
Transmit power &  & $30$ & dBm \\
Number of antennas & & 128 & \\
Chirp period &  & 500 & $\mu$s \\
Samples per chirp &  & 1024 & samples \\
Height of the vehicle radars & & 0.75 & m\\ \hline
\end{tabular}}{}
\end{center}
\end{table}

\begin{example}[Deep learning based radar assisted link configuration]\label{ex:radar-aided-beamtraining}
We consider a radar-aided communication system as the one illustrated in ~\ref{fig:radar-aided-beamtraining}. The parameters of the radar and communication signals are specified in Table~\ref{tab:radar-aided-beamtraining}. The system is simulated with a number of 4 users on the road. Additional details of the urban environment simulated by ray tracing can be found in \cite{Ali2020TVT,Graff2023TVT}. The considered performance metric is the sum rate, defined as $R_{\Sigma} = \sum_{u} R_u$, with the effective rate per user $R_u$
\begin{equation}
    R_u = (1-\frac{T_{\text{train}}}{T_{\text{coh}}}) \deltaf s_u,
\end{equation}
with $\deltaf$ the subcarrier spacing and $s_u$ the spectral efficiency for the $u$-th link.
\begin{equation}
    s_u = \sum_{k=1}^{N} \log_2 \left( 1 + \text{SINR}_{u}[k] \right),
\end{equation}
with $N$ the number of subcarriers. The initial estimation of the radar covariance matrix at the radar arrays of the BS is performed using the mutiuser separation and covariance estimation strategies designed in \cite{Ali2020TVT,Graff2023TVT}. 
A fully connected network  trained with a set of 9600 samples and evaluated with a test set of 2400 samples is used to map the radar covariance to the communication covariance. Radar aided beam training is implemented using a reduced codebook composed of 4 beams around the directions suggested by radar.
Fig.~\ref{fig:radar-aided-beamtraining} shows the sum-rate for exhaustive search, radar covariance-aided beam training without compensating for mismatches, and communication covariance-aided beam training after ML-based mapping of the radar covariance to the communication covariance. The performance gain of radar-aided beam training strategies varies with the coherence time, but it is significant in all cases. The benefit of ML-based mismatch prediction is higher at smaller coherence times.
\begin{figure}[t!]
\begin{center}
			 \includegraphics[width=0.45\textwidth]{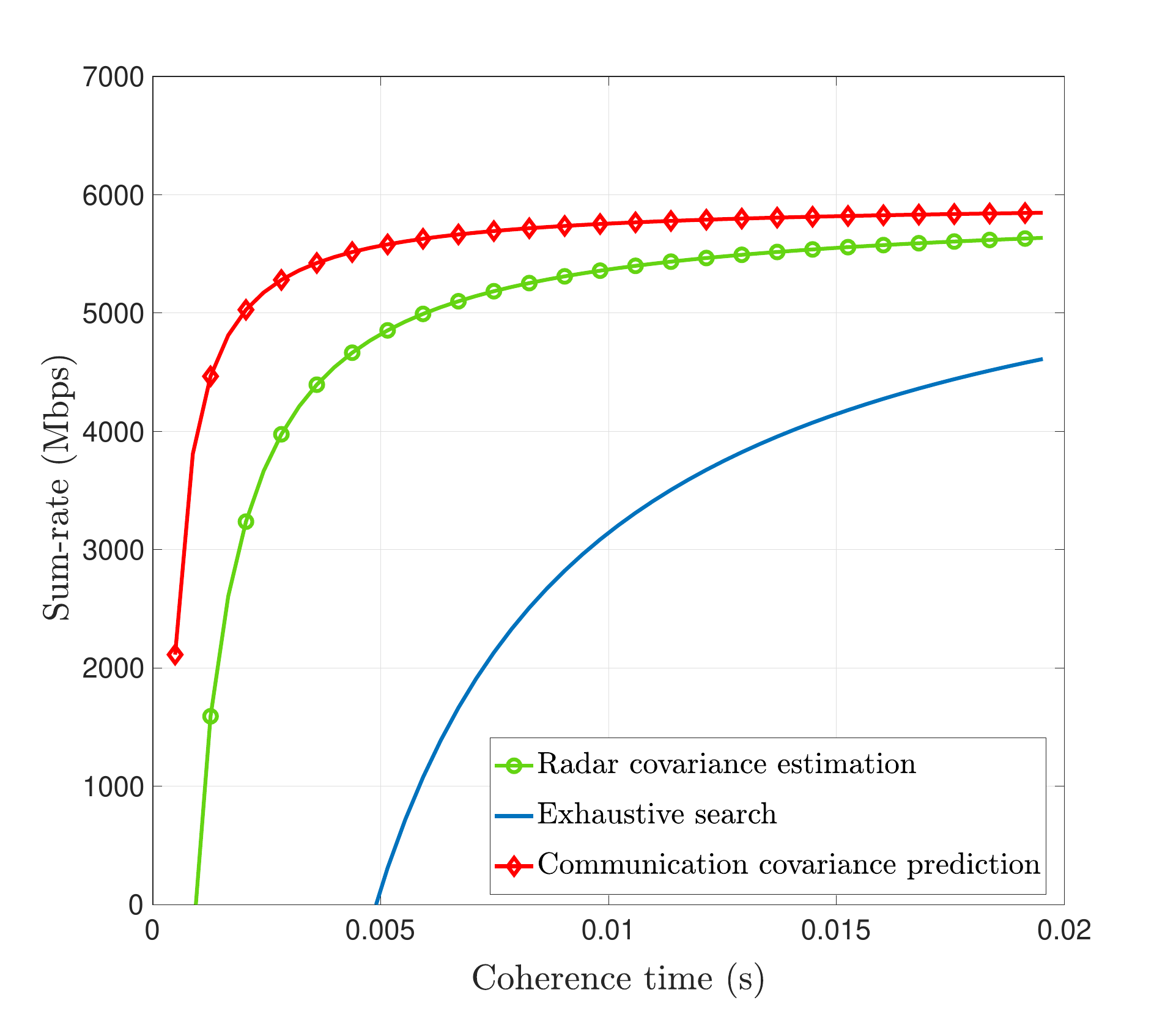}
\caption{Sum-rate versus coherence time for different beam training strategies including exhaustive search (blue), radar covariance-aided beam training without compensating for mismatches (green), and communication covariance-aided beam training after ML-based mapping of the radar covariance to the communication covariance (red).}
\label{}
\end{center}
\end{figure}
\end{example}

Other recent strategies for radar-aided communication exploit the radar information obtained with a joint radar and communication system to speed up beam training. For example, \cite{Liu2020TWC} considers the beam tracking stage of a vehicular communication system, exploiting a BS operating as a radar to estimate the angle, distance and velocity of a 
given vehicle. The beamformers are then updated based on the predicted angle for  the communication link, providing a relevant performance gain with respect to other benchmarks. In a similar setting, \cite{Yuan2021TWC} proposes a variation of this idea, exploiting the delay and Doppler parameters estimated at the BS for the link corresponding to a particular vehicle. Then, a message passing algorithm based on factor graphs is derived to estimate the unknown range, speed, angle of arrival and path loss, to finally leverage the angular information for beam tracking as in \cite{Liu2020TWC}.

\subsubsection{Vision-aided beam training}
The images obtained from cameras can be exploited to infer decisions related to beam selection for mmWave systems, which is motivated by the recent development of deep learning powered computer vision methods such as object detection and 3D scene reconstruction \cite{Tian2021OJCS}. Moreover, the advancement of camera technology allows the adoption of high resolution cameras at both \acp{BS} and \acp{UE} at a low cost. The features extracted from the images may indicate the locations of the \acp{UE} and reflectors which would also help determine the existence of the LoS path. The images can be fused with other information, such as the position of the \acp{UE}, to enhance the beam selection performance. Furthermore, images can be useful to track the users in dynamic environments, which would be beneficial for beam prediction. Vision-aided beam training methods have the potential to reduce or even completely mitigate the beam training overhead incurred on the communication system.

One of the interesting works considers collecting multiple images in the vicinity of the mmWave \ac{BS} \cite{Xu2020WCL}. The collected images are used to reconstruct the 3D environment which is the input of a deep neural network along with the position of the \ac{UE}. The output of the network indicates the selected beam index. The takeaway from this work is that the reconstructed 3D environment contains spatial information such as the locations of the scatterers. In \cite{Tian2021OJCS}, the authors study the beam prediction problem leveraging previous beam indices and corresponding images in a dynamic vehicular scenario where the roadside units have cameras to the \acp{BS}. The ray tracing-based simulation results show that several future beams can be accurately predicted with various deep learning architectures such as LSTM. Since these algorithms utilize deep learning methods, it is imperative to have a dataset that contains both communication channels and corresponding images. Although one such dataset is provided in \cite{Alrabeiah2020VTCa}, where the authors obtained the images of the scenes from ray tracing environments, creating datasets in realistic environments is an open challenge.

Another approach is to exploit the cameras mounted on the \acp{UE}, which is more suitable to vehicular settings. It is shown that it is possible to detect the surrounding objects and vehicles if the images obtained from the cameras mounted on an autonomous vehicle are exploited in \cite{Xu2023TWC}. This approach not only provides the beam decisions, but also predicts the duration of the beam coherence interval. The simulation results obtained by using a 3D modeling and a ray tracing software show that utilizing the images taken at the \ac{UE} is more beneficial than leveraging the cameras at the \ac{BS}. 

\subsubsection{Exploiting LIDAR point clouds and multimodal information}

Sensors like LIDAR or IMU have also been leveraged to reduce beam training overhead \cite{Klautau2019WCL,Zecchin2022TVT,Ali2021Access,Mashhadi2021WCL}. The exploitation of these sensors is particularly interesting in cellular networks supporting vehicles or robots.
The different beam training strategies aided by LIDAR use the LIDAR point cloud to identify potential obstacles in the environment. Proposed designs create different features from the LIDAR point cloud and other potential data. For example, in \cite{Klautau2019WCL}, this feature is created from the positions of the BS and UE, the coverage area and the LIDAR point cloud. This information is the input to a convolutional network that decides about the channel state (LOS or NLOS) and recommends a reduced set of beams for beam training. Subsequent works introduced innovations that improve performance in complex NLOS cases. For example, the approach in \cite{Zecchin2022TVT} considers a curriculum learning strategy that trains with LOS measurements first, and then gradually introduces NLOS observations, resulting in an improved beam classification accuracy. In \cite{Mashhadi2021WCL}, the main novelty is the introduction of a federated learning framework such that a set of connected vehicles use their local LIDAR data to fine tune a shared neural network for beam selection. Since the same network can be utilized by vehicles that join the network later, the overhead is further reduced. LIDAR information is also commonly exploited in a multimodal fashion, combined with position information from GPS, images from cameras, or depth maps \cite{Dias2019SPAWC,Xu2021TCOM,Salehi2022TVT,Zhang2022INFOCOM}. 

\subsection{Blockage prediction and management}
Wireless communication systems often experience blockages that could lead to unreliable communication links, even link failure. This problem is more pronounced at \ac{mmWave} and THz bands where the \ac{LoS} path is usually much stronger than the other paths. Work has been done to characterize the impact of blockage, especially at millimeter wave frequencies. There have been many measurements of pathloss in the presence of blockage and specific studies describing the impact of a blocked link. For example, pathloss for LOS and NLOS in urban areas can be characterized by different exponents \cite{RappaportEtAlWidebandMillimeterwavePropagationMeasurements2015} possibly with additional corner losses \cite{KarttunenEtAlSpatiallyConsistentStreetbystreetPath2017}. In terms of specific blocking objects, vehicles may incur $20$ dB of attenuation \cite{BobanEtAlMultibandVehicletovehicleChannelCharacterization2019}, people $40$ dB  \cite{MacCartneyEtAlMillimeterwaveHumanBlockage732016}, and hand $20$ dB \cite{RaghavanEtAlStatisticalBlockageModelingRobustness2019}. The impact the blockage and the duration depends critically on the mobility of the transmitter, receiver, and blockage as well as the corresponding distances, with impact generally increasing with frequency \cite{BobanEtAlMultibandVehicletovehicleChannelCharacterization2019}. It seems likely the trend will continue for frequencies above $100$ GHz.  

In terms of blockage management, 5G NR primarily relies on high \ac{BS} density to facilitate macro-diversity and increase the probability that at least one link is unblocked. Its operation is primarily reactive, e.g. repeating the beam training phase after a link failure. In particular, it does not make special accommodations for the impact of blockage despite the potential for harvesting a tremendous amount of environmental information that could be available in a network with integrated sensing and communications. 

One of the earliest works on sensor-aided blockage management utilizes RGB and depth (RGB-D) cameras that are located close to the \ac{mmWave} \acp{BS} to predict the human blockages \cite{Oguma2015Globecom}. This work employs a centralized proactive \ac{BS} selection algorithm which is shown to be effective with experimental results. The work in \cite{Alrabeiah2020VTC} utilizes a dual-band \ac{BS} supporting both sub-6GHz and mmWave frequencies at the roadside unit of a vehicular communication system. Additionally, the roadside unit is equipped with a camera to improve the blockage detection performance. The image feed provided by the camera and the \ac{CSI} from the sub-6GHz system are used as the input of a deep neural network for blockage detection. Ray tracing-based simulation results show that the blockages can be detected with high accuracy.
In \cite{Charan2021TVT}, a bimodal machine learning algorithm that makes use of the temporal beam index sequence and the image feed from a camera equipped at the \ac{BS} is used to predict future blockages with the help of a deep learning model that detects objects in the images. Furthermore, this work also describes  a \ac{BS} handover framework that is shown to be effective in ray tracing-based simulations.
The work in \cite{Reus-Muns2021MSN} utilizes the GPS information at the vehicles and images provided by the camera at the \ac{mmWave} \ac{BS} to detect blockages. A feature extraction and a fusion network are used to process the vehicle location and camera images to detect the blockages at the edge.

LIDAR is another sensor that can provide physical information related to the environment, which can be useful for blockage prediction as shown in \cite{Marasinghe2021Globecom} where a LIDAR is used to scan an indoor environment. Trajectories of the users are predicted with an LSTM network. Then, the trajectories are used to construct the 3D environment that can be used to trace the rays, referred as ray casting, which are converted to the human blockage predictions.
Similarly, the work in \cite{Nerini2023WCL} proposes to convert the LIDAR data to 3D scenes. The created scene is used for the ray tracing simulations to obtain the path information between the \ac{BS} and the \ac{UE}, which is converted to an initial channel estimate and blockage detection without any communication overhead.
In \cite{Zhang2022INFOCOM}, an experimental setup with a dual-band \ac{BS} unit that is also equipped with a camera and a LIDAR. The authors use the camera images if the environment is bright, and the LIDAR output if it is dark. Object detection algorithms and a recurrent neural network are utilized for blockage prediction. The experimental results show that successful frequency handover is achieved with the described system.   
In \cite{Wu2023OJCS}, a LIDAR is equipped alongside a \ac{mmWave} \ac{BS} at the roadside unit of a vehicular communication system. The LIDAR data and power sequence, i.e., received power sequence obtained with the transmitted beams, are used as the inputs of a deep neural network to predict blockages. The experimental results verify the effectiveness of the considered approach.

Finally, the work in \cite{Demirhan2022ICC} uses range-angle maps obtained with the radar at the roadside unit. The range-angle maps are utilized as the input of a LSTM network to predict blockages. The authors show the effectiveness of the proposed approach with an experimental setup.

All this recent work shows that blockage detection and proactive blockage prediction can be achieved with the aid of sensors in mmWave systems. Moreover, machine learning-based solutions are shown to be effective for converting the data obtained from the sensor(s) to blockage detection and/or prediction.

\section{Conclusion}
Cellular networks are integrating sensing and communication functionalities. This integration enables new sensing services and also provides data which can be exploited to enhance network operation. This article has presented a vision of the ISAC cellular network and a comprehensive overview of methods for network sensing and also sensing-aided communication.  Regarding technologies for network sensing, we have introduced strategies for bistatic and multistatic sensing---including positioning as a special case---, monostatic sensing, and also specific designs to operate with wide apertures.  We have also described approaches for radio SLAM in bistatic and monostatic scenarios. In the overview of the opposite setting, network operation assisted by sensing, we have focused on the problems of sensor aided beam training for overhead reduction and  blockage prediction and management. We have introduced multiple examples which illustrate the performance that these different technologies can achieve in practical scenarios. We have made the case that communication and sensing must be considered together and that this integration will be one of the key features of 6G.

\section{Acknowledgements}
The authors would like to thank Dr. Satyam Dwivedi from Ericsson research for his invaluable feedback and discussions related to the evolution of sensing services in cellular networks and the current discussions on sensing use cases and prioritizations  in  the context of 3GPP standardization for 6G.

%


\end{document}